\documentclass[11pt]{article}

\usepackage{fancyheadings}
\usepackage{times}
\usepackage{fullpage}
\usepackage{amsmath}
\usepackage{amsbsy}
\usepackage{amsfonts}
\usepackage{supertabular}
\usepackage{graphicx}
\usepackage{subfigure}
\usepackage{natbib}
\usepackage{epsfig}
\usepackage{xspace}
\usepackage{euscript}
\newcommand{\mbs}[1]{\boldsymbol{#1}}
\newcommand{\mbb}[1]{\mathbb{#1}}

\def\ba{{\mbs{a}}} \def\bb{{\mbs{b}}} 
  
\def\bg{{\mbs{g}}}  
 \def\bk{{\mbs{k}}} 
\def\bm{{\mbs{m}}}  
  
\def\bs{{\mbs{s}}}  
  \def\bx{{\mbs{x}}}
\def\by{{\mbs{y}}}

\def\bg{{\mbs{g}}}
\def\blambda{{\mbs{\lambda}}}

\title{An Exactly Solvable Phase-Field Theory of Dislocation Dynamics,
Strain Hardening and Hysteresis in Ductile Single Crystals}

\author{
        {\bf M.~Koslowski${}^\dagger$, A.~M.~Cuiti\~no${}^\ddagger$ and
        M.~Ortiz${}^\dagger$} \\[.5em]
        ${}^\dagger$Graduate Aeronautical Laboratories \\
        California Institute of Technology \\
        Pasadena, CA 91125, USA \\[.5em]
        ${}^\ddagger$Department of Mechanical and Aerospace Engineering \\
        Rutgers University \\
        Piscataway, NJ 08854, USA
        }

\date{September 14, 2001 \\ 
      Submitted to: {\it Journal of the Mechanics and Physics of Solids}}

\begin{document}
\maketitle

\begin{abstract}
An \emph{exactly solvable} phase-field theory of dislocation
dynamics, strain hardening and hysteresis in ductile single
crystals is developed. The theory accounts for: an arbitrary
number and arrangement of dislocation lines over a slip plane;
the long-range elastic interactions between dislocation lines;
the core structure of the dislocations resulting from a piecewise
quadratic Peierls potential; the interaction between the
dislocations and an applied resolved shear stress field; and the
irreversible interactions with short-range obstacles and lattice
friction, resulting in hardening, path dependency and hysteresis.
A chief advantage of the present theory is that it is
analytically tractable, in the sense that the complexity of the
calculations may be reduced, with the aid of closed form
analytical solutions, to the determination of the value of the
phase field at point-obstacle sites. In particular, no numerical
grid is required in calculations. The phase-field representation
enables complex geometrical and topological transitions in the
dislocation ensemble, including dislocation loop nucleation,
bow-out, pinching, and the formation of Orowan loops. The theory
also permits the consideration of obstacles of varying strengths
and dislocation line-energy anisotropy. The theory predicts a
range of behaviors which are in qualitative agreement with
observation, including: hardening and dislocation multiplication
in single slip under monotonic loading; the Bauschinger effect
under reverse loading; the fading memory effect, whereby reverse
yielding gradually eliminates the influence of previous loading;
the evolution of the dislocation density under cycling loading,
leading to characteristic `butterfly' curves; and others.

\end{abstract}

\newpage
\section{Introduction}
\label{Introduction}

This paper is concerned with the formulation of an \emph{exactly
solvable} phase-field theory of dislocation dynamics, strain
hardening and hysteresis in ductile single crystals. The present
theory is an outgrowth of the phase-field model of
crystallographic slip of Ortiz \cite{Ortiz1999}, which was
restricted to a single dislocation loop moving through point
obstacles of uniform strength and was based on a line-tension
approximation. The present theory accounts for: an arbitrary
number and arrangement of dislocation lines over a slip plane;
the long-range elastic interactions between dislocation lines;
the core structure of the dislocations described by means of a
Peierls potential; the interaction between the dislocations and
an applied resolved shear stress field; and the irreversible
interactions with short-range obstacles and lattice friction,
resulting in hardening, path dependency and hysteresis.

A chief advantage of the present theory is that it is
analytically tractable, in the sense that the complexity of the
calculations may be reduced, with the aid of closed form
analytical solutions, to the determination of the value of the
phase field at point-obstacle sites. In particular, \emph{no
numerical grid is required in calculations}. On the sole basis of
the value of the phase field at the obstacle sites, the theory
reconstructs, analytically and in closed form, the location
of---possibly large numbers of---discrete dislocation lines, often
in complex arrangements and undergoing intricate topological
transitions, and endows each dislocation line with a well-defined
core. Furthermore, the theory characterizes the equilibrium
configurations of the dislocation ensemble directly, with the
result that no short transients associated with artificial
kinetics need to be resolved in calculations. By virtue of these
attributes, the theory would appear to be advantageous in
comparison to other recent phase-field models of dislocation
dynamics \cite{Wang:2001,Ghoniem:2000}, which require the use
 of large computational
grids and time integration over large numbers of time steps.

For definiteness, in this paper we restrict our attention to
systems consisting of a dislocation ensemble moving within a
single slip plane through a random array of discrete forest
dislocations under the action of an applied shear stress. In the
present theory, the dislocation ensemble populating the slip
plane is represented by means of a scalar phase field.
Specifically, the value of the phase field at a point of the slip
plane simply records the extent of slip in \emph{quanta} of
Burgers vector. The phase field is, therefore, integer valued,
and its value at a point may alternatively be regarded as
recording the number of dislocations, with proper accounting for
their sign, which have crossed the point. Thus, the individual
dislocation lines may be identified with the lines over which the
phase field jumps by one. The phase-field approach furnishes a
simple yet efficient means of representing arbitrary dislocation
geometries, possibly involving very large numbers of individual
dislocations, as well as enabling the tracking of intricate
topological transitions, including loop nucleation, pinching and
the formation of Orowan loops.

The central objective of the theory is to characterize the
evolution of the phase field. The theory is couched within a
general variational framework for dissipative systems developed by
Ortiz {\it et al.} \cite{OrtizRepetto1999, RadovitzkyOrtiz1999,
OrtizRepettoStainier2000, OrtizStainier1999,
KaneMarsdenOrtizWest:2000}, and accounts for energetic and kinetic
effects. The energy terms contemplated by the theory include: the
core energy of the dislocations, represented by a
piecewise-quadratic Peierls potential \cite{OrtizPhillips:1999};
the long-range elastic interactions between primary dislocations
and between primary and forest dislocations; and the energy of
interaction with the applied resolved shear stress. The
particular piecewise quadratic form of the Peierls potential
adopted here lends itself to an effective analytical treatment
based on the Fourier transform \cite{OrtizPhillips:1999}. The
main effect of the Peierls potential is to endow the dislocations
with a well-defined core energy and to coarse-grain the slip
plane by suppressing wavelengths shorter than the lattice
parameter.

At zero temperature and in the absence of irreversible processes,
the equilibrium configurations of the dislocation ensemble follow
directly from energy minimization. The resulting variational
problem is strongly \emph{nonlinear} owing to the all-important
constraint that the phase field be an integer-valued function.
Furthermore, the minimization problem is \emph{nonconvex} owing
to the multiwell structure of the Peierls potential. In addition,
the variational problem is \emph{nonlocal}, owing to the presence
of long-range elastic interactions. These attributes render the
energy minimization problem mathematically non-trivial and confer
its solutions a rich structure. Despite these difficulties, the
choice of a piecewise quadratic Peierls potential lends the
problem analytical tractability and, remarkably, a general
solution to the problem can be found analytically in closed form.

In the present theory, hysteresis arises from the assumed
irreversible short-range interactions between primary and forest
dislocations and from lattice friction. The strength of some of
these interactions has recently been investigated using atomistic
and continuum models \cite{baskes:1998, RodneyPhillips1999,
ShenoyKuktaPhillips2000}. Thus, we assume that the crossing of a
forest-dislocation site by a primary dislocation always costs
energy, regardless of the direction of the crossing. In the
context of a phase-field representation, this is tantamount to
assuming that any variation in the phase field at sites occupied
by obstacles requires the supply of a certain amount of work,
regardless of the sign of the variation. Physically, this work
may be identified with the energy required to dissolve the
dislocation--obstacle reaction product. Evidently, this form of
interaction is irreversible and dissipative, and thus cannot be
described by means of an energy function. It is possible,
however, to develop an incremental variational framework which
characterizes the evolution of the system by means of a sequence
of minimization problems \cite{OrtizRepetto1999,
RadovitzkyOrtiz1999, OrtizRepettoStainier2000, OrtizStainier1999,
KaneMarsdenOrtizWest:2000}. The function to be minimized over a
given step includes the incremental work of dissipation incurred
as a result of obstacle crossings, and depends on the state of
the system at the beginning of the step, which results in
path-dependent and hysteretic behavior.

The theory predicts a range of behaviors which are in qualitative
agreement with observation, including: hardening and dislocation
multiplication in single slip under monotonic loading; the
Bauschinger effect under reverse loading; the fading memory
effect, whereby reverse yielding gradually eliminates the
influence of previous loading. Simultaneously with the
deformation and hardening characteristics of the system, the
theory naturally predicts the evolution of the dislocation
density. In particular, no independent equation of evolution for
the dislocation density needs to be supplied.

\section{Dislocation energies}
\label{Sec:DislocationEnergies}

In this section we derive explicit expressions for the energy of a
dislocation ensemble contained within a slip plane in an elastic
crystal. The dislocation ensemble is described by means of a
phase field defined over the slip plane. In addition to the
elasticity of the crystal, the slip plane is endowed with a
piecewise-quadratic Peierls interplanar potential. The elastic
interaction energy formulated in this section is a special case
of a general class of energies for continuously distributed
dislocation loops in isotropic elastic crystals derived by Ortiz
and Xu \cite{xu:1993}. Extensions to anisotropic crystals have
also been given by Xu \cite{Xu:2000}.

We consider a crystal undergoing deformations characterized by a
displacement field $u_i$. Following Kr\"oner \cite{Kroner:1958}, we
begin by decomposing the displacement gradient in the additive form:
\begin{equation}\label{Eq:BeBp}
u_{i,j} = \beta^e_{ij} + \beta^p_{ij}
\end{equation}
where $u_{i,j}$ is the displacement gradient, or distortion
field, and, here and subsequently, commas are used to denote
partial differentiation. For simplicity, we shall specifically
focus on crystallographic slip occurring on a single slip plane
$S$. Under these conditions, the plastic distortion
$\beta^p_{ij}$ is supported on $S$ and has the form
\begin{equation}\label{Eq:Bp}
\beta^p_{ij} = \delta_i \, m_j \, \delta_S
\end{equation}
where $\delta_i$ is the displacement jump across $S$, $\bm$ is
the unit normal to $S$, and $\delta_S$ is the Dirac distribution
supported on $S$. The elastic distortion $\beta^e_{ij}$ is
assumed to be continuous. Neither the elastic nor the plastic
distortion is required to be compatible, i.~e., to be a gradient.

We shall additionally assume that the energy of the dislocation
ensemble may be written as the sum of three terms: a core energy
expressible in terms of a Peierls interplanar potential; the
elastic interaction energy of the dislocations; and the energy of
interaction with the applied stress field. These assumptions lead
to the consideration of an energy functional of the form:
\begin{equation}\label{Eq:Energy}
E = \int_S \phi(\mbs{\delta}) dS + \int \frac{1}{2} c_{ijkl}
\beta^e_{ij} \beta^e_{kl} d^3x - \int_S t_i \delta_i dS \equiv
E^{\rm core} + E^{\rm int} + E^{\rm ext}
\end{equation}
Here, $\phi$ denotes the Peierls interplanar potential and the
first term in (\ref{Eq:Energy}) represents the misfit or core
energy of the dislocations. In the second term, which represents
the elastic interaction energy, $c_{ijkl}$ are the elastic moduli
of the crystal. Finally, the third term represents the interaction
energy between the dislocations and a self-equilibrated applied
stress field resulting in a distribution of tractions $t_i$ over
the slip plane. The aim now is to derive a phase-field
representation of each of the terms in (\ref{Eq:Energy}).

\subsection{Elastic interaction energy}

Suppose that the slip distribution $\delta_i$ over $S$, or
equivalently, $\beta^p_{ij}$, is prescribed. The corresponding
elastic interaction energy may then be computed as follows.
Insertion of (\ref{Eq:BeBp}) into (\ref{Eq:Energy}) and
minimization with respect to the displacement field yields the
equilibrium equations
\begin{equation}\label{Eq:Equilbrium}
(c_{ijkl} u_{k,l}),_j - (c_{ijkl}\beta^p_{kl}),_j = 0
\end{equation}
The corresponding displacement field is:
\begin{equation}\label{Eq:u}
u_k = - G_{ki} \star (c_{ijmn}\beta^p_{mn}),_j
\end{equation}
where $G_{ki}$ is the Green's function and $(\star)$ denotes the
convolution operator. The distortion field then follows as
\begin{equation}\label{Eq:Du}
u_{k,l} = - G_{ki,l} \star (c_{ijmn}\beta^p_{mn}),_j
\end{equation}
and the elastic distortion as
\begin{equation}\label{Eq:Be}
\beta^e_{kl} = - G_{ki,l} \star (c_{ijmn}\beta^p_{mn}),_j -
\beta^p_{kl}
\end{equation}
Finally, the elastic interaction energy $E^{\rm int}$ is obtained
by inserting (\ref{Eq:Be}) into the second term of
(\ref{Eq:Energy}).

A convenient explicit expression for $E^{\rm int}$ may be
obtained by recourse to the Fourier transform. Using the
convolution theorem, the Fourier transform of the elastic
distortion follows from (\ref{Eq:Be}) as
\begin{equation}\label{Eq:Behat}
\hat{\beta}^e_{kl} = \hat{G}_{ki} k_j k_l c_{ijmn}
\hat{\beta}^p_{mn} - \hat{\beta}^p_{kl}
\end{equation}
where a superposed ($\,\hat{}\,$) denotes the Fourier transform of
a function, $k_i$ is the wavenumber vector, and the Fourier
transform of the Green's function is determined by the relation:
\begin{equation}\label{Eq:Ghat}
\hat{G}_{ik}^{-1} = c_{ijkl} k_j k_l
\end{equation}
Finally, the interaction energy follows from an application of
Parseval's identity, with the result:
\begin{equation}\label{Eq:Eint}
E^{\rm int} = \frac{1}{(2\pi)^3} \int \frac{1}{2}
\hat{A}_{mnuv}(\bk) \hat{\beta}^p_{mn}(\bk)
\hat{\beta}^{p\ast}_{uv}(\bk) d^3k
\end{equation}
Here the symbol ($\ast$) denotes complex conjugation and we write
\begin{eqnarray}
\hat{A}_{mnuv}(\bk) &=& c_{klrs} c_{ijmn} c_{pquv} ( \hat{G}_{ki}
k_j k_l - c^{-1}_{klij} ) ( \hat{G}_{rp} k_q k_s - c^{-1}_{rspq}
) \label{Eq:A1} \\
&=& c_{mnuv} - c_{kluv} c_{ijmn} \hat{G}_{ki} k_j k_l
\label{Eq:A2}
\end{eqnarray}
Mura \cite{mura:1987} has shown that $E^{\rm int}$ can also be
expressed in terms of Nye's \cite{Nye:1953} dislocation density
tensor.  Indeed, it is readily verified from (\ref{Eq:Behat}) and
(\ref{Eq:Ghat}) that, if the plastic distortion $\beta^p_{ij}$ is
compatible, i.~e., if it is a gradient, then the elastic distortions,
and hence the elastic interaction energy, vanish identically.

In order to facilitate analysis we shall resort to several
simplifying assumptions. Thus, for the special case of an
isotropic crystal the Green's function reduces to
\begin{equation}\label{Eq:GhatIsotropic}
\hat{G}_{ij} = \frac{1}{2\mu} \left( \frac{2\delta_{ij}}{k^2} -
\frac{1}{1-\nu} \frac{k_i k_j}{k^4} \right)
\end{equation}
where $\mu$ is the shear modulus of the crystal, $\nu$ is its
Poisson's ratio, and we write $k = |\bk|$. The interaction energy
(\ref{Eq:Eint}) then simplifies to
\begin{equation}\label{Eq:EintIsotropic}
E^{\rm int} = \frac{1}{(2\pi)^3} \int \frac{\mu}{2} \left\{ [ 1 -
(\mbs{\eta}\cdot\bm)^2 ] |\hat{\mbs{\delta}}|^2 -
(\hat{\mbs{\delta}}\cdot\mbs{\eta})^2 + \frac{2}{1-\nu}
(\hat{\mbs{\delta}}\cdot\mbs{\eta})^2 (\mbs{\eta}\cdot\bm)^2
\right\} d^3k
\end{equation}
where we write $\eta_i = k_i/k$. In addition, we shall adopt the
constrained displacement hypothesis of Rice \cite{Rice1992,
SunBeltzRice1993} according to which displacements, and the
attendant shear resistance, take place predominantly in the
direction of the dominant Burgers vector, i.~e.,
\begin{equation}\label{Eq:ConstrainedSlip}
\delta_i(\bx) = \delta(\bx) \, s_i
\end{equation}
where $\bs = \bb/|\bb|$ is a constant unit vector and $\delta$ is
a scalar function. For certain crystals, this conjecture has
found support in atomistic calculations
\cite{YamaguchiVitekPope1981, SunRiceTruskinovsky1991,
JuanKaxiras1996, JuanSunKaxiras1996}. By virtue of these
assumptions, (\ref{Eq:EintIsotropic}) simplifies to
\begin{equation}\label{Eq:EintIsotropic2}
E^{\rm int} = \frac{1}{(2\pi)^3} \int \frac{\mu}{2} \left(
\eta_2^2 + \frac{2}{1-\nu} \eta_1^2 \eta_3^2 \right) |
\hat{\delta} |^2 d^3k
\end{equation}
where we have chosen axes such that the slip plane is parallel to
the $(x_1, x_2)$-plane and the Burgers vector points in the
direction of the $x_1$-axis. Finally, for the particular case in
which the slip distribution is confined to the plane $x_3 = 0$,
the elastic interaction energy further reduces to
\begin{equation}\label{Eq:EintIsotropicSinglePlane}
E^{\rm int} = \frac{1}{(2\pi)^2} \int \frac{\mu}{4} \left(
\frac{k_2^2}{\sqrt{k_1^2 + k_2^2}} + \frac{1}{1-\nu}
\frac{k_1^2}{\sqrt{k_1^2 + k_2^2}} \right) |\hat{\delta}|^2 d^2k
\end{equation}
which is the sought expression.

By way of a simple illustrative and verification example we may
consider the case of an edge dislocation pile-up, for which the
slip distribution is of the form
\begin{equation}
\hat{\delta} = 2\pi b \, \delta(k_2)f(k_1)
\end{equation}
whence the elastic energy per unit length of dislocation follows
from (\ref{Eq:EintIsotropicSinglePlane}) as
\begin{equation}\label{Eq:EintIsotropicEdgePileUp}
\frac{E^{\rm int}}{L} = \frac{\mu b^2}{4 \pi (1-\nu)}
\int_{-\infty}^\infty |k_1| |\hat{f}(k_1)|^2 dk_1
\end{equation}
For a single Volterra dislocation at the origin, $\hat{f}(k_1) =
1/ik_1$ and the elastic energy per unit length evaluates to
\begin{equation}\label{Eq:EintIsotropicEdge}
\frac{E^{\rm int}}{L} = \frac{\mu b^2}{4 \pi (1-\nu)}
\int_{\pi/R}^{\pi/r_0} \frac{dk_1}{k_1} = \frac{\mu b^2}{4\pi
(1-\nu)} \log \frac{R}{r_0}
\end{equation}
where, in order to avoid logarithmic divergences, we have
introduced lower and upper cutoff radii. We verify that
(\ref{Eq:EintIsotropicEdge}) coincides with the well-known
expression for the energy per unit length of a straight edge
dislocation in an isotropic crystal \cite{hirth:1968}. For a screw
pile-up, a similar derivation yields:
\begin{equation}\label{Eq:EintIsotropicScrewPileUp}
\frac{E^{\rm int}}{L} = \frac{\mu b^2}{4 \pi}
\int_{-\infty}^\infty |k_2| |\hat{f}(k_2)|^2 dk_2
\end{equation}
For a single screw dislocation at the origin the dislocation
energy per unit length evaluates to
\begin{equation}\label{Eq:EintIsotropicScrew}
\frac{E^{\rm int}}{L} = \frac{\mu b^2}{4 \pi}
\int_{\pi/R}^{\pi/r_0} \frac{dk_2}{k_2} = \frac{\mu b^2}{4\pi}
\log \frac{R}{r_0}
\end{equation}
which again coincides with the classical result \cite{hirth:1968}.

\subsection{Core energy}

In the Peierls theory of the dislocation core, the interplanar
potential $\phi(\mbs{\delta})$ is identified with the energy per
unit area that results when two semi-infinite crystals are taken
through a relative rigid displacement $\delta_i$. From symmetry
considerations it follows that, in the absence of an applied
field, the energy of the crystal attains minima when the
displacement jump $\delta_i$ is an integral multiple of a Burgers
vector $b_i$ of the lattice, i.~e., at
\begin{equation}\label{Eq:Wells}
\delta_i = \xi \, b_i, \quad \xi \in {\mbb{Z}}
\end{equation}
where ${\mbb{Z}}$ denotes the set of all integer numbers. These
special slips determine the location of the wells of $\phi$. With
a view to enabling the application of Fourier transform methods,
we shall assume that $\phi$ is piecewise quadratic
\cite{OrtizPhillips:1999}. In this model, the interplanar potential is
taken to be of the form
\begin{equation}\label{Eq:Quadratic}
\phi(\mbs{\delta}) = \min_{\xi \in {\mbb{Z}}} \frac{1}{2} C_{jl}
(\delta_j - \xi b_j)(\delta_l - \xi b_l)
\end{equation}
The moduli $C_{jl}$ may be determined by equating
(\ref{Eq:Quadratic}) with the energy per interatomic plane of a
crystal undergoing a simple shear deformation of the form:
\begin{equation}\label{Eq:C1}
\beta_{ij} = \frac{1}{d} \delta_i m_j
\end{equation}
where $d$ is the interplanar distance. The result is:
\begin{equation}
C_{ik} = \frac{1}{d} c_{ijkl} m_j m_l
\end{equation}
By way of example we may consider the case of crystallographic
slip on a $\{111\}$-plane of an fcc crystal. Then a simple
calculation gives $d = a/\sqrt{3}$, with $a$ the cubic lattice
parameter, and
\begin{equation}\label{Eq:Cfcc}
C_{ik} = \frac{1}{\sqrt{3} a} (c_{11} + c_{44} - c_{12})
\delta_{ik}
\end{equation}
In this expression $c_{11}$, $c_{44}$ and $c_{12}$ are the three
independent cubic elastic moduli.

\begin{figure}
    \centerline{ \hbox{
    \subfigure[Interplanar potential]{
    \epsfig{file=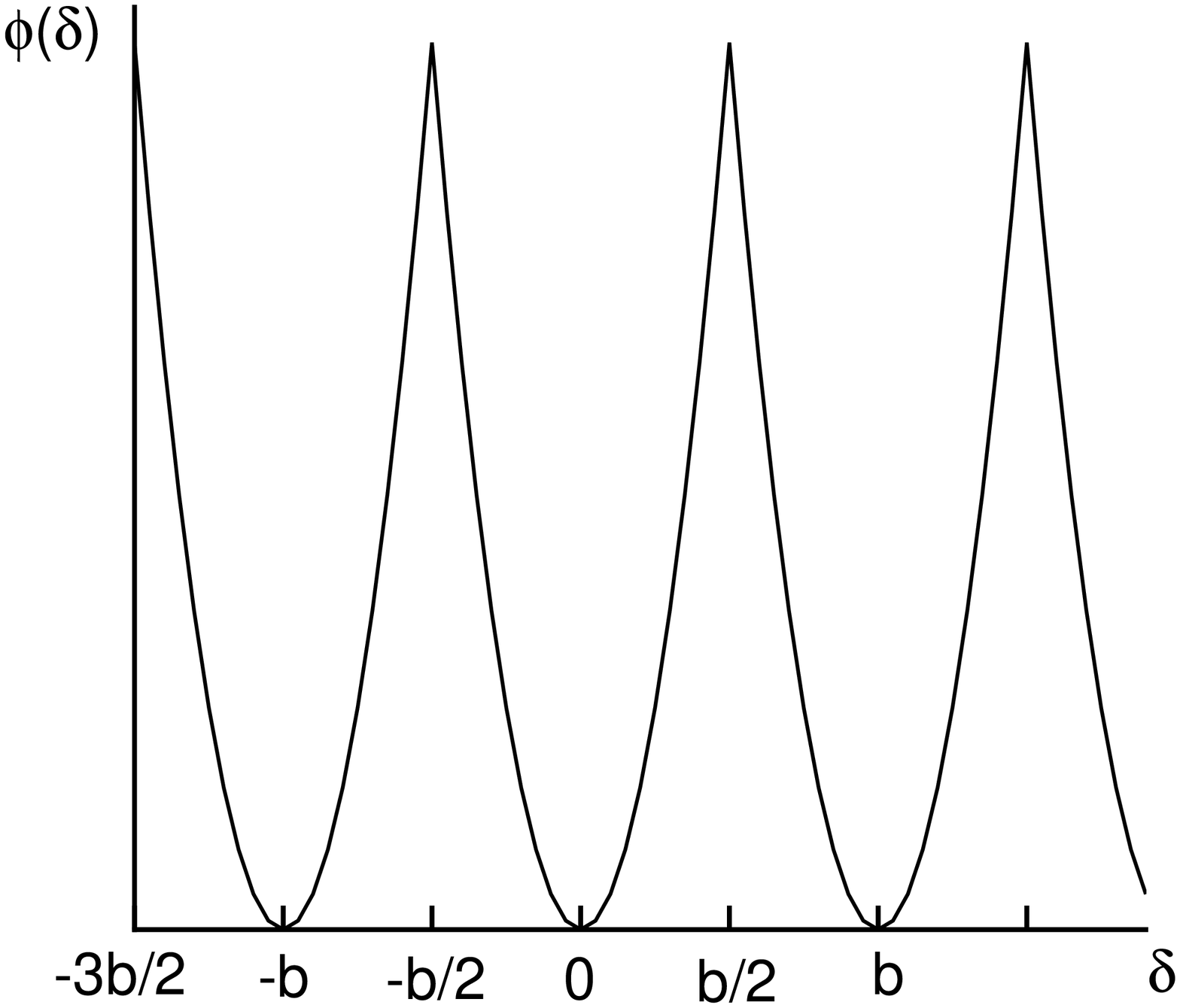,height=0.35\textheight} }
    \hglue -0.2in
    \subfigure[Resolved shear stress]{
    \epsfig{file=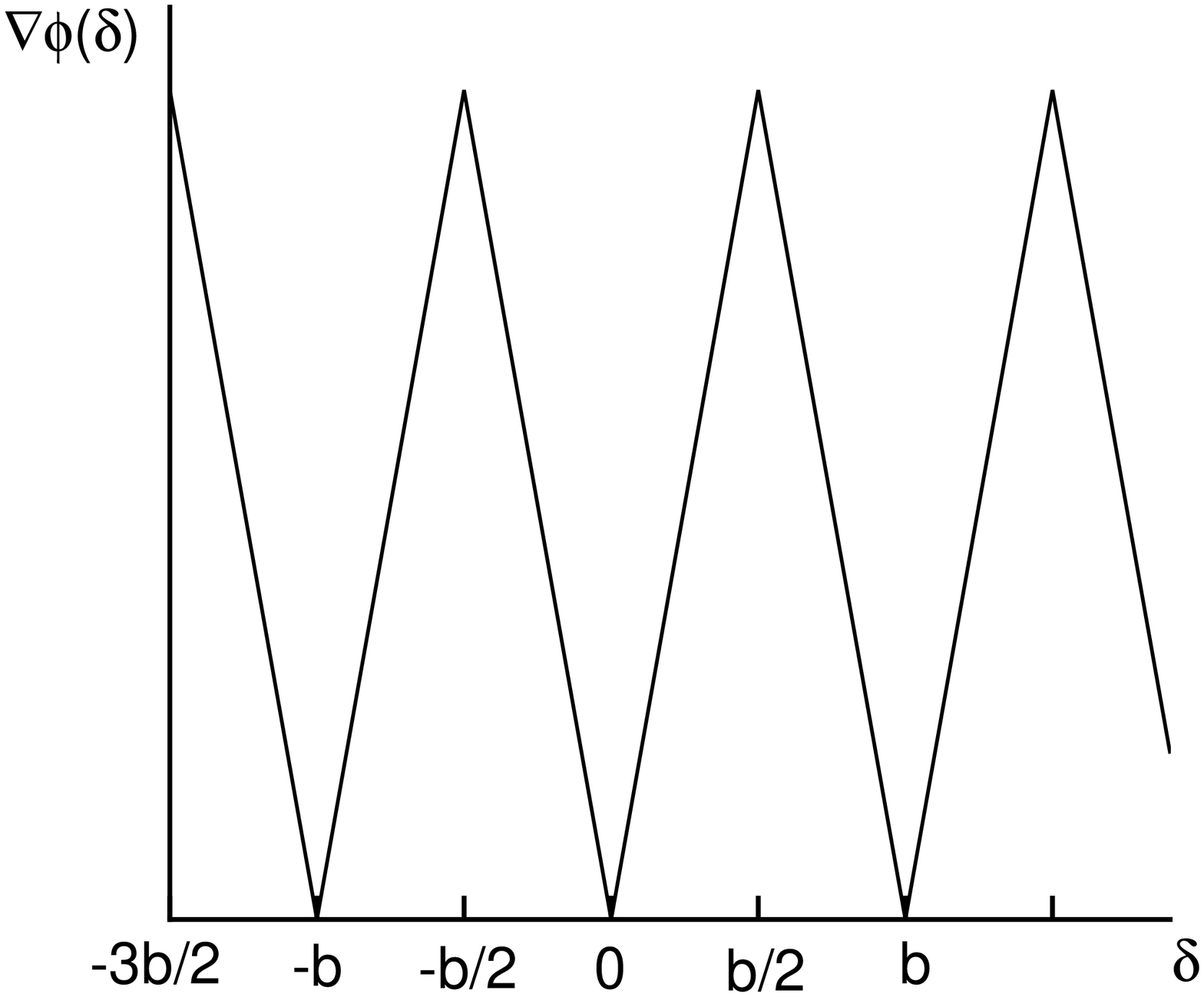,height=0.35\textheight} }}}
    \caption{Piecewise-quadratic interplanar potential and its
    derivative.}
    \label{fig:potential}
\end{figure}

If the slip is additionally constrained to take place in the
direction of the Burgers vector, as in
eq.~(\ref{Eq:ConstrainedSlip}), then the piecewise quadratic
interplanar potential (\ref{Eq:Quadratic}) reduces to the form
\begin{equation}\label{Eq:QuadraticConstrained}
\phi(\delta) = \min_{\xi \in {\mbb{Z}}} \frac{C}{2} | \delta - \xi
b |^2
\end{equation}
where $\delta = \mbs{\delta}\cdot \bs$ and
\begin{equation}\label{Eq:C2}
C = C_{ik} s_i s_k
\end{equation}
The function $\phi(\delta)$ and its derivative, which gives the
resolved shear stress as a function of slip, are shown in
Fig.~\ref{fig:potential}.

\subsection{Computation of the energy from the phase field}
\label{Sec:EnergyComputation}

At zero temperature, the stable configurations of the dislocation
ensemble may be identified with the relative minima of the energy
(\ref{Eq:Energy}). We begin by attempting a characterization of
the absolute mimimizers of (\ref{Eq:Energy}), and defer the more
ambitious program of understanding the relative energy minimizers
as well. Inserting (\ref{Eq:EintIsotropicSinglePlane}) and
(\ref{Eq:QuadraticConstrained}) into (\ref{Eq:Energy}) leads to
the energy:
\begin{equation}\label{Eq:EnergyZeta}
E[\zeta] = \inf_{\xi \in X} E[\zeta | \xi]
\end{equation}
where
\begin{equation}
\zeta = \delta/b
\end{equation}
is a normalized slip function, and
\begin{equation}\label{Eq:EnergyZetaXi}
E[\zeta | \xi] = \int \frac{\mu b^2}{2 d} | \zeta - \xi |^2 d^2x
+ \frac{1}{(2\pi)^2} \int \frac{\mu b^2}{4} K |\hat{\zeta}|^2
d^2k - \int b s \zeta d^2x
\end{equation}
In this expression,
\begin{equation}\label{Eq:Tau}
s = t_i m_i
\end{equation}
is the resolved shear stress field, and we write
\begin{equation}\label{Eq:K}
K = \frac{k_2^2}{\sqrt{k_1^2 + k_2^2}} + \frac{1}{1-\nu}
\frac{k_1^2}{\sqrt{k_1^2 + k_2^2}}
\end{equation}
The normalized slip distribution $\zeta$ of interest now follows
from the problem:
\begin{equation}\label{Eq:MinZeta}
\inf_{\zeta \in Y} E[\zeta]
\end{equation}
In definition (\ref{Eq:EnergyZeta}), the function $\xi$ may be
regarded as an integer-valued \emph{phase field} defined on the
slip plane $S$, with the property that $\xi(\bx)$ equals the
number of dislocations which have passed over the point $\bx \in
S$. Thus, the quantized phase field $\xi$ describes an ensemble of
perfect, or Volterra, dislocations on $S$. In writing
(\ref{Eq:EnergyZeta}) and (\ref{Eq:MinZeta}), $X \times Y$
denotes the space of slip distributions and integer-valued phase
fields of finite energy, i.~e., the linear space of functions
$\zeta: \mbb{R}^2 \to \mbb{R}$ and $\xi: \mbb{R}^2 \to \mbb{Z}$
such that $E[\zeta | \xi] < \infty$, which is the physically
relevant space of solutions.

Next we seek to express the energy of the dislocation ensemble
directly in terms of $\xi$. To this end, we begin by inserting
definition (\ref{Eq:EnergyZeta}) into the variational problem
(\ref{Eq:MinZeta}), with the result:
\begin{equation}\label{Eq:MinZetaXi}
\inf_{\zeta \in Y} \inf_{\xi \in X} E[\zeta | \xi]
\end{equation}
At this point, we may invert the order of minimization in
(\ref{Eq:MinZetaXi}), which results in the reduced minimum
problem:
\begin{equation}\label{Eq:MinXi}
\inf_{\xi \in X} E[\xi]
\end{equation}
where
\begin{equation}\label{Eq:EnergyConstrained3}
E[\xi] =  \inf_{\zeta\in Y} \left\{ \int \frac{\mu b^2}{2 d} |
\zeta - \xi |^2 d^2x + \frac{1}{(2\pi)^2} \int \frac{\mu b^2}{4}
K |\hat{\zeta}|^2 d^2k - \int b s \zeta d^2x \right\}
\end{equation}
This minimization leads to a linear problem in $\zeta$ which may
be solved analytically by an application of the Fourier transform.
To this end, we apply Parseval's identity to obtain:
\begin{equation}\label{Eq:EnergyConstrained4}
E[\xi] =  \inf_{\zeta \in Y} \left\{ \frac{1}{(2\pi)^2} \int
\left( \frac{\mu b^2}{2 d} | \hat{\zeta} - \hat{\xi} |^2 +
\frac{\mu b^2}{4} K |\hat{\zeta}|^2 - b \hat{s}^\ast \hat{\zeta}
\right) d^2k \right\}
\end{equation}
Minimization with respect to $\zeta$ gives:
\begin{equation}\label{Eq:EL}
\hat{\zeta} = \hat{\zeta}_0 + \frac{\hat{\xi}}{1 + K d/2}
\end{equation}
where
\begin{equation}
\hat{\zeta}_0 = \frac{d}{\mu b}\frac{\hat{s}}{1 + K d/2}
\end{equation}
is the slip distribution for $\xi=0$, i.~e., the slip
distribution due to the elasticity of the interatomic plane.
Insertion of (\ref{Eq:EL}) into (\ref{Eq:EnergyConstrained3})
gives the sought energy:
\begin{equation}\label{Eq:EnergyXi}
E[\xi] = E_0 + \frac{1}{(2\pi)^2} \int \left( \frac{\mu b^2}{4}
\frac{K}{1 + K d/2} |\hat{\xi}|^2 - \frac{b \hat{s} \hat{\xi}}{1
+ K d/2} \right) d^2k
\end{equation}
where
\begin{equation}\label{Eq:E0}
E_0 = \frac{1}{(2\pi)^2} \int \frac{d}{2\mu} \frac{\hat{s}^2}{1 +
K d/2} d^2k
\end{equation}
is independent of $\xi$ and represents the elastic energy of the
interatomic in the absence of dislocations, i.~e., at $\xi=0$.
Thus, for instance, if $s = \tau = \text{constant}$,
(\ref{Eq:E0}) corresponds to an elastic energy density per unit
area of $(d/2\mu) \tau^2$, which indeed coincides with the elastic
energy density of the interatomic plane. Therefore, the excess
energy
\begin{equation}
E^{\rm dis}[\xi] = E[\xi] - E_0
\end{equation}
may be identified with the energy proper of the dislocation
ensemble. By virtue of (\ref{Eq:EnergyXi}), this energy is now
expressed entirely in terms of the phase field $\xi$. If is
evident from the form of (\ref{Eq:EnergyXi}) that, in the absence
of an applied field, the dislocation energy vanishes for uniform
slip, as required by symmetry. Indeed, in this case $\hat{\xi} =
n \delta_D(\bk)$, where $n \in \mbb{Z}$ and $\delta_D(\bk)$
denotes the Dirac delta at $\bk = {\bf 0}$, and, consequently, $K
\hat{\xi}$ vanishes identically in the sense of distributions.

As already noted, the phase field $\xi$ in (\ref{Eq:EnergyXi})
takes integer values and, therefore, represents a distribution of
Volterra or perfect dislocations. It is evident from the form of
(\ref{Eq:EnergyXi}) that the Peierls potential has the effect of
regularizing linear elasticity so as to render the energy of
Volterra dislocations finite. The lattice parameter is retained
in the energy functional (\ref{Eq:EnergyXi}) through the
interplanar distance $d$. The effect of the core regularization
resides in the factor
\begin{equation}\label{Eq:Mollifier}
\hat{\varphi}_d(\bk) = \frac{1}{1 + K d/2}
\end{equation}
appearing in both terms of the energy (\ref{Eq:EnergyXi}). This
factor tends to suppress wavelengths on the scale of the lattice
parameter or shorter. The same factor relates the integer-valued
phase field $\xi$ to the core-regularized slip distribution
$\zeta$, eq.~(\ref{Eq:EL}). In this relation, the function
$\varphi_d(\bx)$ acts as a mollifier: the integral of
$\varphi_d(\bx)$ over the entire plane is one, or, equivalently,
$\hat{\varphi}_d({\bf 0}) = 1$; and $\varphi_d(\bx)$ defines a
Dirac-delta sequence, i.~e.,
\begin{equation}\label{Eq:DiracSequence}
\lim_{d\to 0}\varphi_d(\bx) = \delta_D(\bx)
\end{equation}
in the sense of distributions. In real space eq.~(\ref{Eq:EL})
takes the convolution form:
\begin{equation}\label{Eq:EL2}
\zeta = \zeta_0 + \varphi_d \star \xi
\end{equation}
which shows that the effect of the piecewise-quadratic Peierls
interplanar potential is to smooth out the phase field $\xi$. A
consequence of this smoothing is that the perfect dislocations
described by $\xi$ acquire a core of width commensurate with $d$.

It should be carefully noted that, despite the quadratic
appearance of the energy functional (\ref{Eq:EnergyXi}), the
attendant variational problem is strongly \emph{nonlinear} owing
to the all-important constraint that $\xi$ be an integer-valued
function. Furthermore, the minimization problem is
\emph{nonconvex} owing to the lack of convexity of the set $X$.
This nonlinear and nonconvex structure of the variational problem
is inherited from the---similarly nonlinear and
nonconvex---structure of the Peierls potential
(\ref{Eq:QuadraticConstrained}). In addition, the variational
problem is \emph{nonlocal}, owing to the presence of long-range
elastic interactions. These attributes render the problem
(\ref{Eq:MinXi}) mathematically non-trivial and confer its
solutions a rich structure. Despite these difficulties, the
choice of a piecewise quadratic Peierls potential lends the
problem analytical tractability, as demonstrated subsequently.

\subsection{Illustrative examples}
\label{Sec:Examples}

\begin{figure}
\begin{center}
\epsfig{file=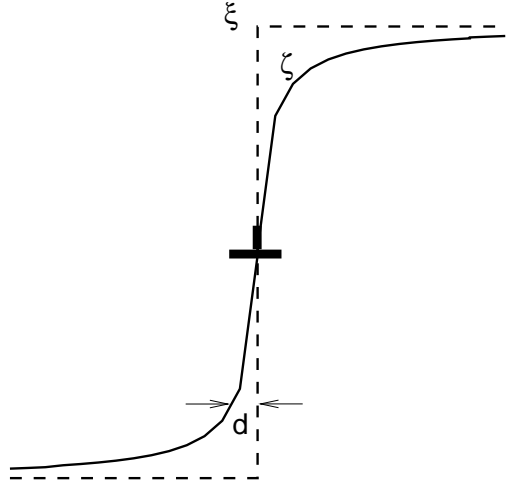,height=0.35\textheight} \caption{Core
structure of an infinite straight dislocation predicted by the
piecewise quadratic model.} \label{fig:dis}
\end{center}
\end{figure}

As an illustration of the type of core structure predicted by the
piecewise quadratic model, we may consider the case of a general
straight dislocation. For simplicity, we restrict our attention
to the isotropic case. It is convenient to introduce auxiliary
orthonormal axes $(x'_1, x'_2)$ with $x'_1$ normal to the
dislocation line. In this coordinate frame one has:
\begin{equation}\label{Eq:CoreXi}
\hat{\xi} =  \frac{2 \pi \delta_D(k'_2)}{ i k'_1}
\end{equation}
corresponding to uniform slip over the half-plane $x'_1 < 0$.
Here, $\delta_D$ denotes the Dirac-delta distribution, and $i$ is
the unit imaginary number. A straightforward calculation gives
the slip distribution as
\begin{equation}\label{Eq:CoreZeta}
\hat{\zeta} = \frac{ 2 \pi \delta_D(k'_2)}{i k'_1(1 + c k')}
\end{equation}
where
\begin{equation}\label{Eq:CoreC}
c = \left(\sin^2\theta + \frac{\cos^2\theta}{1-\nu} \right)
\frac{d}{2}
\end{equation}
is a characteristic core width, and $\theta$ is the angle
subtended by the normal to the dislocation line and the Burgers
vector. Thus, the case $\theta = 0$ corresponds to an edge
dislocation, whereas the case $\theta = \pi/2$ corresponds to a
screw dislocation. A plot of the core profile is shown in
Fig.~\ref{fig:dis}. In addition, the energy per unit dislocation
length follows in the form
\begin{equation}\label{Eq:CoreEnergy}
\frac{E}{L} = \frac{\mu b^2}{4\pi}\left(\sin^2\theta +
\frac{\cos^2\theta}{1-\nu} \right) \log\frac{R}{c}
\end{equation}
where $R$ is an upper cut-off radius. It is clear from this
expression that $c$ plays the role of a core cut-off radius. The
width of the core depends on the orientation of the dislocation
line relative to the Burgers vector, and it attains its minimum
(maximum) value for edge (screw) dislocations.

This example illustrates how the introduction of a Peierls
interplanar potential renders the dislocation core structure,
including the core width, well-defined. In particular, the high
wave number divergence in the dislocation energy is eliminated.
The specific core structure (\ref{Eq:CoreZeta}) is, of course, a
result of the assumed piecewise quadratic form of the Peierls
interplanar potential. It bears emphasis that here the interplanar
potential is regarded simply as a convenient device for
regularizing the equations of elasticity, and no attempt is made
to model real dislocation core structures. However, it seems
reasonable to expect that, for sufficiently well-spaced
dislocations, the only macroscopically relevant core parameter is
the core energy per unit length, and that the details of the core
structure play a limited role as regards the overall energetics of
the dislocation ensemble.

The Peierls potential introduces a lengthscale $d$ into the
formulation of the order of the crystal lattice parameter. The
behavior of dislocation loops may therefore be expected to differ
sharply according as to whether their size is much larger or,
contrariwise, much smaller, than $d$. These regimes are exhibited
in the simple example of a circular dislocation loop of radius
$R$. In this case, $\xi$ is the characteristic function of the
circle of radius $R$, e.~g., centered at the origin, and its
Fourier transform is
\begin{equation}\label{Eq:LoopXi}
\hat{\xi} = \int_0^R \int_0^{2\pi} {\rm e}^{- i k r \cos(\theta -
\phi)} r dr d\theta = 2 \pi \frac{R}{k}J_1(kR)
\end{equation}
where $J_1$ is the Bessel function of the first kind, and we have
written $(x_1, x_2) = (r \cos\theta, r \sin\theta)$ and $(k_1,
k_2) = (k \cos\phi, k \sin\phi)$. Insertion of (\ref{Eq:LoopXi})
into (\ref{Eq:EnergyXi}) gives
\begin{equation}\label{Eq:LoopEnergy}
E =  \frac{\mu b^2}{2 d \pi^{3/2}}\pi R^2
G^{3,2}_{2,4}\left (\frac{4 R^2}{d^2}
\left|\begin{array}{cccc}1/2 & 1/2 &   &  \\
0   & 1/2 & 1 & -1  \end{array} \right. \right)
\end{equation}
where $G^{3,2}_{2,4} $ is the Meijer's $G$ function \cite{Mathai:1973},
and we have set $\nu=0$ for simplicity. From the asymptotic
behavior of the Meijer function it follows that
\begin{equation}\label{Eq:SmallLoopEnergy}
E \sim \frac{\mu b^2}{d} \pi R^2, \qquad R \to 0
\end{equation}
and
\begin{equation}\label{Eq:BigLoopEnergy}
E \sim \frac{\mu b^2}{2}R  \log \frac{4R}{d} ,
\qquad R \to \infty
\end{equation}
As expected, for fixed $d$ the energy of the loop tends to zero as
$R \to 0$. Conversely, for fixed $R$, the energy diverges as $d
\to 0$, which corresponds to the limiting case of a perfect
Volterra loop in a linear elastic crystal. Once again, this
divergence illustrates the breakdown of linear elasticity in the
presence of perfect dislocations and the crucial role played by
the core regularization in eliminating that breakdown. The form
(\ref{Eq:BigLoopEnergy}) for the energy of large loops is
consistent with expressions derived from linear elasticity
\cite{hirth:1968}. However, it should be noted that the core
width $c = d/4$ implied by (\ref{Eq:BigLoopEnergy}) is predicted
by the present theory, which stands in contrast to the theory of
linear-elastic dislocations wherein the core cut-off radius is an
{\it ad-hoc} and extraneous parameter.

The issue of metastability and the existence of relative energy
minimizers may be illustrated simply by the stable configurations
of a dislocation dipole under the action of a resolved shear
stress $\tau$. In order to preclude uniform slip distributions,
$\xi$ may be required to decay to zero at infinity. Suppose that
slip occurs on the interval $-r/2 < x_1 < r/2$, where $r$ is
dipole width. Then, the corresponding phase field is
\begin{equation}\label{Eq:DipoleXi}
\hat{\xi} = \frac{2}{k_1} \sin \frac{k_1 r}{2}
\end{equation}
For purposes of the present discussion it suffices to consider the
asymptotic limit of $r \gg d$.  In this regime
(\ref{Eq:EnergyXi}) yields the relation
\begin{equation}
\frac{\partial E}{\partial r} \sim \frac{\mu b^2}{2\pi(1-\nu)}
\int_0^\infty \sin k_1 r dk_1 - b \tau = \frac{\mu
b^2}{2\pi(1-\nu)} \frac{1}{r} - b \tau, \quad r \to \infty
\end{equation}
and
\begin{equation}
\frac{\partial E}{\partial c} \sim - \frac{\mu b^2}{2\pi(1-\nu)}
\frac{1}{c}, \quad r \to \infty
\end{equation}
From these two expressions we find:
\begin{equation}\label{Eq:DipoleEnergy}
E \sim E_0 + \frac{\mu b^2}{2\pi(1-\nu)} \log \frac{r}{c} - b \tau
r, \quad r \to \infty
\end{equation}
for some constant $E_0$ independent of $r$ and $c$. Equilibrium
now demands:
\begin{equation}\label{Eq:DipoleEquilibrium}
\frac{\partial E}{\partial r} = 0 \Rightarrow \frac{\mu b^2}{2\pi
(1-\nu)} \frac{1}{r} = b \tau
\end{equation}
The dipole energy at equilibrium is, therefore,
\begin{equation}\label{Eq:DipoleEnergy2}
E \sim E_0 + \frac{\mu b^2}{2\pi (1-\nu)} \left( \log \frac{r}{c}
- 1 \right), \quad r \to \infty
\end{equation}
where $r$ is the solution of (\ref{Eq:DipoleEquilibrium}). The
preceding analysis shows that, in the regime under consideration,
there exist stable dipoles of arbitrary width $r$, with
Eq.~(\ref{Eq:DipoleEquilibrium}) supplying the resolved shear
stress which equilibrates the dipole. It is evident from
(\ref{Eq:DipoleEnergy2}) that, for sufficiently large $r$, the
dipole energy is positive and, therefore, greater than the
dislocation energy corresponding to a phase field $\xi=0$, which
is zero. Hence, the dipole configuration is metastable and
constitutes a relative minimum of the dislocation energy.

\section{Energy-minimizing phase fields}

The variational problem (\ref{Eq:MinXi}) which characterizes the
stable equilibrium configurations of the slip plane may be
regarded as a constrained minimization problem, in which the
unknown field $\xi$ is constrained to take integer values
everywhere on the slip plane.  This constraint is nonlinear and
nonconvex, which renders the problem (\ref{Eq:MinXi})
mathematically nontrivial. In addition, the long-range elastic
interactions between dislocations render the energy nonlocal.
These difficulties notwithstanding, in this section we obtain a
distinguished set of minimizers of the energy (\ref{Eq:EnergyXi})
analytically in closed form.

\subsection{Constrained minimizers and energy-norm projection}

A compelling characterization of the solutions of the constrained
minmization problem (\ref{Eq:MinXi}) may be derived as follows.
Let $\eta$ be the unconstrained energy minimizer. Thus, $\eta$ is
the solution of the unconstrained minimization problem:
\begin{equation}\label{Eq:EnergyUnconstrained}
\inf_{\eta \in Y} E[\eta]
\end{equation}
In particular, $\eta$ is allowed to take arbitrary real values.
Then, the constrained phase field is the solution of the problem
(e.~g., \cite{Clarke:1983}):
\begin{equation}\label{Eq:EnergyConstrained5}
\inf_{\xi\in X} \frac{1}{2} \parallel \xi - \eta \parallel^2
\end{equation}
where
\begin{equation}\label{Eq:EnergyNorm}
\parallel u \parallel = \sqrt{\langle u, u \rangle}
\end{equation}
is an energy norm corresponding to the inner product
\begin{equation}\label{Eq:InnerProduct}
\langle u, v \rangle = \frac{1}{(2\pi)^2} \int \frac{\mu b^2}{2}
\frac{K }{1 + K d/2} \, \hat{u}^* \, \hat{v} \, d^2k
\end{equation}
The solution of (\ref{Eq:EnergyConstrained5}) may be expressed as
\begin{equation}\label{Eq:Projection}
\xi = P_X \eta
\end{equation}
where $P_X$ denotes the closest-point projection of $Y$ onto $X$
in the sense of norm (\ref{Eq:EnergyNorm}).

The solution procedure leading to the core-regularized slip
distribution $\zeta$ may now be broken down into three steps. The
first step consists of the solution of the unconstrained
minimization problem (\ref{Eq:EnergyUnconstrained}), resulting in
the real-valued unconstrained minimizer $\eta$. The second step
consists of the solution of the constrained minimization problem
(\ref{Eq:EnergyConstrained5}), resulting in the projection of
$\eta$ onto its closest integer-valued phase-field $\xi$. The
third step consists of the smoothing of $\xi$ according to
(\ref{Eq:EL}), leading to the introduction of smooth cores around
all dislocation lines.

The exact evaluation of the projection
(\ref{Eq:EnergyConstrained5}) entails some difficulty. The
corresponding stationarity condition consists of the requirement
that the functional be stable with respect to a unit increment or
decrement of $\xi$ over an arbitrary region of the slip plane.
Thus, if $R \subset \mbb{R}^2$ is one such region and $\chi_R$
denotes its characteristic function, stationarity demands that the
inequalities
\begin{eqnarray}
\parallel \xi + \chi_R - \eta \parallel^2 & \geq &
\parallel \xi - \eta \parallel^2 \label{Eq:In1} \\
\parallel \xi - \chi_R - \eta \parallel^2 & \geq &
\parallel \xi - \eta \parallel^2 \label{Eq:In2}
\end{eqnarray}
be simultaneously satisfied. After some straightforward
manipulations (\ref{Eq:In1}-\ref{Eq:In2}) may be recast in the
form
\begin{equation}\label{Eq:Bounds}
\frac{1}{2} \parallel \chi_R \parallel^2 \, \geq \, \langle \xi -
\eta, \, \chi_R \rangle \, \geq \, - \frac{1}{2} \parallel \chi_R
\parallel^2
\end{equation}
These inequalities set bounds on the possible values of the
resultant of the residual tractions acting over arbitrary regions
of the slip plane.

We proceed to proof the remarkable result that $\xi(\bx)$
\emph{is the closest integer to} $\eta(\bx)$ pointwise on the
slip plane, i.~e.,
\begin{equation}\label{Eq:PZ}
\xi(\bx) = P_{\mbb{Z}} \eta(\bx), \quad \forall \bx\in\mbb{R}^2
\end{equation}
where $P_{\mbb{Z}}$ denotes the closest-point projection of
$\mbb{R}$ onto $\mbb{Z}$, i.~e., $\forall x \in \mbb{R}$,
$P_{\mbb{Z}} x$ is the integer closest to $x$. In order to verify
the bounds (\ref{Eq:Bounds}), we choose circles of radius
$\epsilon \to 0$ as the test regions $R$. Let $B_\epsilon$ denote
the circle of radius $\epsilon$ centered at the origin, and set
\begin{equation}\label{Eq:ChiR}
\chi_R(\by) = \chi_{B_\epsilon}(\by-\bx)
\end{equation}
where $\bx$ ranges over all of ${\mbb{R}}^2$. Given a function $u
: {\mbb{R}}^2 \to \mbb{R}$, we have from (\ref{Eq:InnerProduct})
\begin{equation}\label{Eq:InnerProduct2}
\langle u, \chi_R \rangle = \frac{1}{(2\pi)^2} \int \frac{\mu
b^2}{2} \frac{K }{1 + K d/2} \, \hat{u}^* (\bk)\, {\rm e}^{i
\bk\cdot\bx} \hat{\chi}_{B_\epsilon}(\bk) \, d^2k
\end{equation}
In order to elucidate the structure of this inner product as
$\epsilon \to 0$, we may introduce the normalized variables:
\begin{equation}
\bx' = \frac{\bx}{\epsilon}, \quad \bk' = \epsilon \bk
\end{equation}
whereupon (\ref{Eq:InnerProduct2}) becomes
\begin{equation}\label{Eq:InnerProduct3}
\langle u, \chi_R \rangle = \frac{1}{(2\pi)^2} \int \frac{\mu
b^2}{2} \frac{K'/\epsilon}{1 + (K'/\epsilon) d/2} \, \hat{u}^*
(\bk') \, {\rm e}^{i \bk'\cdot\bx'} \hat{\chi}_{B_\epsilon}(\bk')
\, \frac{d^2k'}{\epsilon^2}
\end{equation}
In the limit of $\epsilon \to 0$ we have, asymptotically,
\begin{equation}\label{Eq:InnerProduct4}
\langle u, \chi_R \rangle \sim \frac{1}{(2\pi)^2} \int \frac{\mu
b^2}{d} \, \hat{u}^* (\bk') \, {\rm e}^{i \bk'\cdot\bx'}
\hat{\chi}_{B_\epsilon}(\bk') \, \frac{d^2k'}{\epsilon^2}
\end{equation}
or, in terms of the original set of variables,
\begin{equation}\label{Eq:InnerProduct5}
\langle u, \chi_R \rangle \sim \frac{1}{(2\pi)^2} \int \frac{\mu
b^2}{d} \, \hat{u}^* (\bk) \, {\rm e}^{i \bk\cdot\bx}
\hat{\chi}_{B_\epsilon}(\bk) \, d^2k
\end{equation}
In addition, distributionally as $\epsilon \to 0$ we have
$\chi_{B_\epsilon} \sim \pi\epsilon^2 \delta_D$, and
(\ref{Eq:InnerProduct5}) simplifies to
\begin{equation}\label{Eq:InnerProduct6}
\langle u, \chi_R \rangle \sim \frac{1}{(2\pi)^2} \int \frac{\mu
b^2}{d} \, \hat{u}^* (\bk) \, {\rm e}^{i \bk\cdot\bx} \pi
\epsilon^2 \, d^2k
\end{equation}
This expression finally evaluates to
\begin{equation}\label{Eq:InnerProduct7}
\langle u, \chi_R \rangle \sim \frac{\mu b^2}{d} \pi \epsilon^2
u(\bx)
\end{equation}
Setting $u = \xi - \eta$ we have
\begin{equation}\label{Eq:InnerProduct8}
\langle \xi - \eta, \chi_R \rangle \sim \frac{\mu b^2}{d} \pi
\epsilon^2 \big(\xi(\bx) - \eta(\bx)\big)
\end{equation}
whereas choosing $u = \chi_R$, with $\chi_R$ as in
(\ref{Eq:ChiR}), yields
\begin{equation}\label{Eq:InnerProduct9}
\parallel \chi_R \parallel^2 \sim \frac{\mu b^2}{d} \pi \epsilon^2
\end{equation}
in agreement with (\ref{Eq:SmallLoopEnergy}). Inserting
(\ref{Eq:InnerProduct8}) and (\ref{Eq:InnerProduct9}) into
(\ref{Eq:Bounds}) finally gives the stationarity condition
\begin{equation}\label{Eq:Bounds2}
\frac{1}{2}  \, \geq \, \xi(\bx) - \eta(\bx) \, \geq \, -
\frac{1}{2}, \quad \forall \bx \in {\mbb{R}}^2
\end{equation}
whose unique solution is (\ref{Eq:PZ}), \emph{q.~e.~d.}

This result is remarkable in several notable respects. Firstly,
given the nonlocal character of the energy, the local character of
the projection (\ref{Eq:PZ}) is unexpected. Secondly, as we have
already noted, problem (\ref{Eq:EnergyConstrained3}) is
non-convex owing to the non-convexity of the admissible set $X$.
It is therefore remarkable that the problem is amenable to an
exact, and particularly simple, solution. Finally, it is
interesting to note again the crucial role played by the core
regularization of the energy which, provided that $d>0$, results
in finite energies (\ref{Eq:InnerProduct8}) and
(\ref{Eq:InnerProduct9}).

\subsection{Unconstrained equilibrium equations}

Next we turn to the unconstrained minimization problem
(\ref{Eq:EnergyUnconstrained}). The corresponding Euler-Lagrange
equation is:
\begin{equation}\label{Eq:Equilibrium}
\frac{\mu b^2}{2} K \hat{\eta} = b \hat{s}
\end{equation}
It is clear from this equation that all functions of the form
$\hat{\eta} \propto \delta_D$, corresponding to uniform slip
distributions over the entire slip plane, are solutions of the
homogeneous equation. Therefore, it follows from the Fredholm
alternative theorem that a necessary condition for
(\ref{Eq:Equilibrium}) to have solutions is that $\hat{s}({\bf
0}) = 0$, i.~e.,
\begin{equation}\label{Eq:ZeroResultant}
\int s(\bx) d^2 x = 0
\end{equation}
This condition expresses the requirement that the resultant of
the resolved shear-stress field be zero. If this condition is met,
then $\hat{\eta}$ follows from (\ref{Eq:Equilibrium}) as
\begin{equation}\label{Eq:Eta}
\hat{\eta} = \frac{2}{K b} \frac{\hat{s}}{\mu} + 2 \pi C \delta_D
\end{equation}
where $C$ is an arbitrary constant. An application of the inverse
Fourier transform to this expression gives
\begin{equation}\label{Eq:Eta2}
\eta = G \star s + C
\end{equation}
where the fundamental solution
\begin{equation}\label{Eq:Eta3}
G(\bx) = \frac{1}{(2\pi)^2} \int \frac{2}{\mu b} \frac{1}{K} {\rm
e}^{i \bk\cdot\bx} d^2k = \frac{1}{\mu \pi b}
\frac{\sqrt{x_1^2+x_2^2}}{x_1^2 + x_2^2/(1-\nu)}
\end{equation}
represents an unconstrained slip distribution which decays to zero
at infinity and is in equilibrium with a resolved shear stress in
the form of a Dirac delta applied at the origin.

\subsection{The general solution}

We proceed to collect all the solution steps outlined in the
foregoing and to provide an explicit expression for the slip
distribution $\zeta$ as a function of the applied field $s$.
Thus, the solution $\zeta$ follows in three steps, namely: $s$
$\to$ $\eta$ $\to$ $\xi$ $\to$ $\zeta$. The first step yields the
unconstrained phase field $\eta$, the second the integer-valued
phase field $\xi$, and the third step returns the core-regularized
slip distribution $\zeta$. Gathering the relations
(\ref{Eq:Eta2}), (\ref{Eq:Projection}) and (\ref{Eq:EL})
corresponding to each of the three steps just enunciated gives,
explicitly,
\begin{equation}\label{Eq:GeneralSolution}
\zeta = \varphi_d \star P_X (G \star s) + C
\end{equation}
where the projection $P_X$ is given by (\ref{Eq:PZ}) and, as noted
earlier, we require $s$ to have zero mean, whereupon the solution
is determined up to an arbitrary integer $C \in \mbb{Z}$.

At this point, it is illuminating to revisit the question of
uniqueness and metastability, especially in view of the existence
of relative energy minimizers such as presented in
Section~\ref{Sec:Examples}. In this respect, it should be noted
that the solution procedure presented in the foregoing returns a
\emph{unique solution}, modulo uniform slips. Indeed, the
unconstrained equilibrium equations are linear and, within the
constraints imposed by the Fredholm alternative, determine a
\emph{unique} unconstrained slip distribution $\eta$. The local
truncation (\ref{Eq:PZ}) in turn yields a \emph{unique}
integer-valued phase field $\xi$. Finally, smoothing as in
(\ref{Eq:EL2}) results in a \emph{unique} core-regularized phase
field $\zeta$. It is therefore clear that the solution procedure
outlined above selects a `preferred' slip distribution among a
vast array of competitors. Thus, for instance, in regions of the
slip plane where the dislocation density is small, i.~e., where
the distance between neighboring dislocations is large compared to
the core size $d$, metastable slip distributions may be obtained
by inserting small loops in equilibrium with the local resolved
shear stress within the intervening area between dislocations.
Examples of metastable solutions of this type have been discussed
in Section~\ref{Sec:Examples}. It is intriguing that, in those
examples, e.~g., in the case of a plane slipping under the action
of a uniform resolved shear stress, the unique preferred solution
$\zeta=0$ delivers the minimum attainable energy, i.~e., it is an
absolute energy minimizer. An open mathematical question of some
interest is whether the preferred solution
(\ref{Eq:GeneralSolution}) does indeed always deliver an absolute
energy minimizer.

\subsection{Averages and macroscopic variables}

Many of the macroscopic quantities of interest pertaining to the
behavior of a slip plane may be recovered by taking the
appropriate averages. In order to render the operation of taking
averages over the slip plane well defined, we may simply assume
that the phase field is periodic with unit cell $\Omega$, i.~e.,
\begin{equation}
\zeta(\bx) = \zeta(\bx + l^1 \ba_1 + l^2 \ba_2), \quad ( l^1, l^2
) \in \mbb{Z}^2
\end{equation}
for some basis vectors $\{ \ba_1, \ba_2 \}$. Under these
conditions, an application of Orowan's relation gives the
macroscopic slip strain as
\begin{equation}\label{Eq:Orowan1}
\gamma = \frac{\gamma_0}{|\Omega|} \int_\Omega \zeta d^2 x =
\gamma_0 \langle \zeta \rangle
\end{equation}
or, in view of (\ref{Eq:EL}), alternatively as
\begin{equation}\label{Eq:Orowan2}
\gamma = \frac{\gamma_0}{|\Omega|} \int_\Omega \xi d^2 x =
\gamma_0 \langle \xi \rangle
\end{equation}
where
\begin{equation}
\gamma_0 = \frac{b}{l} \label{eq:SlipStrainSat}
\end{equation}
is a reference slip strain. Here $l$ is the slip-plane spacing,
which is assumed known, and $\gamma_0$ is the slip strain which
is attained when one loop sweeps over the entire slip plane. The
dislocation line density per unit volume is related to the phase
field as
\begin{equation}\label{Eq:Rho}
\rho = \frac{1}{l} \langle |\nabla \xi| \rangle
\end{equation}
Indeed, the integral of $|\nabla \xi|$ over a region of the slip
plane simply measures the total length of all dislocation lines
contained in that region. Finally, the global equilibrium
condition (\ref{Eq:ZeroResultant}) may equivalently be expressed
in the form:
\begin{equation}\label{Eq:SAverage}
\langle s \rangle = 0
\end{equation}
which requires that the mean resolved shear stress be zero.

\section{Irreversible processes and kinetics}
\label{sec:Irreversible_processes}

The preceding developments have focused on the energetics of a
dislocation ensemble in an elastic crystal endowed with a
piecewise quadratic Peierls interplanar potential. In addition to
this energetics, the dislocation ensemble may undergo inelastic
interactions with the lattice, resulting in lattice friction; and
short-range inelastic interactions with an assortment of
obstacles, such as second-phase particles, forest dislocations,
and others. We assume that these interactions are {\it
irreversible} and, therefore, kinetic in nature. The essential
assumption is that the crossing of an obstacle by a dislocation
`costs' a certain energy, regardless of the direction of
crossing. The energy toll depends on the strength of the
interaction and is dissipated, e.~g., as heat, and irreversibly
lost to the system.

\subsection{Variational formulation}

The irreversible dislocation-obstacle interactions may be built
into the variational framework developed previously by recourse to
time discretization \cite{RadovitzkyOrtiz1999, OrtizRepetto1999,
OrtizRepettoStainier2000, OrtizStainier1999,
KaneMarsdenOrtizWest:2000}. Thus, henceforth we consider a
sequence of discrete times $t_0$, $t_1$, $\dots$, $t_n$,
$t_{n+1}$, $\dots$, presume the slip distribution $\zeta_0$ at
time $t_0$ to be known, and seek to compute the slip
distributions $\zeta^1$, $\dots$, $\zeta^n$, $\zeta^{n+1}$,
$\dots$, at all subsequent times. The central problem is to
determine the slip distribution $\zeta^{n+1}$ at time $t_{n+1}$
given the solution $\zeta^n$ at time $t_n$ and the applied field
$s_{n+1}$ at time $t_{n+1}$. To this end, following
\cite{RadovitzkyOrtiz1999, OrtizRepetto1999,
OrtizRepettoStainier2000, OrtizStainier1999,
KaneMarsdenOrtizWest:2000} we introduce the incremental work
function:
\begin{equation}\label{Eq:Winc}
W[\zeta^{n+1} | \zeta^n] = E[\zeta^{n+1}] - E[\zeta^n] + \int
f(\bx) |\zeta^{n+1}(\bx) - \zeta^n(\bx)| d^2x
\end{equation}
where the elastic energy $E[\zeta]$ is given by
(\ref{Eq:EnergyZeta}), and the second term represents the
incremental work of dissipation. In this term, the field $f(\bx)
\geq 0$ represents the energy cost per unit area associated with
the passage of one dislocation over the point $\bx$, i.~e., with a
transition of the form $\zeta(\bx) \to \zeta(\bx) \pm 1$. Thus,
the field $f(\bx)$ represents the distribution of obstacles over
the slip plane and is assumed known. By the work and energy
identity, it follows that $W$ equals the total work supplied to
the system during the interval $[t_n, t_{n+1}]$.
Eq.~(\ref{Eq:Winc}) simply states that part of this work is
invested in raising the energy of the crystal, whereas the
remainder of the external work supplied is invested in overcoming
the obstacle resistance. The updated slip distribution now follows
from the minimum principle:
\begin{equation}\label{Eq:Mininc}
\inf_{\zeta^{n+1} \in Y} W[\zeta^{n+1} | \zeta^n]
\end{equation}
It should be carefully noted that the work function
(\ref{Eq:Winc}) depends on the initial conditions $\zeta^n$ for
the time step, which allows for irreversibility and hysteresis.

\subsection{Min-max formulation}

In order to enable the application of the general solution
procedure outlined in the foregoing, we proceed to rephrase
problem (\ref{Eq:Mininc}) as a min-max problem (e.~g.,
\cite{Rockafellar1970}, \S~36). We begin by noting the identity:
\begin{equation}
|x| = \max_{\Lambda\in [-1, 1]} \Lambda x, \quad \forall x \in
\mbb{R}
\end{equation}
Since $f(\bx) \geq 0$, it follows from this identity that
\begin{equation}\label{Eq:MaxDiss}
\int f(\bx) |\zeta^{n+1}(\bx) - \zeta^n(\bx)| d^2x =
\sup_{|g^{n+1}| \leq f} \int g^{n+1}(\bx) \big( \zeta^{n+1}(\bx) -
\zeta^n(\bx) \big) d^2x
\end{equation}
which may be viewed as a statement of the principle of maximum
dissipation. The field $g^{n+1}(\bx)$, which is required to lie
within the bounds $\pm f(\bx)$, may be regarded as the
reaction-force field exerted by the obstacles on the dislocation
ensemble at time $t_{n+1}$. Inserting representation
(\ref{Eq:MaxDiss}) into (\ref{Eq:Mininc}) and using standard
properties of saddle points of concave-convex functions (e.~g.,
\cite{Rockafellar1970}, Theorem~36.3) leads to the problem
\begin{equation}\label{Eq:MinMaxZetaG}
\sup_{|g^{n+1}| \leq f} \ \inf_{\zeta^{n+1} \in Y} \left\{
E[\zeta^{n+1}] - E[\zeta^n] + \int g^{n+1}(\bx) \big(
\zeta^{n+1}(\bx) - \zeta^n(\bx) \big) d^2x \right\}
\end{equation}
For fixed $g^{n+1}(\bx)$ the central minimization problem stated
in (\ref{Eq:MinMaxZetaG}) is now of the form (\ref{Eq:MinZeta})
analyzed earlier, and the solution is given by
(\ref{Eq:GeneralSolution}). The problem which then remains is to
optimize the field $g^{n+1}(\bx)$ in order to maximize the
dissipation at the obstacles.

The optimization of $g^{n+1}(\bx)$ may conveniently be reduced to
a problem with linear constraints by the introduction of Lagrange
multipliers (e.~g., \cite{Rockafellar1970}, \S~28), which leads to
the Lagrangian:
\begin{eqnarray}
L[g^{n+1}, \lambda^\pm] = \inf_{\zeta^{n+1} \in Y} \left\{
E[\zeta^{n+1}] - E[\zeta^n] + \int g^{n+1}(\bx) \big(
\zeta^{n+1}(\bx) - \zeta^n(\bx)\big) d^2 x \right\} \nonumber \\
+ \int \lambda^+(\bx) \big( g^{n+1}(\bx) - f(\bx) \big) d^2x -
\lambda^-(\bx) \big( g^{n+1}(\bx) + f(\bx) \big) d^2x
\label{Eq:Lagrangian}
\end{eqnarray}
The corresponding Kuhn-Tucker optimality conditions are (e.~g.,
\cite{Rockafellar1970}, Theorem 28.3):
\begin{eqnarray}
& \zeta^{n+1}(\bx) - \zeta^n(\bx) = \lambda^+(\bx) -
\lambda^-(\bx) &
\label{Eq:KT1} \\
& g^{n+1}(\bx) - f(\bx) \leq 0, \quad -
g^{n+1}(\bx) - f(\bx) \leq 0 & \label{Eq:KT2} \\
& \lambda^+(\bx) \geq 0, \quad \lambda^-(\bx) \geq 0 &
\label{Eq:KT3} \\
& \big(g^{n+1}(\bx) - f(\bx)\big) \lambda^+(\bx) = 0, \quad
\big(g^{n+1}(\bx) + f(\bx)\big) \lambda^-(\bx) = 0 &
\label{Eq:KT4}
\end{eqnarray}
which must be satisfied simultaneously by the solution. In
(\ref{Eq:KT1}), the updated slip distribution $\zeta^{n+1}$
follows from the general solution (\ref{Eq:GeneralSolution}) as
\begin{equation}
\zeta^{n+1} = \varphi_d \star P_X \big( G \star (s_{n+1} -
g^{n+1}/b) \big) + C_{n+1}
\end{equation}
and the global equilibrium condition (\ref{Eq:ZeroResultant}) now
becomes
\begin{equation}\label{Eq:ZeroResultant2}
\int b \, s_{n+1}(\bx) d^2x = \int g^{n+1}(\bx) d^2x
\end{equation}
which requires that the total applied force be exactly
equilibrated by the resultant of the obstacle reactions.

It is interesting to note how the optimality conditions
(\ref{Eq:KT1}-\ref{Eq:KT4}) give mathematical expression to the
assumed frictional interaction between the phase and obstacle
fields, including \emph{stick-slip} behavior. Thus, the
equalities (\ref{Eq:KT4}) prevent $\lambda^+(\bx)$ and
$\lambda^-(\bx)$ from being nonzero simultaneously. If, for
instance, $\zeta^{n+1}(\bx) - \zeta^n(\bx) > 0$, then by virtue of
(\ref{Eq:KT3}) and (\ref{Eq:KT1}) it necessarily follows that
$\lambda^-(\bx) = 0$ and $\lambda^+(\bx)
> 0$, and, by the first of (\ref{Eq:KT4}), $g^{n+1}(\bx) =
f(\bx)$. Conversely, if $\zeta^{n+1}(\bx) - \zeta^n(\bx) < 0$,
then (\ref{Eq:KT3}) and (\ref{Eq:KT1}) require that
$\lambda^-(\bx) > 0$ and $\lambda^+(\bx) = 0$, and, by the second
of (\ref{Eq:KT4}), $g^{n+1}(\bx) = - f(\bx)$. These two cases
correspond to \emph{slip} conditions at $\bx$, or yielding. If the
conditions (\ref{Eq:KT2}) are satisfied as strict inequalities,
then it follows from (\ref{Eq:KT2}) that, necessarily,
$\lambda^+(\bx) = 0$ and $\lambda^-(\bx) = 0$, which, in view of
(\ref{Eq:KT1}), requires that $\zeta^{n+1}(\bx) - \zeta^n(\bx) =
0$. This case corresponds to \emph{stick} conditions, or elastic
unloading. The theory is thus capable of describing the
loading-unloading irreversibility characteristic of plastic
materials.

\subsection{Special case of short-range obstacles}
\label{Sec:ShortRangeObstacles}

Next we consider the special case of an obstacle field consisting
of a uniform Peierls stress and a distribution of short-range
obstacles. This case merits special attention owing to the fact
that the problem can conveniently be reduced to the determination
of the phase field at the obstacles, which greatly facilitates
calculations.

For definiteness, we shall assume the resolved shear-stress field
to be of the form
\begin{equation}
s(\bx, t) = \tau(t) + s_0(\bx)
\end{equation}
where $\tau(t)$ is the applied resolved shear stress at time $t$
and $s_0(\bx)$ is a self-equilibrated field representing the
long-range elastic stresses induced by the obstacles. For
instance, in calculations of forest hardening $s_0(\bx)$ may
describe the action on the slip plane of the long-range stress
field of the forest dislocations. In this case, in the vicinity of
a secondary dislocation $s_0(\bx)$ tends asymptotically to the
field of the osculating straight dislocation, and, therefore, has
the form
\begin{equation}\label{Eq:LongRangeStress}
s_0 \sim b \mu \frac{A(\theta)}{r}, \qquad r \to 0
\end{equation}
for some function $A(\theta)$ which depends on the geometry of the
secondary dislocation. Here $(r, \theta)$ are polar coordinates
centered at the secondary dislocation. The corresponding
unconstrained phase field $\eta_0$ diverges logarithmically,
i.~e.,
\begin{equation}\label{Eq:LongRangeStress2}
\eta_0 \sim B(\theta) \log r, \qquad r \to 0
\end{equation}
for some function $B(\theta)$. The attendant phase field $\xi_0$
represents an infinite pile up of perfect dislocation loops encircling
the secondary dislocation. The depth of this pile up is truncated and
rendered finite when the core of the secondary dislocations is taken
into account. Since the exact geometry of the secondary dislocations
away from the primary slip plane is unknown within the present
formulation, one possibility that immediately suggests itself is to
model $s_0$ as a random field.  The use of random fields to model
barriers to dislocation slip, or other similar processes such as
first-order phase transitions in magnetic systems, has been proposed
by Hardikar {\it et al.} \cite{Hardikar:2001} and Sethna
\cite{sethna:1993}.

For simplicity, we shall additionally assume that the obstacle
distribution and, correspondingly, the phase field, is periodic
with unit cell $\Omega$. For the obstacle system under
consideration, the obstacle-strength field may be taken to be of
the form:
\begin{equation}\label{Eq:Sobs}
f(\bx) = b \tau^P + \sum_{i=1}^N  f_i \, \psi_d(\bx - \bx_i) ,
\quad \bx \in \Omega
\end{equation}
where $\tau^P \geq 0$ is the Peierls stress, $x_i$ and $f_i \geq
0 $, $i = 1, \dots, N$, are the positions and strengths of the
obstacles in $\Omega$, respectively, and the function
$\psi_d(\bx)$ represents the structure of the obstacles. As noted
earlier, the detailed modeling of the structure of obstacles is
beyond the scope of the present work. However, the treatment of
short-range obstacles as points, corresponding to setting $\psi_d
= \delta_D$ in (\ref{Eq:Sobs}), inevitably leads to logarithmic
divergences of the phase field under the obstacles. In order to
avoid these divergences the obstacles must be endowed with a
finite core. Conveniently, however, for distributions of
well-separated short-range obstacles the precise form of $\psi_d$
plays a role only in the immediate vicinity of the obstacles and
is otherwise largely irrelevant. We shall simply require that the
overlap function $\psi_d\star \varphi_d$ define a Dirac-delta
sequence as $d\to 0$, i.~e., that
\begin{equation}\label{Eq:PsiDelta}
\lim_{d\to 0} (\psi_d \star \varphi_d) (\bx) = \delta_D(\bx)
\end{equation}
in the sense of distributions. Thus, for points such that $|\bx -
\bx_i| \gg d$ the overlap function $\psi_d \star \varphi_d$ may
be treated as a Dirac-delta. In addition,  in order to avoid
divergent integrals we shall require that the product
$\hat{\psi}_d(\bk) \hat{\varphi}_d (\bk)$ decay as $1/k^2$ as
$k\to\infty$. For forest obstacles, a simple choice consistent
with these requirements is
\begin{equation}\label{Eq:PsiPhi}
\psi_d(\bx) = \varphi_d(\bx)
\end{equation}
which simply states that the core structure of the forest
obstacles is identical to the core structure of the primary
dislocations.

The problem to be solved now follows by inserting (\ref{Eq:Sobs})
into the Lagrangian (\ref{Eq:Lagrangian}). In order to facilitate
the solution of this problem we make the {\it ansatz} that the
obstacle reaction field is of the form:
\begin{equation}\label{Eq:Ansatz}
g^{n+1}(\bx) = g^{n+1}_0 + \sum_{i=1}^N  g^{n+1}_i \, \psi_d(\bx
- \bx_i) , \quad \bx \in \Omega
\end{equation}
where $g^{n+1}_i$, $i = 0, 1, \dots, N$, are constants. This {\it
ansatz} may be verified {\it a posteriori} by checking that all
optimality conditions (\ref{Eq:KT1} - \ref{Eq:KT4}) are satisfied.
Inserting (\ref{Eq:Ansatz}) into (\ref{Eq:Lagrangian}) and making
use of identities (\ref{Eq:EL}), (\ref{Eq:Orowan2}) and
(\ref{Eq:PsiDelta}) yields the reduced Lagrangian:
\begin{equation}
\begin{split}
L[\bg^{n+1}, \blambda^\pm] & = \inf_{\xi^{n+1} \in X} \left\{
E[\xi^{n+1}] - E[\xi^n] + |\Omega| \, g^{n+1}_0 ( \gamma_{n+1} -
\gamma_n ) + \sum_{i=1}^N g^{n+1}_i \big(\xi^{n+1}_i -
\xi^n_i\big) \right\}  \\
& + |\Omega| \, \{\lambda^+_0 ( g^{n+1}_0 - b \, \tau^P ) -
\lambda^-_0 ( g^{n+1}_0 + b \, \tau^P ) \} + \sum_{i=1}^N
\lambda^+_i (g^{n+1}_i - f_i ) -
\sum_{i=1}^N \lambda^-_i (g^{n+1}_i + f_i )\\
\end{split}
\label{Eq:Lagrangian2}
\end{equation}
where the energy functional $E[\xi]$ is given by
(\ref{Eq:EnergyXi}). The Kuhn-Tucker optimality conditions
(\ref{Eq:KT1}-\ref{Eq:KT4}) now reduce to:
\begin{eqnarray}
& \gamma^{n+1} - \gamma^n  = \lambda^+_0 - \lambda^-_0 &
\label{Eq:DKT1} \\
& g^{n+1}_0 - b \tau^P \leq 0, \quad -
g^{n+1}_0 - b \tau^P \leq 0 & \label{Eq:DKT2} \\
& \lambda^+_0 \geq 0, \quad \lambda^-_0 \geq 0 & \label{Eq:DKT3} \\
& ( g^{n+1}_0 - b \tau^P ) \lambda^+_0 = 0, \quad ( g^{n+1}_0 + b \tau^P
) \lambda^-_0 = 0 & \label{Eq:DKT4}
\end{eqnarray}
and
\begin{eqnarray}
& \xi^{n+1}_i - \xi^n_i = \lambda^+_i - \lambda^-_i &
\label{Eq:DKT5} \\
& g^{n+1}_i - f_i \leq 0, \quad -
g^{n+1}_i - f_i \leq 0 & \label{Eq:DKT6} \\
& \lambda^+_i \geq 0, \quad \lambda^-_i \geq 0 & \label{Eq:DKT7} \\
& (g^{n+1}_i - f_i) \lambda^+_i = 0, \quad (g^{n+1}_i + f_i)
\lambda^-_i = 0 & \label{Eq:DKT8}
\end{eqnarray}
with $i = 1, \dots, N$.

If $\gamma_{n+1} \neq \gamma_n$, i.~e., in the presence of
macroscopic slip, eqs.~(\ref{Eq:DKT1} - \ref{Eq:DKT4}) simply
require that
\begin{equation}
g^{n+1}_0 = b \, \tau^P {\it sgn}(\gamma_{n+1} - \gamma_n)
\end{equation}
where ${\it sgn}(x) = x/|x|$ is the {\it signum} function. In
addition, if $|g^{n+1}_0| < b \, \tau^P$, then it necessarily
follows from (\ref{Eq:DKT1}-\ref{Eq:DKT4}) that $\gamma_{n+1} =
\gamma_n$, corresponding to elastic unloading. Thus, the Peierls
stress has the effect of introducing an initial threshold, or
yield point, for plastic activity.

In evaluating the optimality conditions (\ref{Eq:DKT5} -
\ref{Eq:DKT8}), the phase field at the obstacles is computed as
\begin{equation}
\xi^{n+1}_i = P_{\mbb{Z}} \eta^{n+1}_i
\end{equation}
which is a special case of the pointwise projection (\ref{Eq:PZ}).
In addition, the unconstrained phase field follows by inserting
(\ref{Eq:Ansatz}) into (\ref{Eq:Eta2}) and taking periodicity
into account, with the result
\begin{equation}
\eta_{n+1}(\bx) = - \sum_{i=1}^N  g_i^{n+1}  \left\{ \int \left[
\sum_{(l^1, l^2) \in \mbb{Z}^2} G(\bx - \bx'+ l^1 \ba_1 + l^2
\ba_2) \right] (\varphi_d\star\psi_d)(\bx' - \bx_i) d^2 x'
\right\} + C_{n+1} \label{Eq:Analyitic_solution}
\end{equation}
Specializing this expression at obstacle sites we obtain the
linear system of equations
\begin{equation}\label{Eq:EtaObstacles}
\eta_i^{n+1} = - \sum_{j=1}^N G_{ij} g_j^{n+1} + C_{n+1}, \quad i =
1, \dots, N
\end{equation}
which directly relates the unconstrained phase field at the
obstacles and the obstacle reactions. In addition, the global
equilibrium condition (\ref{Eq:ZeroResultant2}) further reduces to
\begin{equation}\label{Eq:TauFObstacles}
b \, \tau_{n+1} = g^{n+1}_0 + \frac{1}{|\Omega|} \sum_{i=1}^N
g^{n+1}_i
\end{equation}
where we have made use of the identity $\langle s_0 \rangle = 0$.
Eq.~(\ref{Eq:TauFObstacles}) simply states that the applied
resolved shear stress must be exactly equilibrated by lattice
friction and the resultant of all obstacle reactions.

In view of (\ref{Eq:PsiDelta}), the influence coefficients
coupling pairs of distinct obstacles in (\ref{Eq:EtaObstacles})
may be computed as
\begin{equation}\label{Eq:G}
G_{ij} \sim \sum_{(l^1, l^2) \in \mbb{Z}^2} G(\bx_i - \bx_j + l^1
\ba_1 + l^2 \ba_2), \quad i \neq j
\end{equation}
In arriving at this expression we have made use of the assumption
that the short-range obstacles are well-separated, so that they
effectively interact as point obstacles. The evaluation of the
self-interaction coefficients $G_{ii}$ in (\ref{Eq:G}) requires
some care, as the obstacles can no longer be treated as points
and their assumed structure must be taking into consideration. In
this case we have
\begin{equation}
G_{ii} \sim \int G(\bx) (\varphi_d\star\psi_d)(\bx) d^2 x +
\sum_{(l^1, l^2) \in \mbb{Z}^2-{\bf 0}} G( l^1 \ba_1 + l^2 \ba_2 )
\qquad \text{(no sum in $i$)} \label{Eq:GSelf}
\end{equation}
where the introduction of the function $\psi_d$ describing the
obstacle core is essential in order to avoid a divergent value of
$G_{ii}$. If, by way of example, we choose the obstacle structure
(\ref{Eq:PsiPhi}), then a straightforward calculation gives:
\begin{equation}\label{Eq:GSelf2}
\int G(\bx) (\varphi_d\star\varphi_d)(\bx) d^2 x =
\frac{(2-\nu)\sqrt{1-\nu}} {2 \pi^2 b}\frac{1}{d}
\end{equation}
In general, this constant is well defined and scales as $1/b d$
provided that, as previously assumed, the product
$\hat{\psi}_d(\bk) \hat{\varphi}_d(\bk)$ decays as $1/k^2$ as
$k\to \infty$. It bears emphasis that the sole influence of the
choice of obstacle structure function $\psi_d$ on the entire model
resides in the constant (\ref{Eq:GSelf2}). Thus, difference
choices of $\psi_d$ result in different values of the constant,
but do not otherwise affect the remainder of the model.

\subsection{Algorithmic implementation}

The preceding relations provide a complete basis for updating the
phase field incrementally. The precise algorithm employed in the
calculations reported subsequently proceeds as follows:

\begin{enumerate}

\item {\sl Assembly and factorization}. The calculations are
started by computing the matrix $G_{ij}$ using formulae
(\ref{Eq:G}), (\ref{Eq:GSelf}) and (\ref{Eq:GSelf2}). This matrix
is symmetric and positive-definite and has dimension $N \times
N$. The matrix $G_{ij}$ is then factorized. It should be noted
that, for a constant distribution of obstacles, the matrix
$G_{ij}$ remains constant throughout the entire deformation
process and its factorization may be performed once and for all
at the start of the calculations.

\item {\sl Initialization}. At time $t_0$, the constant $C_0$ in
(\ref{Eq:EtaObstacles}) is set to zero. In addition, the
unconstrained phase field $\eta_0$ induced by the residual
tractions $s_0$ is set up by superposition of fields of the form
(\ref{Eq:LongRangeStress2}), suitably truncated at the origin so
as to account for the obstacle core. The sign of the superposed
fields is alternated randomly in such a way that the mean value of
$s_0$ is zero, as required.

\item {\sl Incremental update}. Assume that the state of the
slip plane is completely known at time $t_n$, and that a new
value $C_{n+1}$ of the constant in (\ref{Eq:EtaObstacles}) is
prescribed at time $t_{n+1}$. Thus, the constant $C$ is chosen as
a convenient control or loading parameter. The state of the slip
plane is updated by means of the following operations:

\begin{enumerate}

\item {\sl Stick predictor}. Set the predictor unconstrained phase
field $\tilde{\eta}^{n+1}_i = \eta^n_i$, and compute the predictor
reactions:
\begin{equation}
\tilde{g}^{n+1}_j = \sum_{i=1}^N G^{-1}_{ji}(C_{n+1} -
\tilde{\eta}^{n+1}_i)
\end{equation}

\item {\sl Reaction projection}. Project $\tilde{g}^{n+1}_j$
onto the closest point of the admissible set: $|g_i| \leq f_i$,
$i=1, \dots, N$, according to:
\begin{equation}
g^{n+1}_i = \left\{
\begin{array}{c}
    f_i, \text{ if} \quad \tilde{g}^{n+1}_i > f_i \\
    \tilde{g}^{n+1}_i, \text{ if} \quad |\tilde{g}^{n+1}_i| \leq f_i \\
    - f_i, \text{ if} \quad \tilde{g}^{n+1}_i < - f_i
\end{array}
\right.
\end{equation}

\end{enumerate}

\item {\sl Phase-field evaluation}. Once the obstacle reactions
$g^{n+1}_i$ are known, the unconstrained phase field
$\eta_{n+1}(\bx)$ may be computed from
(\ref{Eq:Analyitic_solution}). At the obstacles, the
unconstrained phase field follows directly as
\begin{equation}
\eta^{n+1}_i = \sum_{j=1}^N G_{ij} g^{n+1}_j + C_{n+1}
\end{equation}
For $\bx$ sufficiently distant from all obstacles, we may treat
the obstacles as points, whereupon (\ref{Eq:Analyitic_solution})
simplifies to:
\begin{equation}\label{Eq:Analyitic_solution2}
\eta_{n+1}(\bx) = - \sum_{i=1}^N  \left[ \sum_{(l^1, l^2) \in
\mbb{Z}^2} G(\bx - \bx_i + l^1 \ba_1 + l^2 \ba_2) \right]
g_i^{n+1} + C_{n+1}
\end{equation}
Finally, the local projection (\ref{Eq:PZ}) returns the phase
field $\xi_{n+1}(\bx)$, and (\ref{Eq:EL2}) gives the
core-regularized phase field $\zeta_{n+1}(\bx)$.

\item {\sl Macroscopic variables}. Finally, the macroscopic
resolved shear stress $\tau_{n+1}$ and slip strain $\gamma_{n+1}$
may be computed from (\ref{Eq:TauFObstacles}) and
(\ref{Eq:Orowan2}), respectively.

\end{enumerate}

\section{Application to the forest hardening mechanism}
\label{Results}

In this section, we apply the general framework developed in the
foregoing to the forest hardening mechanism. To this end, we consider
a slip plane traversed at random locations by fixed secondary or
forest dislocations and acted upon by an applied resolved shear
stress.  The objective is to characterize the dislocation patterns
which arise in response to the applied loading, and to determine the
effective behavior of the system measured, e.~g., in terms of the
macroscopic slip strain and dislocation line density. In situations
where the forest hardening mechanism is dominant, the kinetics of the
primary dislocation ensemble is primarily governed by: the interaction
between the dislocations and the applied resolved shear stress; the
long-range elastic interactions between the primary dislocations; the
line-tension effect resulting from the core structure of the
dislocations; the interaction between the primary dislocations and the
long-range elastic stress field of the forest dislocations; and the
short-range interactions between primary and forest dislocations such
as may result in jogs, the formation of junctions, and other reaction
products. Thus, the forest hardening mechanism provides a convenient
framework for illustrating the range of behaviors predicted by the
theory.

\subsection{Monotonic Loading} \label{Mono}

\begin{figure}
    \centerline{ \hbox{
    \subfigure[Stress {\it vs.} strain]{
    \epsfig{file=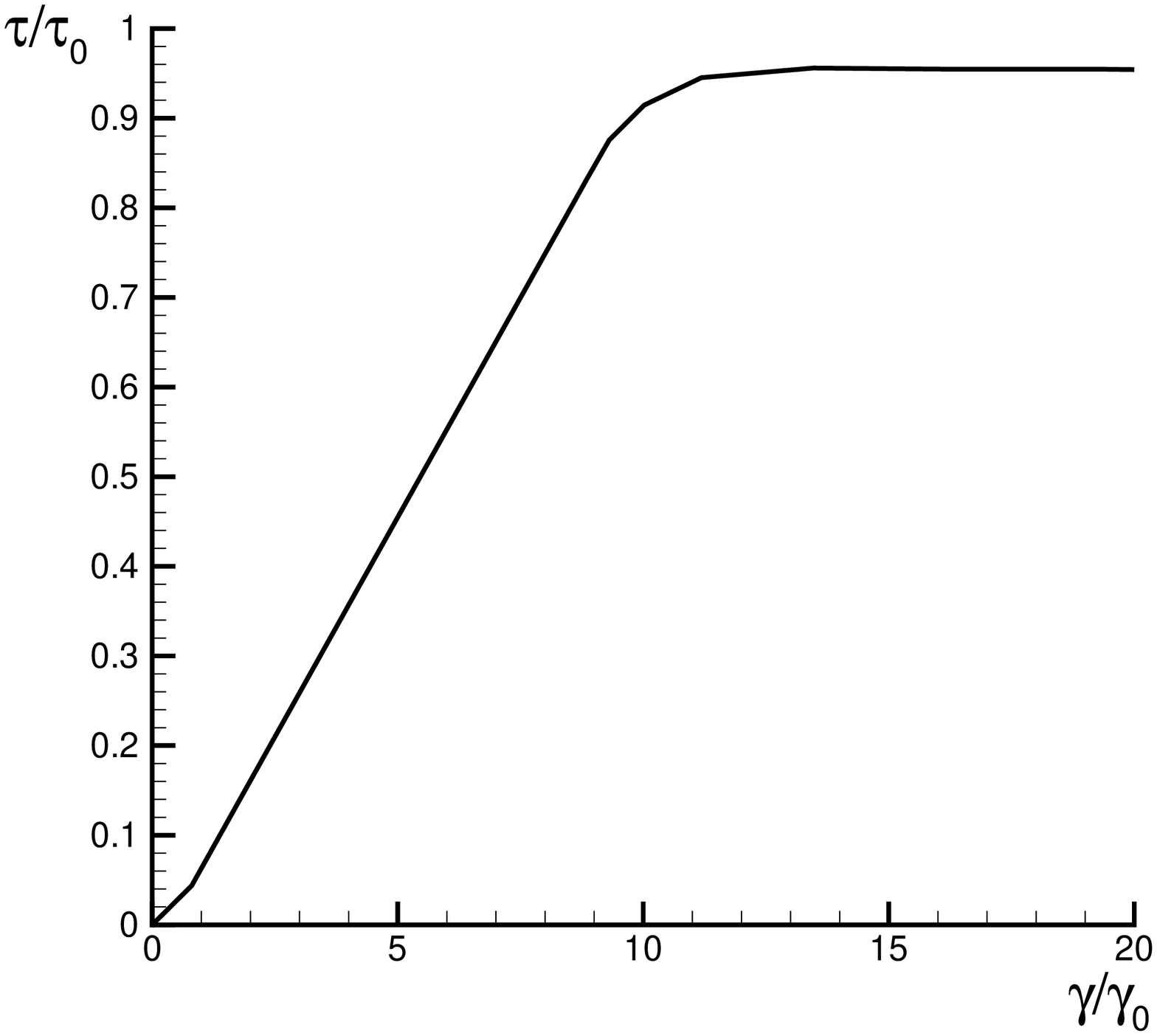,height=0.35\textheight}}
    \hglue -0.2in
    \subfigure[Dislocation density {\it vs.} strain]{
    \epsfig{file=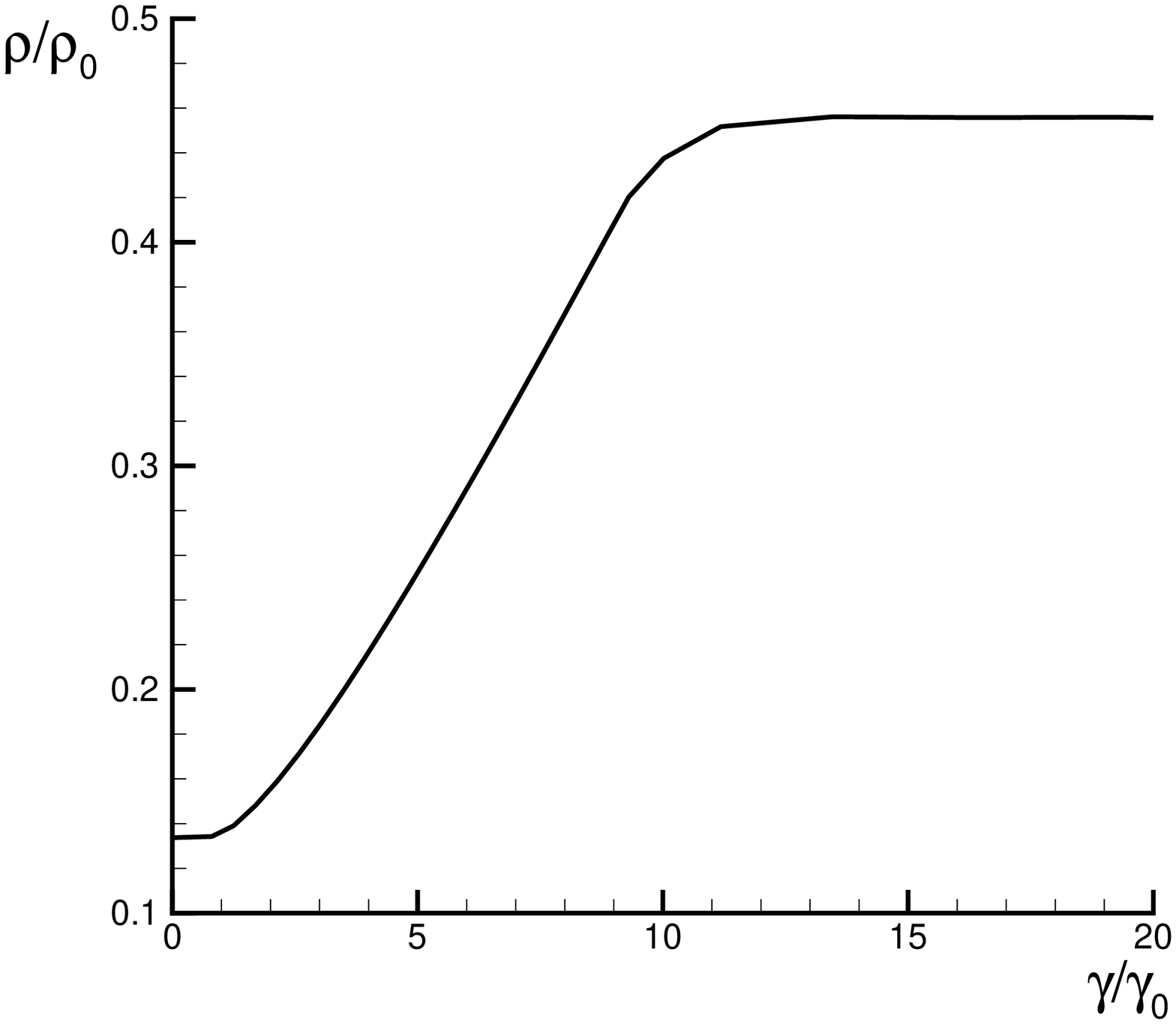,height=0.35 \textheight}}
    } }
    \caption{Monotonic loading. (a) Applied shear stress {\it vs.}
    macroscopic slip strain. (b) Evolution of dislocation density
    with macroscopic slip strain.
    }
    \label{fig:load}
\end{figure}

\begin{figure}
    \centerline{ \hbox{
    \subfigure[$\tau/\tau_0 = 0.00$]{
    \epsfig{file=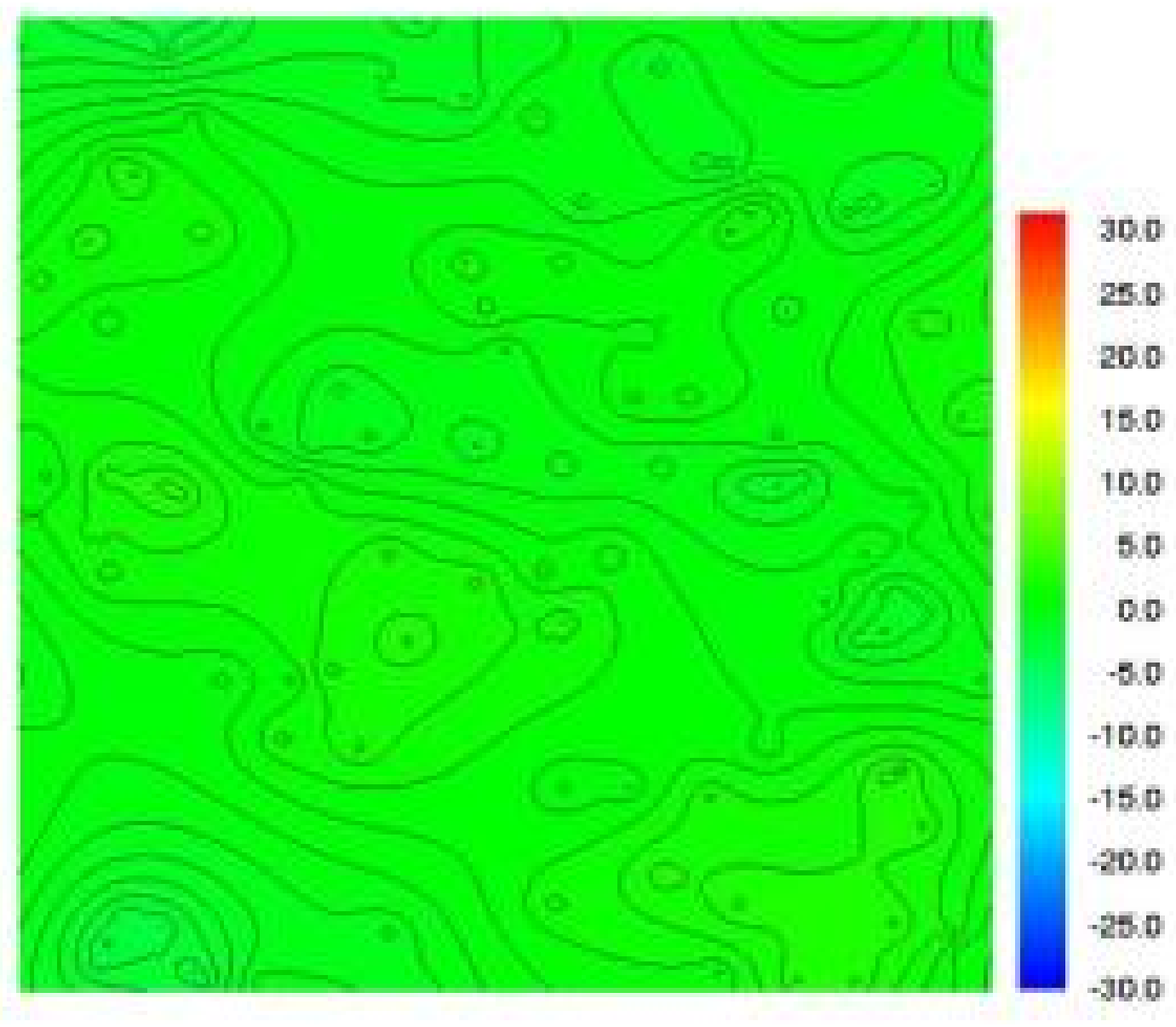,height=0.25\textheight}}
    \hglue -0.7in
    \subfigure[$\tau/\tau_0 = 0.20$] {
    \epsfig{file=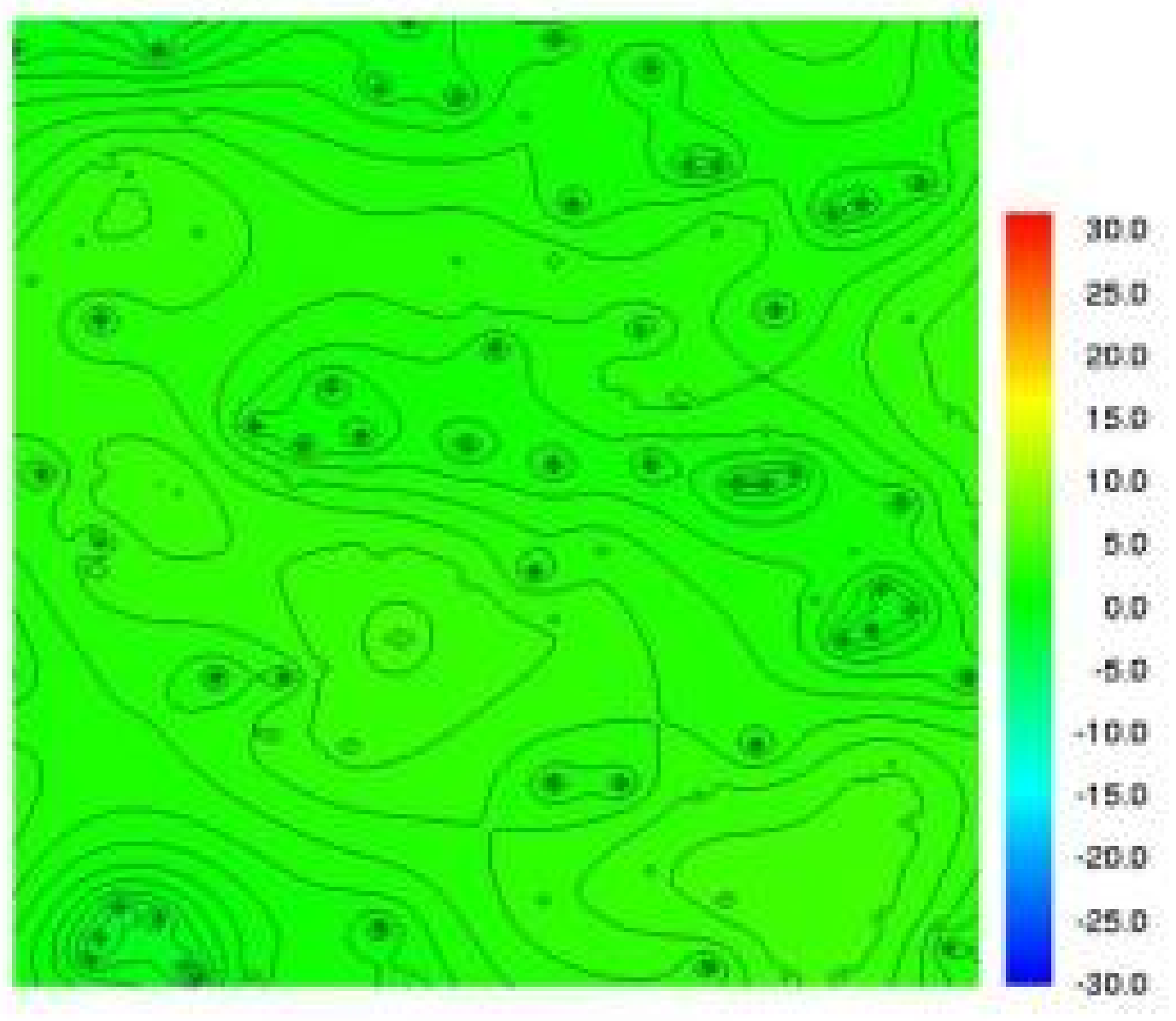,height=0.25\textheight}}
    \hglue -0.7in
    \subfigure[$\tau/\tau_0 = 0.40$] {
    \epsfig{file=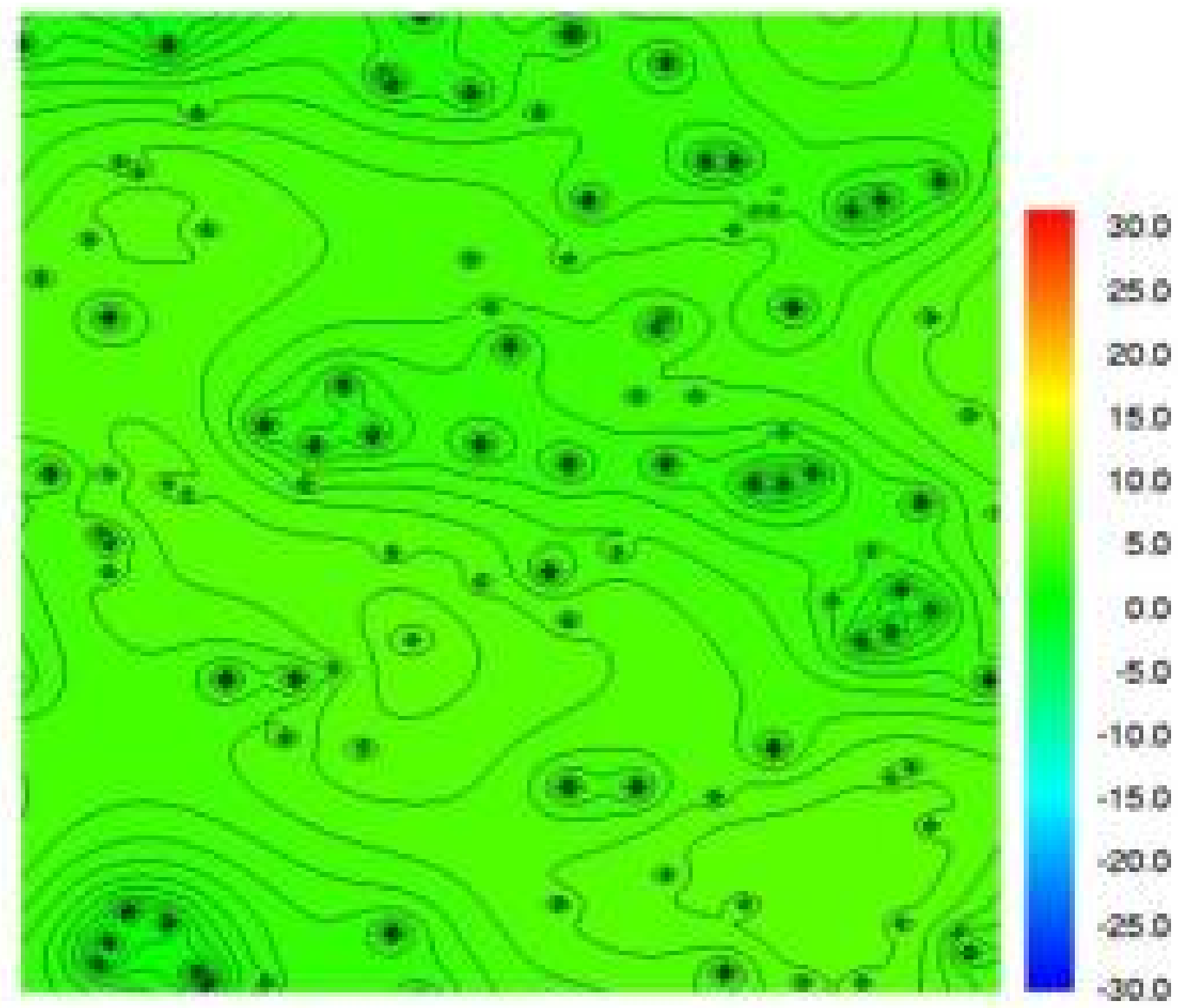,height=0.25\textheight}}
    }}
    \centerline{ \hbox{
    \subfigure[$\tau/\tau_0 = 0.60$]{
    \epsfig{file=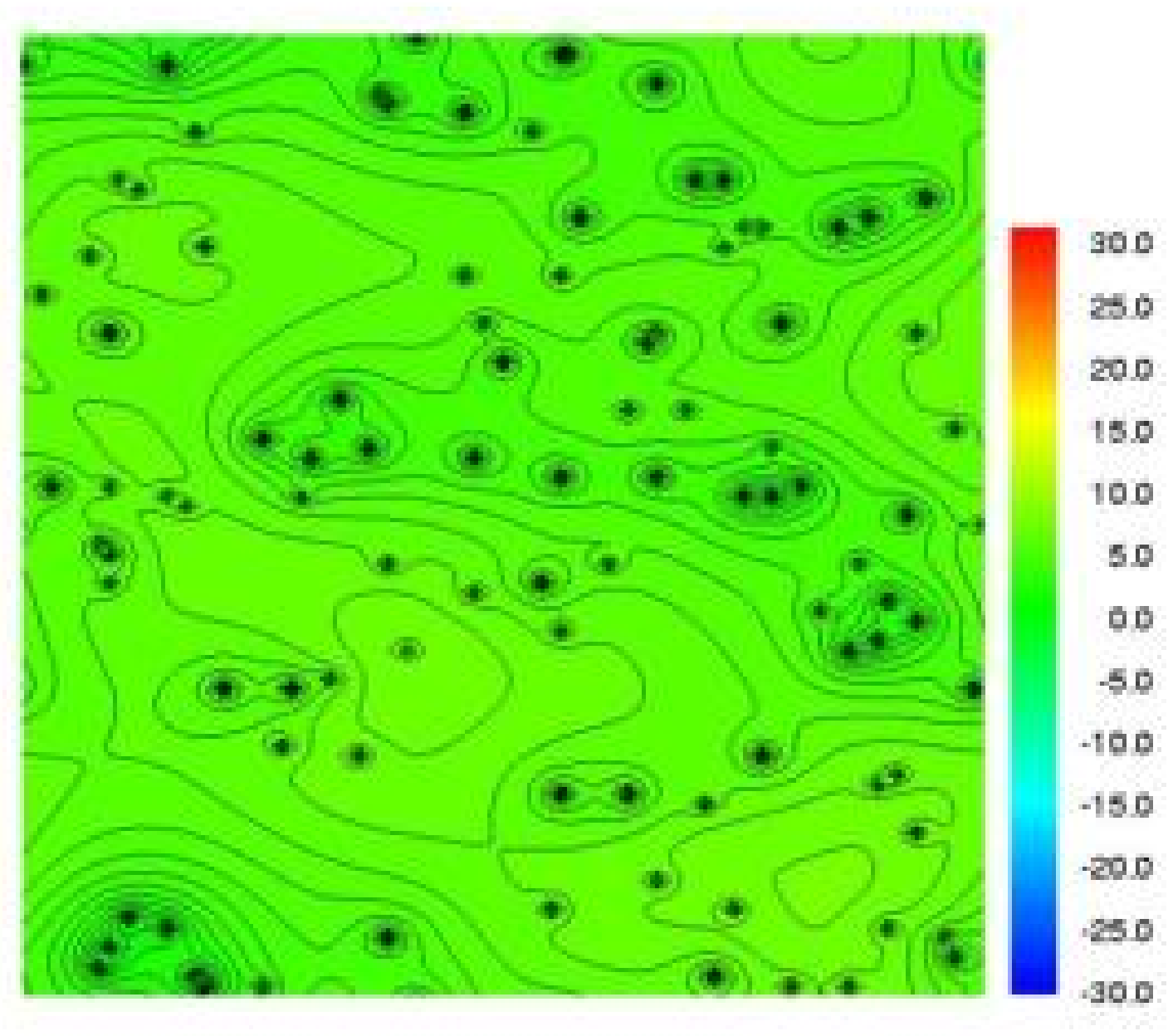,height=0.25\textheight}}
    \hglue -0.7in
    \subfigure[$\tau/\tau_0 = 0.80$]{
    \epsfig{file=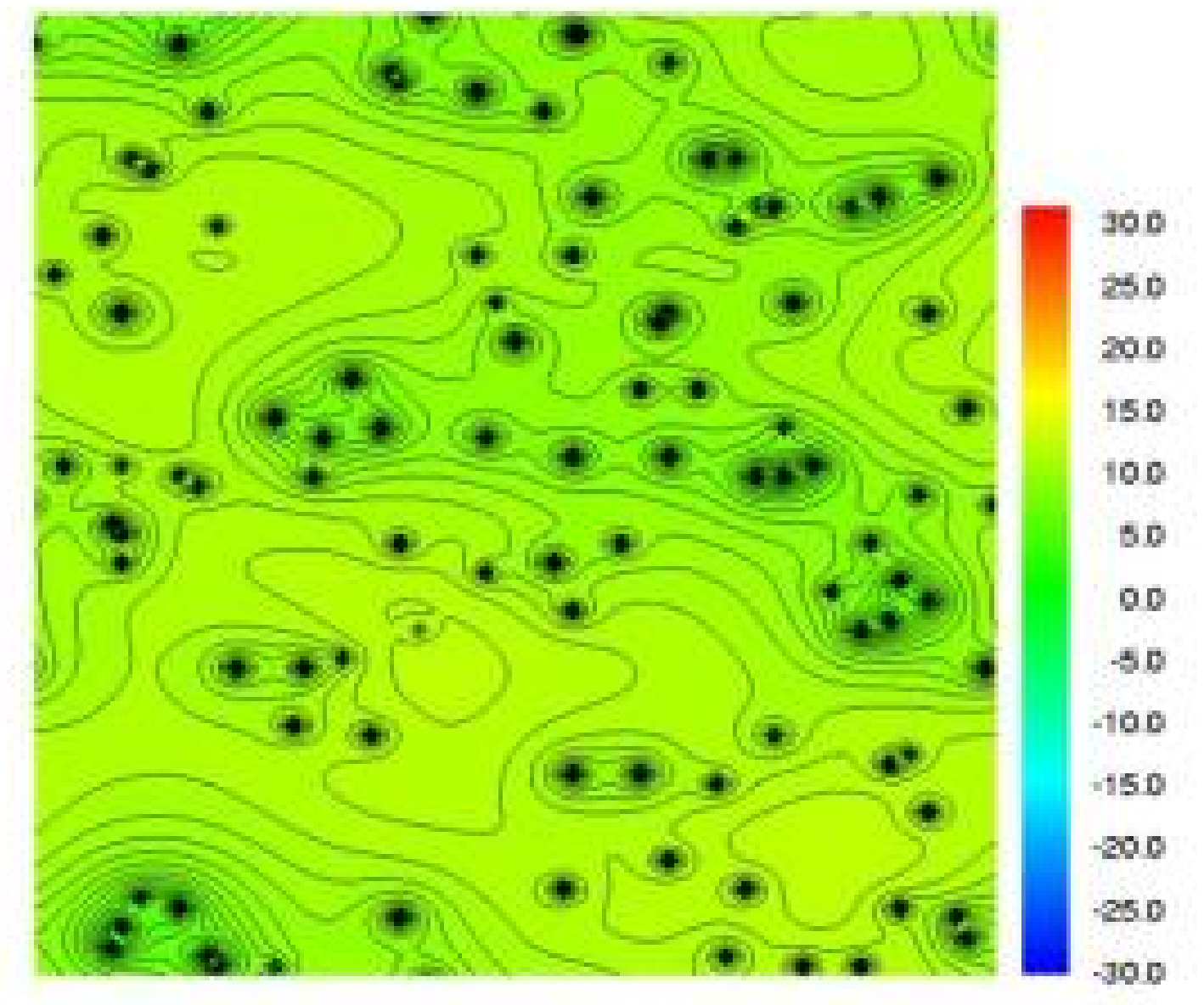,height=0.25\textheight}}
    \hglue -0.7in
    \subfigure[$\tau/\tau_0 = 0.99$]{
    \epsfig{file=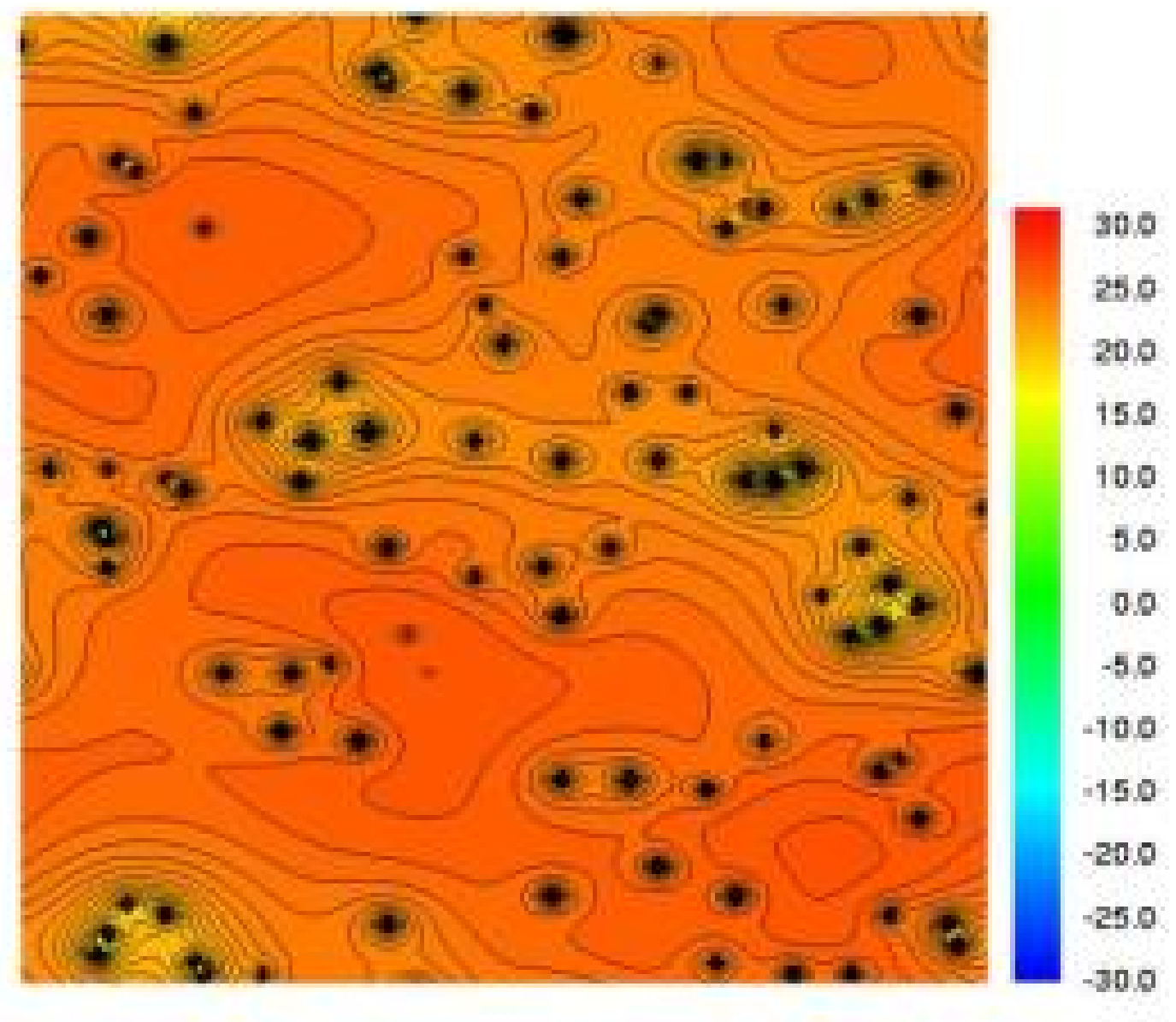,height=0.25\textheight}}
    }}
    \caption{Evolution of the dislocation pattern in response to
    monotonic loading. Figs. (a)--(f) correspond to applied shear
    stresses $\tau/\tau_0$ = 0.00, 0.20, 0.40, 0.60, 0.80 and  0.99,
    respectively.}
    \label{fig:pattern}
\end{figure}

\begin{figure}
    \centerline{ \hbox{
    \subfigure[$\tau/\tau_0 = 0.00$]{
    \epsfig{file=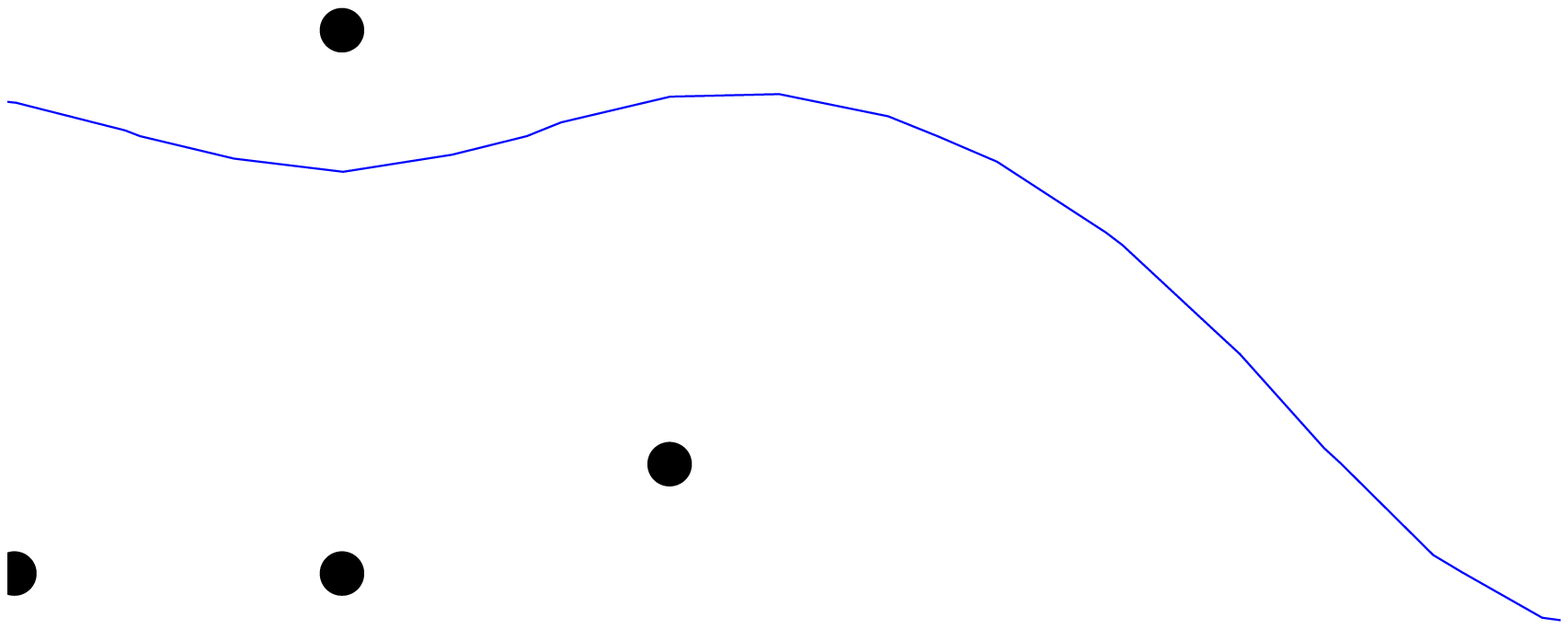,height=0.20\textheight}}
    \hglue -0.2in
    \subfigure[$\tau/\tau_0 = 0.20$]{
    \epsfig{file=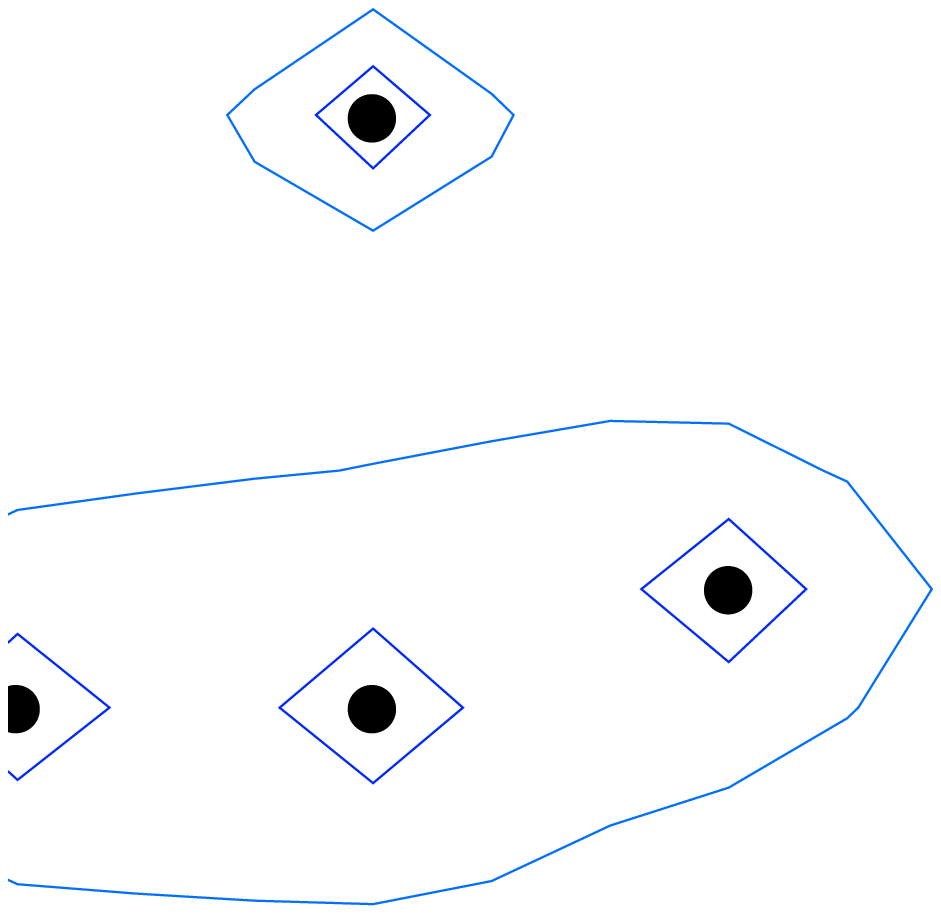,height=0.20\textheight}}
    \hglue -0.2in
    \subfigure[$\tau/\tau_0 = 0.40$]{
    \epsfig{file=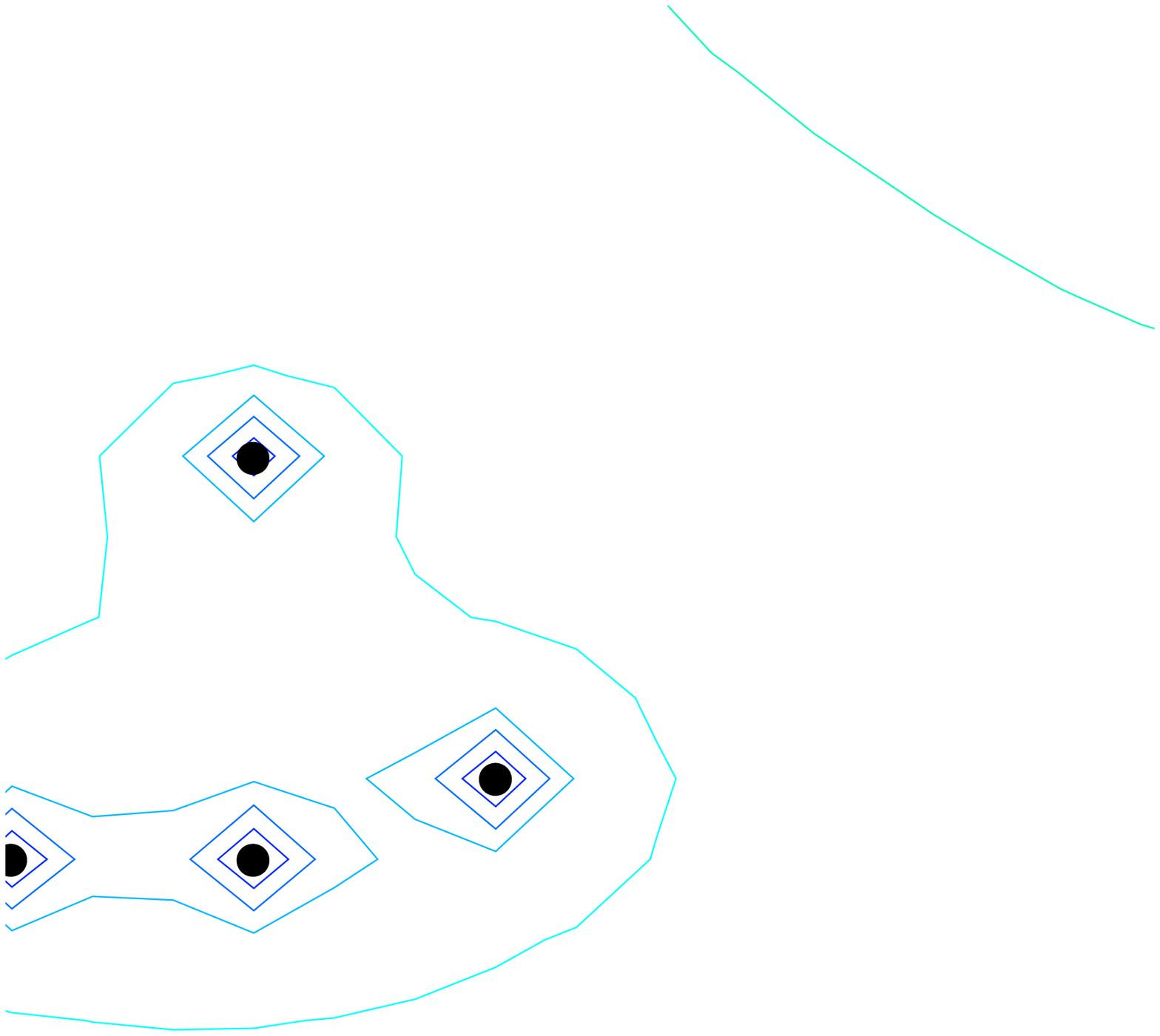,height=0.20\textheight}}
    }}
    \centerline{ \hbox{
    \subfigure[$\tau/\tau_0 = 0.60$]{
    \epsfig{file=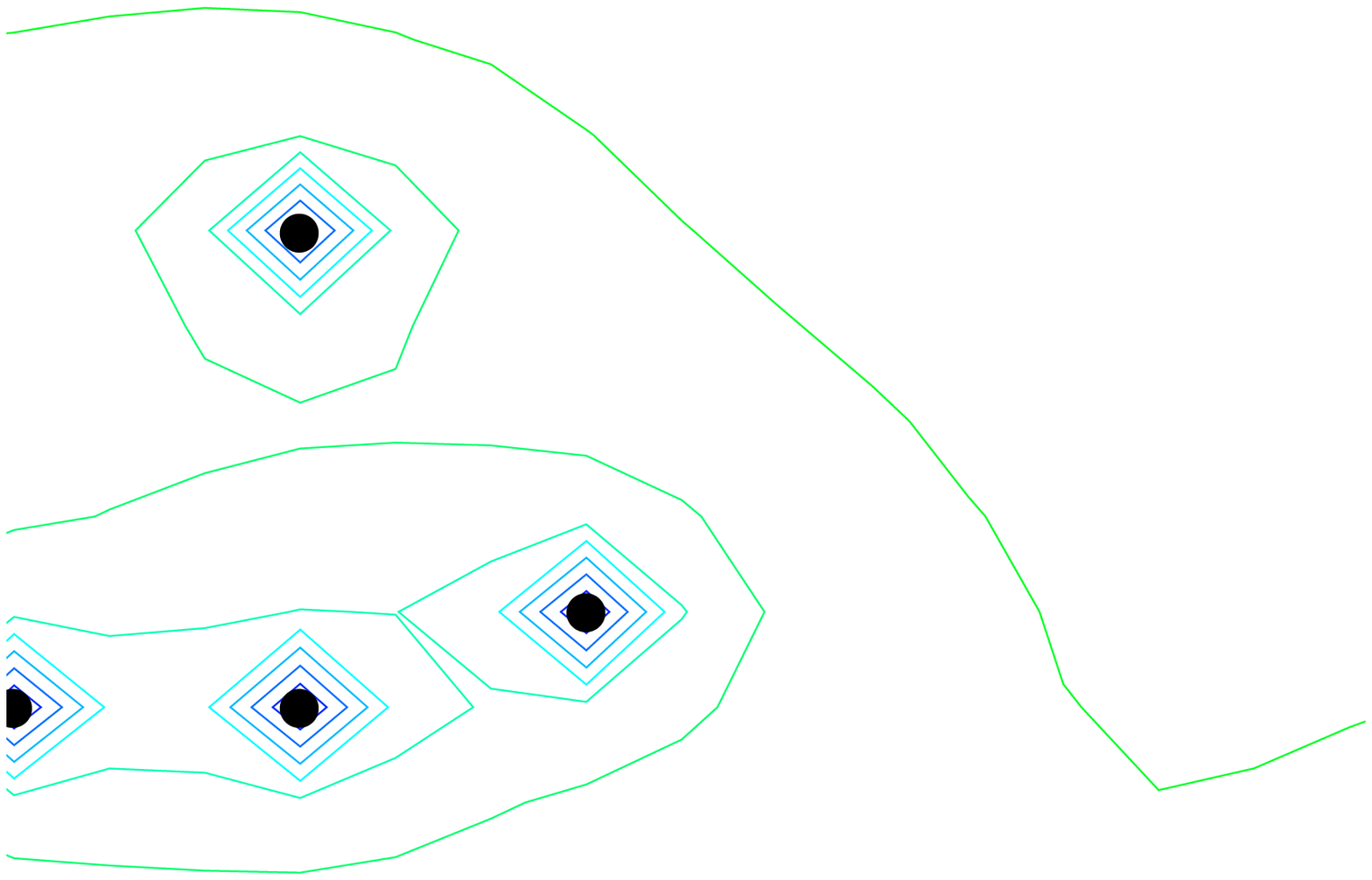,,height=0.20\textheight}}
    \hglue -0.2in
    \subfigure[$\tau/\tau_0 = 0.80$]{
    \epsfig{file=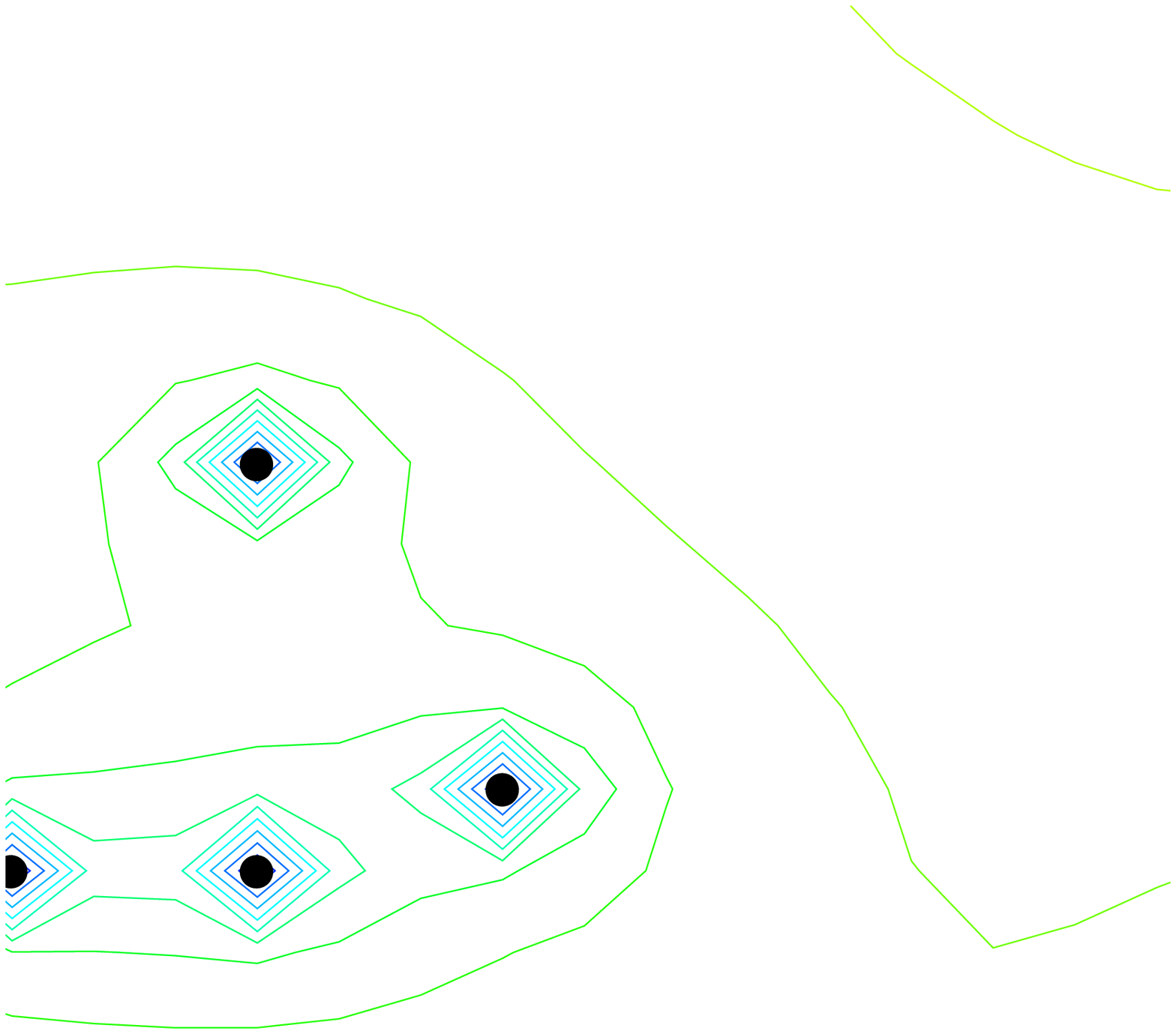,,height=0.20\textheight}}
    \hglue-0.2in \subfigure[$\tau/\tau_0 = 0.99$]{
    \epsfig{file=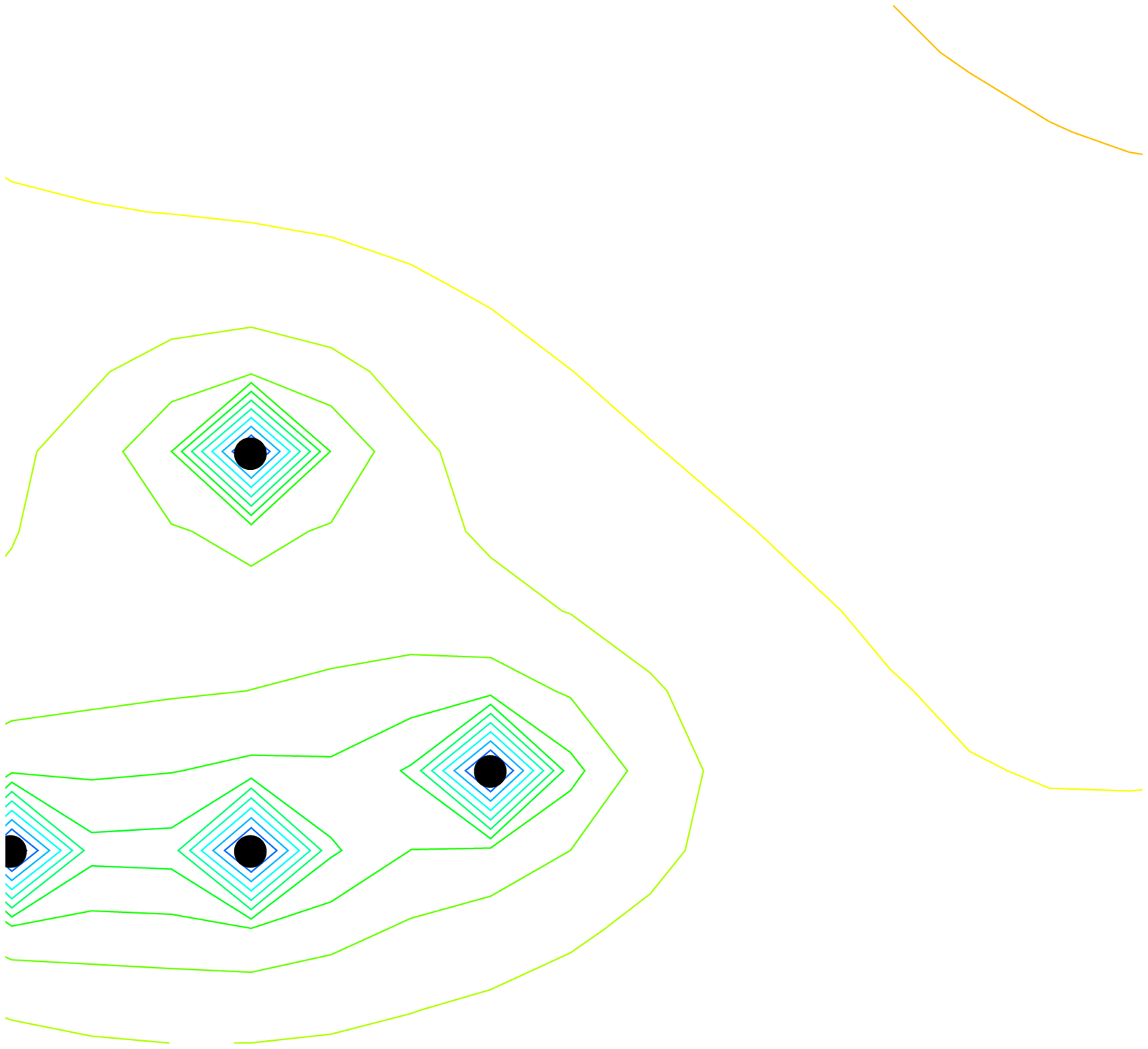,,height=0.20\textheight}}
    }}
    \caption{Detailed view of a pair of obstacles as is bypassed by
    succeeding dislocations. Figs. (a)--(f) correspond to applied
    shear stresses $\tau/\tau_0$ = 0.00, 0.20, 0.40, 0.60, 0.80 and
    0.99, respectively.  Individual dislocations are color coded for
    ease of tracking.}
    \label{fig:loop}
\end{figure}

\begin{figure}
    \centerline{ \hbox{
    \subfigure[Stress {\it vs.} strain]{
    \epsfig{file=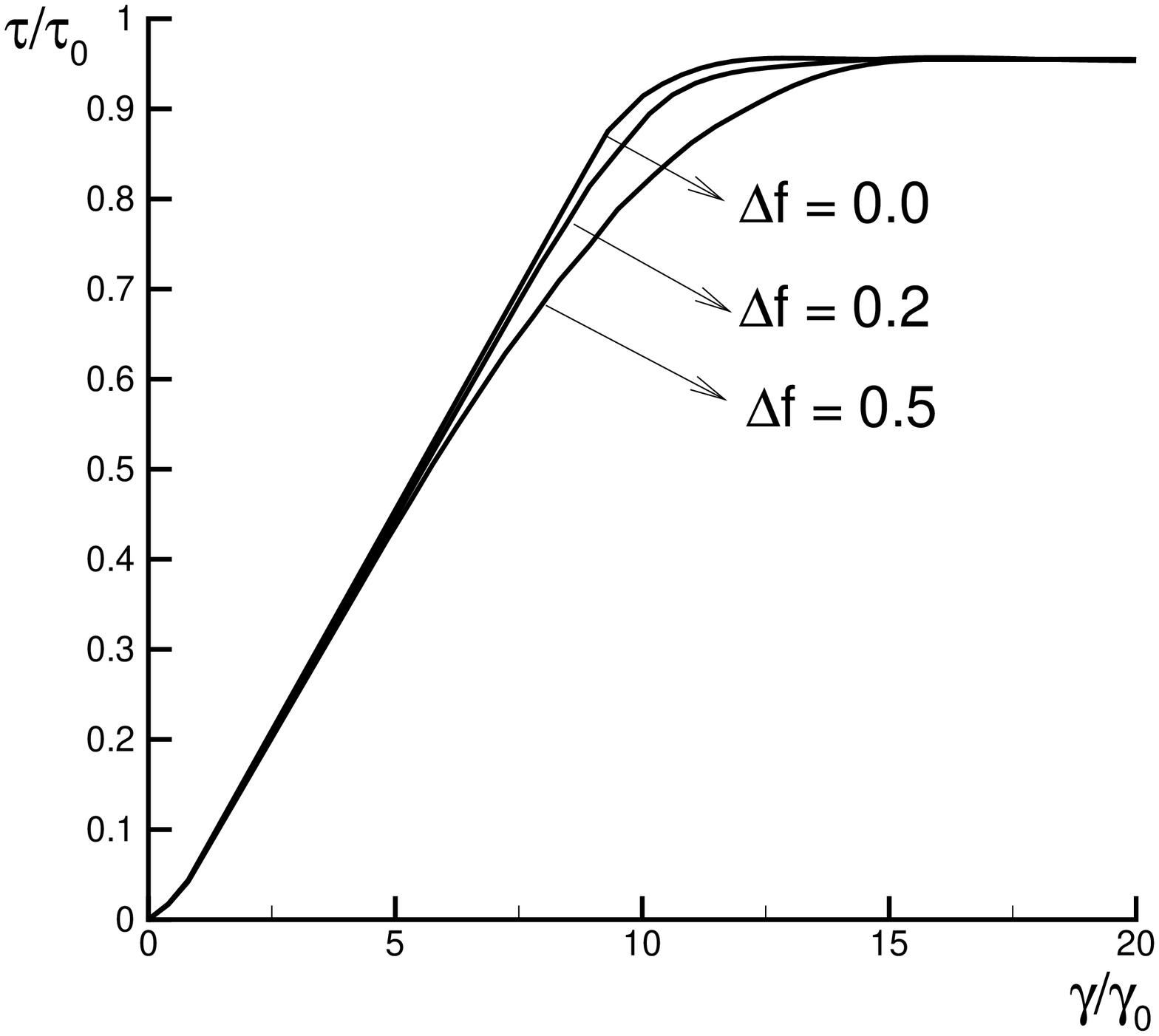,height=0.35\textheight}}
    \hglue -0.2in
    \subfigure[Dislocation density {\it vs.} strain]{
    \epsfig{file=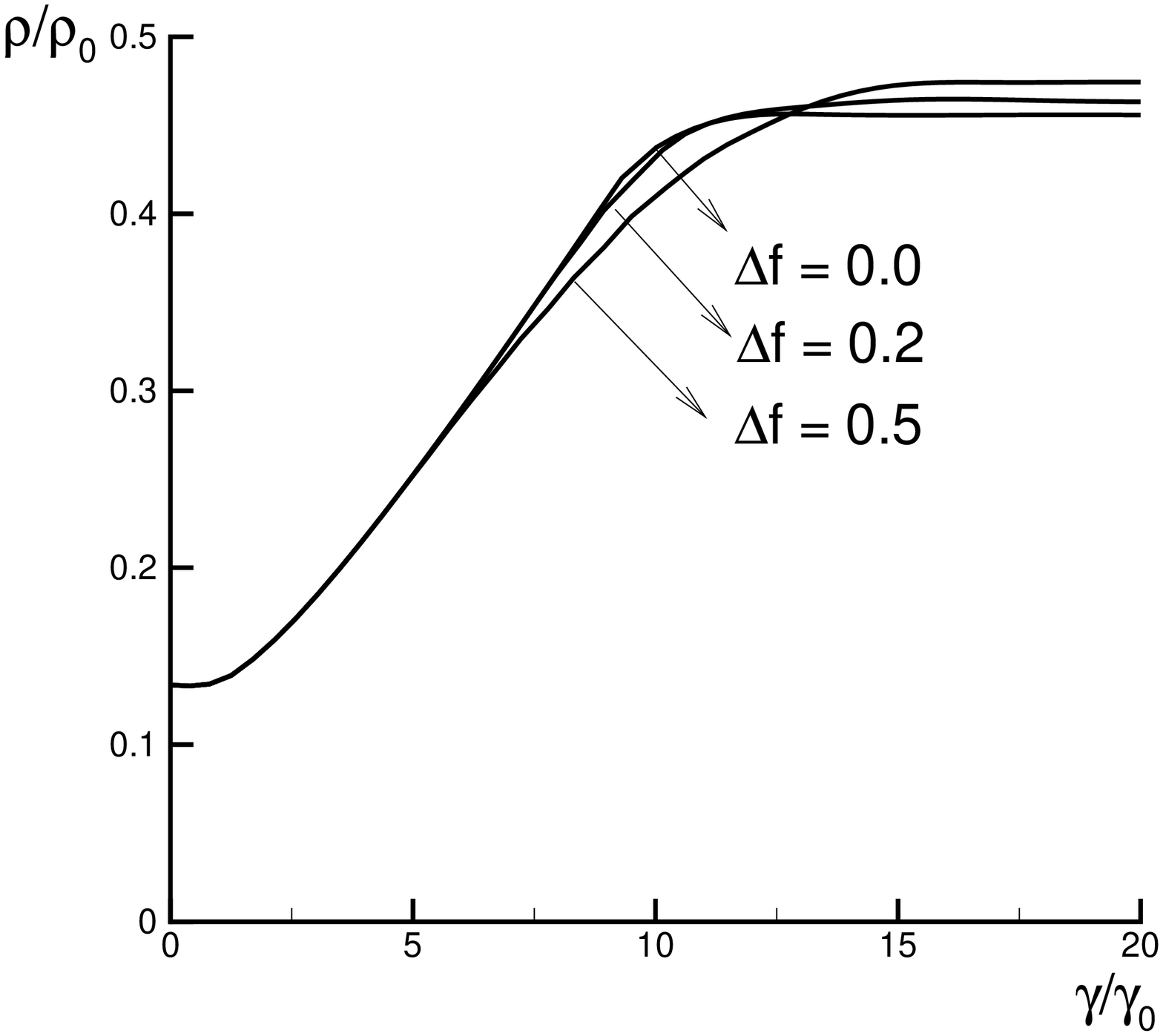,height=0.35 \textheight}}
    } }
    \caption{Monotonic loading, effect of variable obstacle
    strength. (a) Applied shear stress {\it vs.} macroscopic slip
    strain. (b) Evolution of dislocation density with macroscopic
    slip strain.}
    \label{fig:load-disp}
\end{figure}

We begin by considering the base case of a single slip system
containing a fixed concentration of point obstacles of uniform
strength and deforming under the action of a monotonically increasing
shear stress. We assume a periodicity, and the unit cell $\Omega$ is
taken to be a square of dimension $100 b$.  We randomly select $100$
points within the periodic cell as the obstacle sites. To these
obstacles we assign a uniform strengths $f = 10 \mu b^2$. For
simplicity, we take the Peierls stress $\tau^P = 0$, and set Poisson's
ratio $\nu=0.3$.

Fig.~\ref{fig:load} shows the computed stress-strain curve and
dislocation density as functions of the macroscopic slip strain.
Fig.~\ref{fig:pattern} shows the corresponding evolution of the
phase field $\xi$ and the dislocation pattern. In this latter
plot, the dislocation lines are identified with the lines across
which the phase field $\xi$ jumps by one, or equivalently, with
the level contours of $\eta + 1/2$ at integer heights. The phase
field itself counts the number of dislocation which have passed
over a given point of the slip plane, with proper accounting of
the sign of traversal. Two well-differentiated regimes, with a
transition point at $\gamma/\gamma_0 \sim 10$, are clearly
discernible in Fig.~\ref{fig:load}: a first regime of
`micro-slip' dominated by dislocation bow-out; and a second
regime characterized by generalized yielding. The saturation
value of the resolved shear stress in this latter regime follows
from global equilibrium as
\begin{equation}
b \tau_0 = b \tau^P + \frac{1}{|\Omega|} \sum_{i=1}^N f_i
\end{equation}
i.~e., the saturation stress is set by the mean obstacle strength.
During the micro-slip regime, the value of the phase field
remains close to zero at the obstacles, which are essentially
impenetrable to the dislocations. The dislocation loops move
reversibly by bowing through the open spaces between obstacles,
and by subsequently pinching behind the obstacles, leading to the
formation of Orowan loops. As increasing numbers of dislocations
bypass the obstacles by the bow-out mechanism, the number of
Orowan loops surrounding the obstacles correspondingly increases,
Fig.~\ref{fig:pattern}.

The details of this process are illustrated in
Fig.~\ref{fig:loop}, which zooms on a particular pair of
obstacles and shows the evolution of the dislocation/obstacle
interactions with increasing stress. In this figure, individual
dislocations move from left to right and are color coded for ease
of tracking.  Fig.~\ref{fig:loop}a shows the initial
configuration.  In Fig.~\ref{fig:loop}b, the dislocation has
bypassed the obstacles and left Orowan loops in its trail. In
subsequent frames, new dislocations arrive from the left, bow
through the obstacles, and eventually bypass them, leaving behind
additional Orowan loops.  The ease with which the phase-field
representation describes these geometrical and topological
transitions is quite remarkable.

At larger applied stresses, the obstacles gradually yield and are
overcome by the dislocations.  Figs.~\ref{fig:pattern}d-e
correspond to the transitional phase between the micro-slip and
the generalized yielding regimes. At sufficiently large applied
stresses close to the saturation stress $\tau_0$, all obstacles
yield and are crossed by the dislocations. Under these
conditions, the phase field $\xi$ increases uniformly without
change of shape over the entire slip plane. In this regime, the
level contours of $\xi$ remain ostensibly unchanged, even as
their height rises steadily, and the dislocation pattern becomes
frozen in place, Fig.~\ref{fig:pattern}e and
~\ref{fig:pattern}f.  The hardening rate correspondingly drops
and the stress-strain curve saturates asymptotically,
Fig.~\ref{fig:load}a.  The presence of a saturation stage is
consistent with observations of fcc single crystals undergoing
single slip, e.~g., during the easy glide or stage I or hardening.

The evolution of the dislocation line density (\ref{Eq:Rho}) with
slip strain is shown in Fig.~\ref{fig:load}b. The computed
dislocation densities are normalized by the reference density
\begin{equation}
\rho_0 = \frac{1}{b l}
\end{equation}
which is the limiting or saturation density corresponding to an
arrangement of parallel straight dislocations at intervals of $b$.
As may be seen from the figure, the slip plane contains a nonzero
dislocation density at zero slip strain. This initial density is
induced by the long-range elastic-stress field $s_0$ of the
secondary dislocations. The process of slip is accompanied by a
steady increase in the dislocation line density (\ref{Eq:Rho}).
Dislocation multiplication and the proliferation of dislocation
loops with increasing slip strain are clearly evident in
Figs.~\ref{fig:pattern}a-f.  In the micro-slip regime, the
increase in macroscopic slip is accompanied by a steady supply of
dislocations bowing between the obstacles. This results in an
initial parabolic growth rate, $\rho \sim \gamma^2$,
Fig.~\ref{fig:load}b. By contrast, as already noted when $\tau$
approaches the saturation stress $\tau_0$ the dislocation pattern
becomes frozen, e.~g., Figs.~\ref{fig:pattern}e and f, and the
dislocation density remains ostensibly constant.  This parabolic
growth regime and the subsequent saturation phase are indeed
consistent with observation \cite{ashby:1972, livingston:1962}.
The ability of the theory to predict the evolution of the
dislocation density, in addition to predicting hardening rates,
is noteworthy.

Fig.~\ref{fig:load-disp} shows the effect of a variable obstacle
strength on the computed stress-strain curve and dislocation
density evolution during monotonic loading. In order to
investigate this effect we consider obstacle strengths of the
form $f = 10 \mu b^2 (1 + \omega)$, with the random variable
$\omega$ distributed uniformly within the intervals $[-0.2, 0.2]$
and $[-0.5, 0.5]$. When the obstacle strengths exhibit
variability, the weakest obstacles tend to yield first and are
overcome by dislocations, which are pinned at the stronger
obstacles. Consequently, the transition between the micro-slip and
generalized yielding regimes becomes more gradual. It is
interesting to note that, since the average obstacle strength is
the same in all cases, the saturation stress is not affected by
the obstacle strength variability. This example illustrates the
ability of the theory to account for the combined effect of
obstacles of different species.

\subsection{Cyclic Loading}

\begin{figure}
    \centerline{\hbox{
    \subfigure[Stress]{
    \epsfig{file=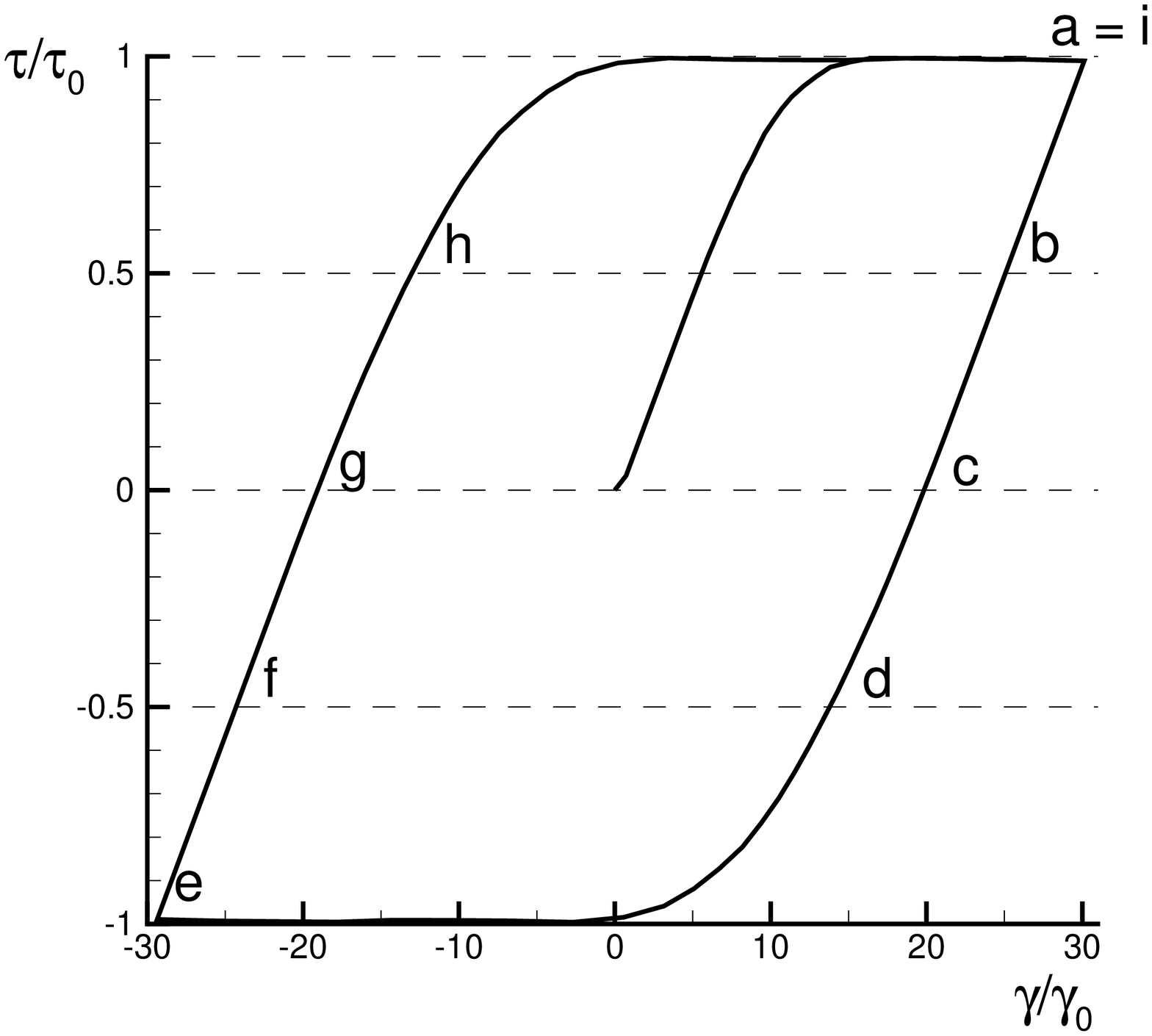,height=0.35\textheight} }
    \hglue -0.2in
    \subfigure[Dislocation density]{
    \epsfig{file=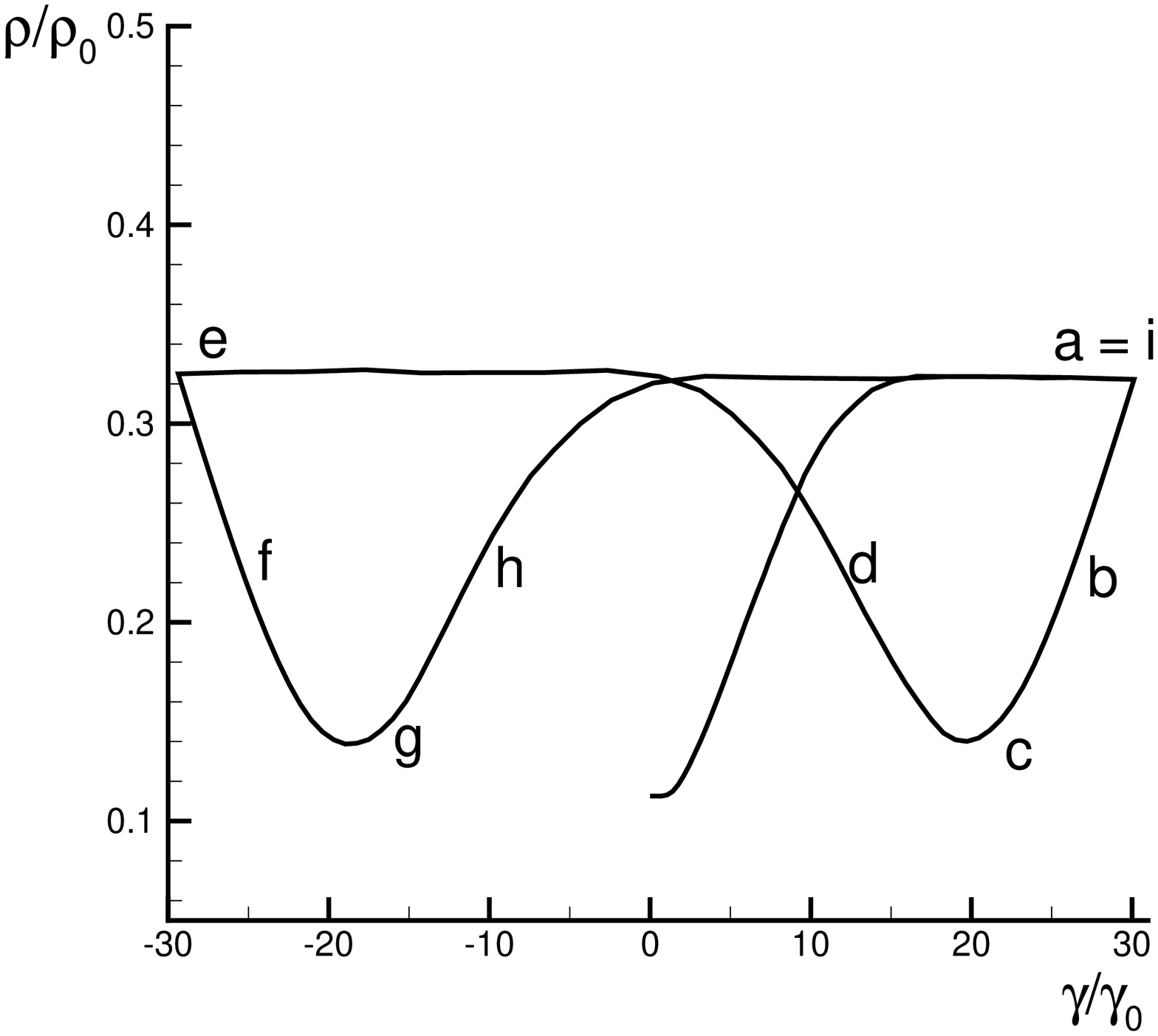,height=0.35\textheight} }
    }}
    \caption{Cyclic behavior. Labels a--i indicate the loading
    sequence. (a) Applied resolved shear stress {\it vs.} average
    slip (b) Evolution of dislocation density {\it vs.} average slip.}
    \label{fig:cycle}
\end{figure}

\begin{figure}
    \centerline{ \hbox{
    \subfigure[$\tau/\tau_0 = 0.99$ ]{
    \epsfig{file=./fig/pattern-nu03-t095.eps,height=0.2\textheight}}
    \hglue -0.2in \subfigure[$\tau/\tau_0 = 0.50$]{
    \epsfig{file=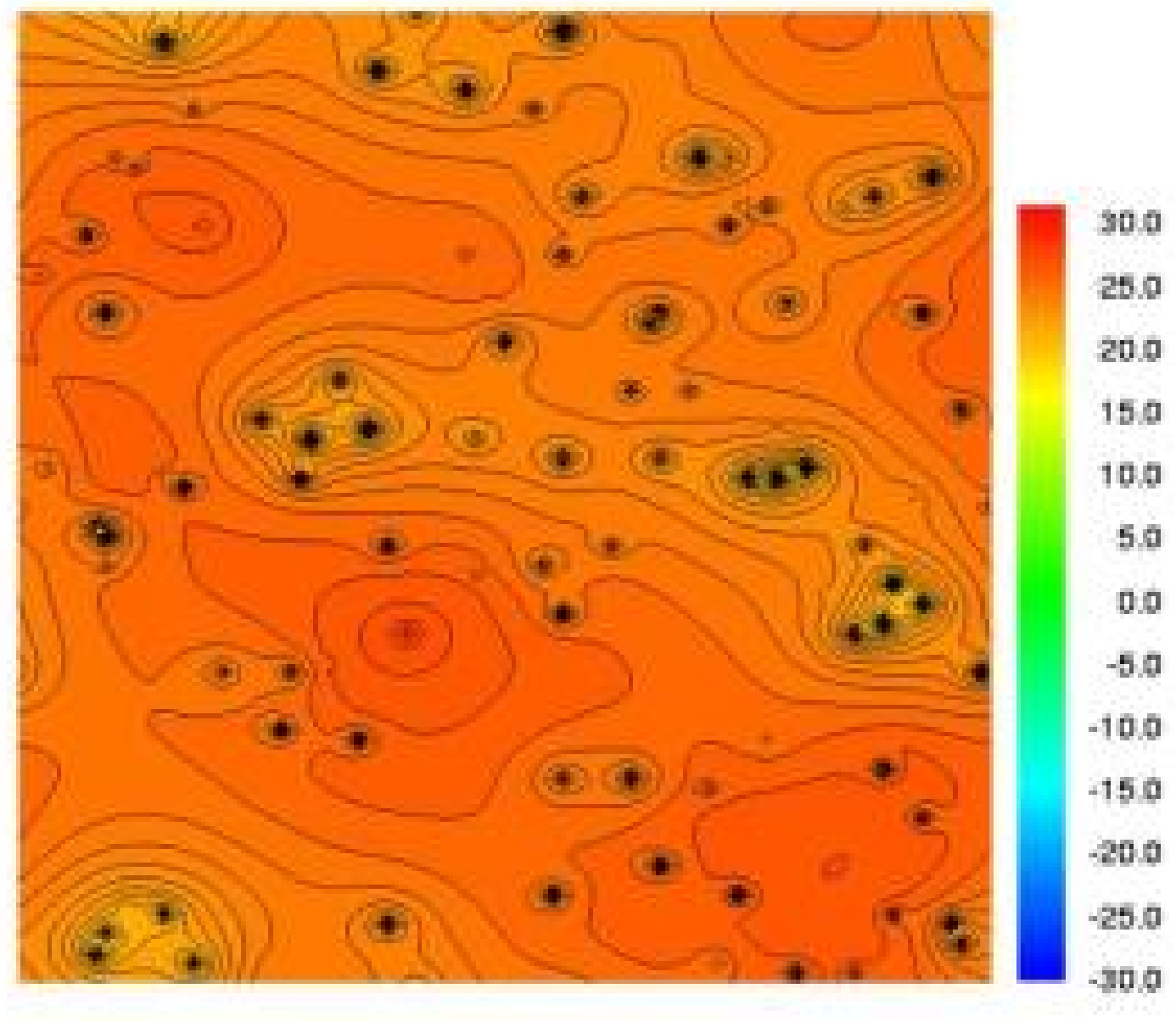,height=0.2\textheight}}
    \hglue -0.2in \subfigure[$\tau/\tau_0 = 0.00$]{
    \epsfig{file=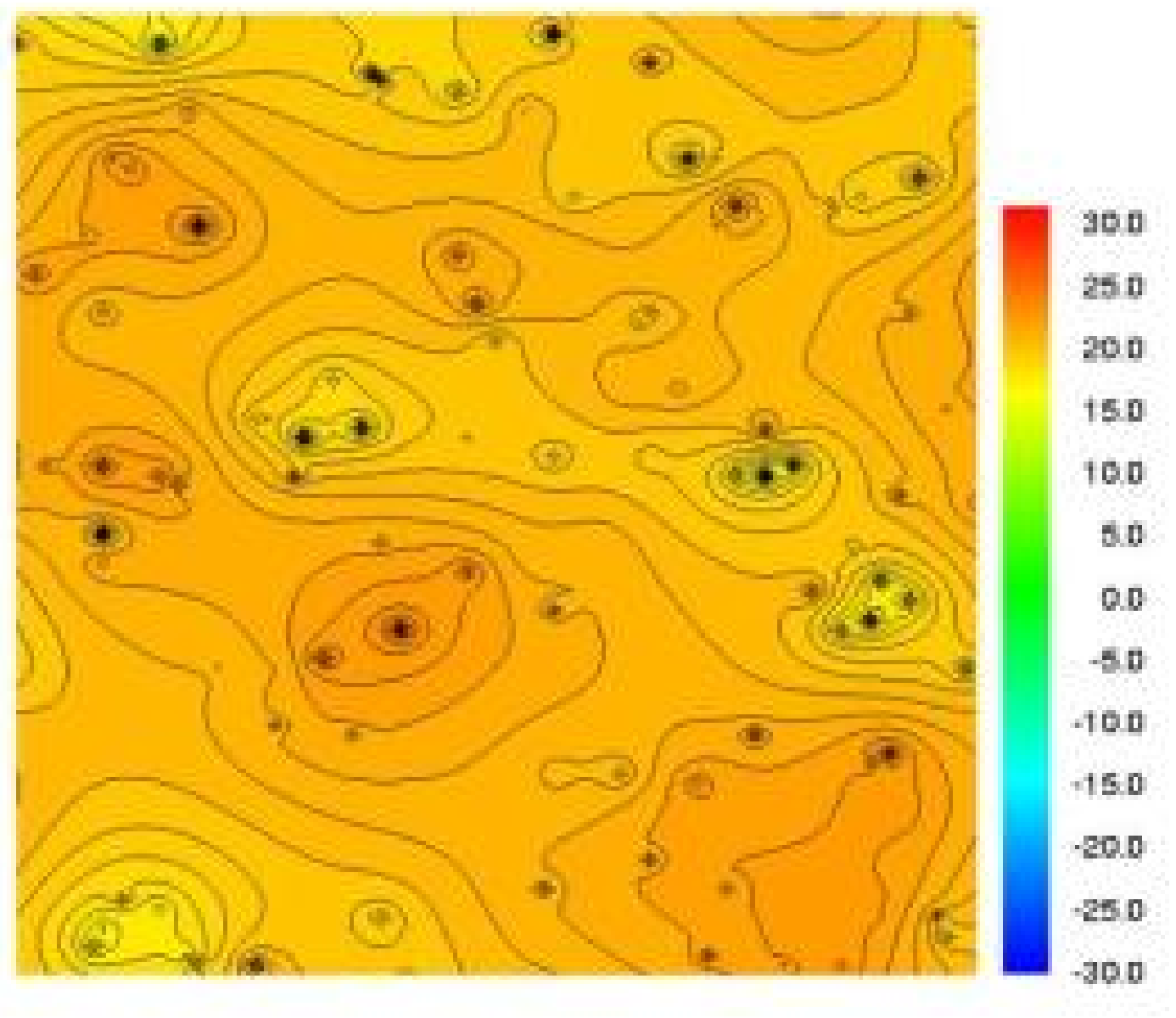,height=0.2\textheight}}
    }}
    \centerline{ \hbox{
    \subfigure[$\tau/\tau_0 = -0.50$]{
    \epsfig{file=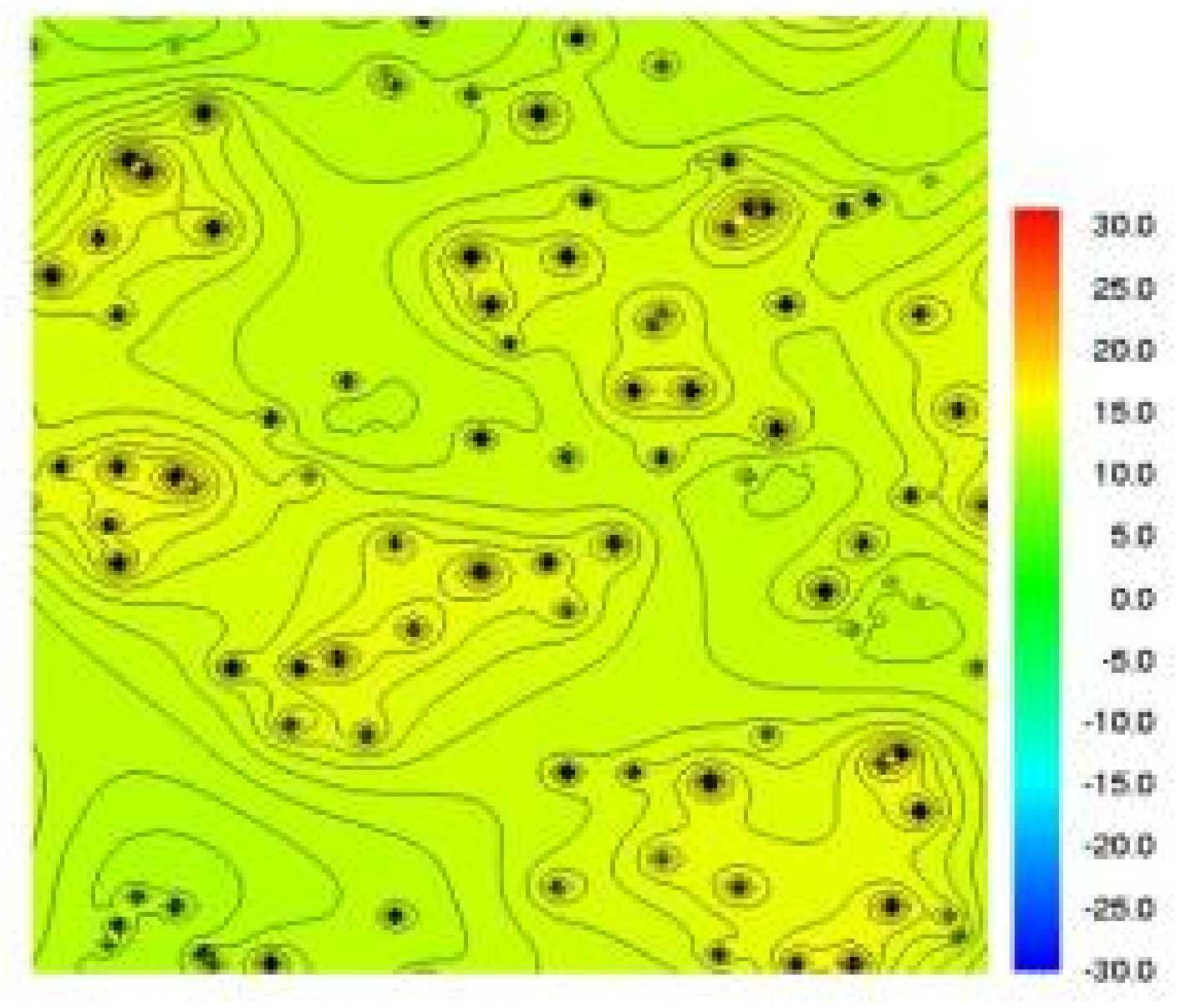,height=0.2\textheight}}
    \hglue -0.2in \subfigure[$\tau/\tau_0 = -0.99$ ]{
    \epsfig{file=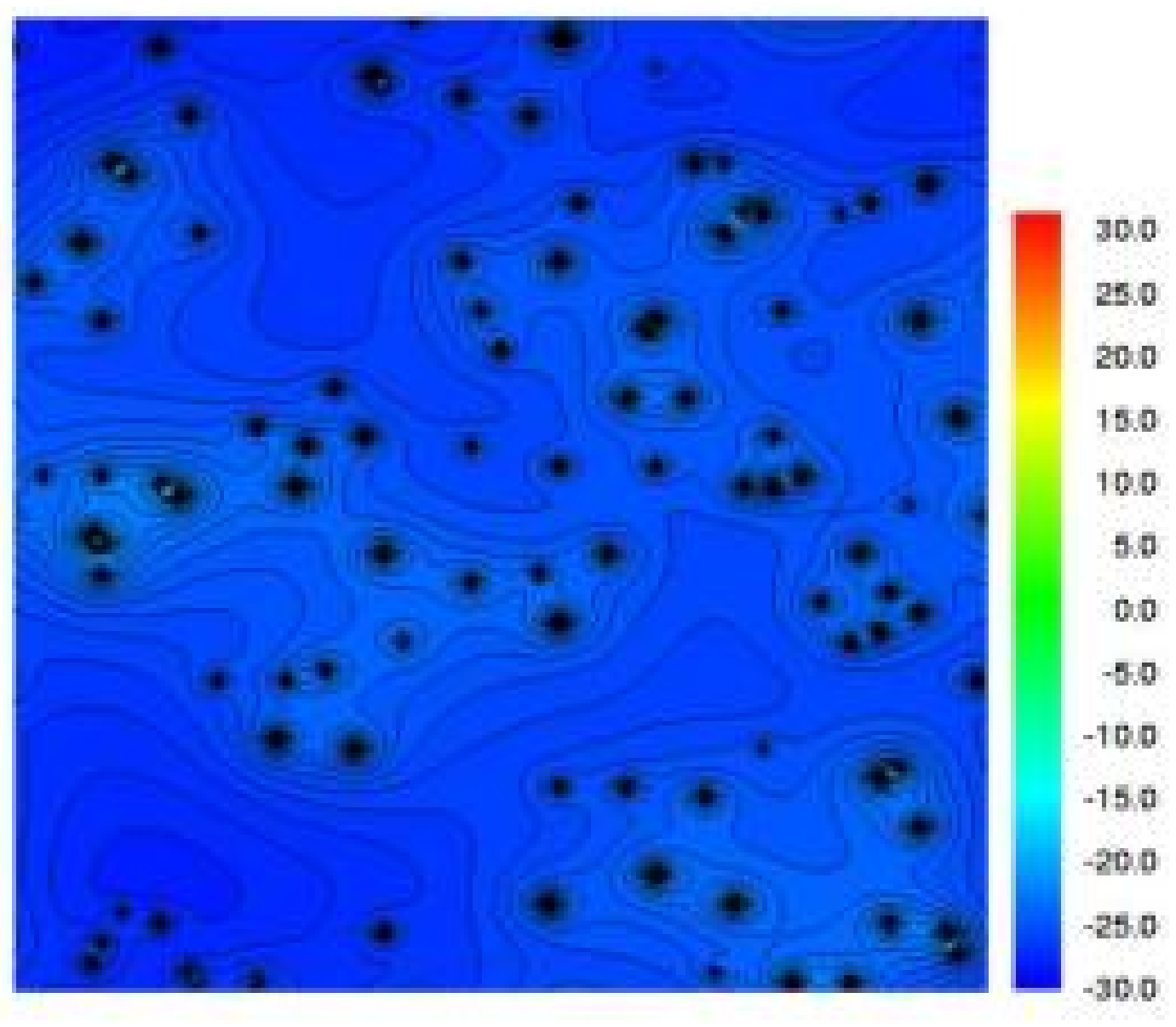,height=0.2\textheight}}
    \hglue -0.2in \subfigure[$\tau/\tau_0 = -0.50$]{
    \epsfig{file=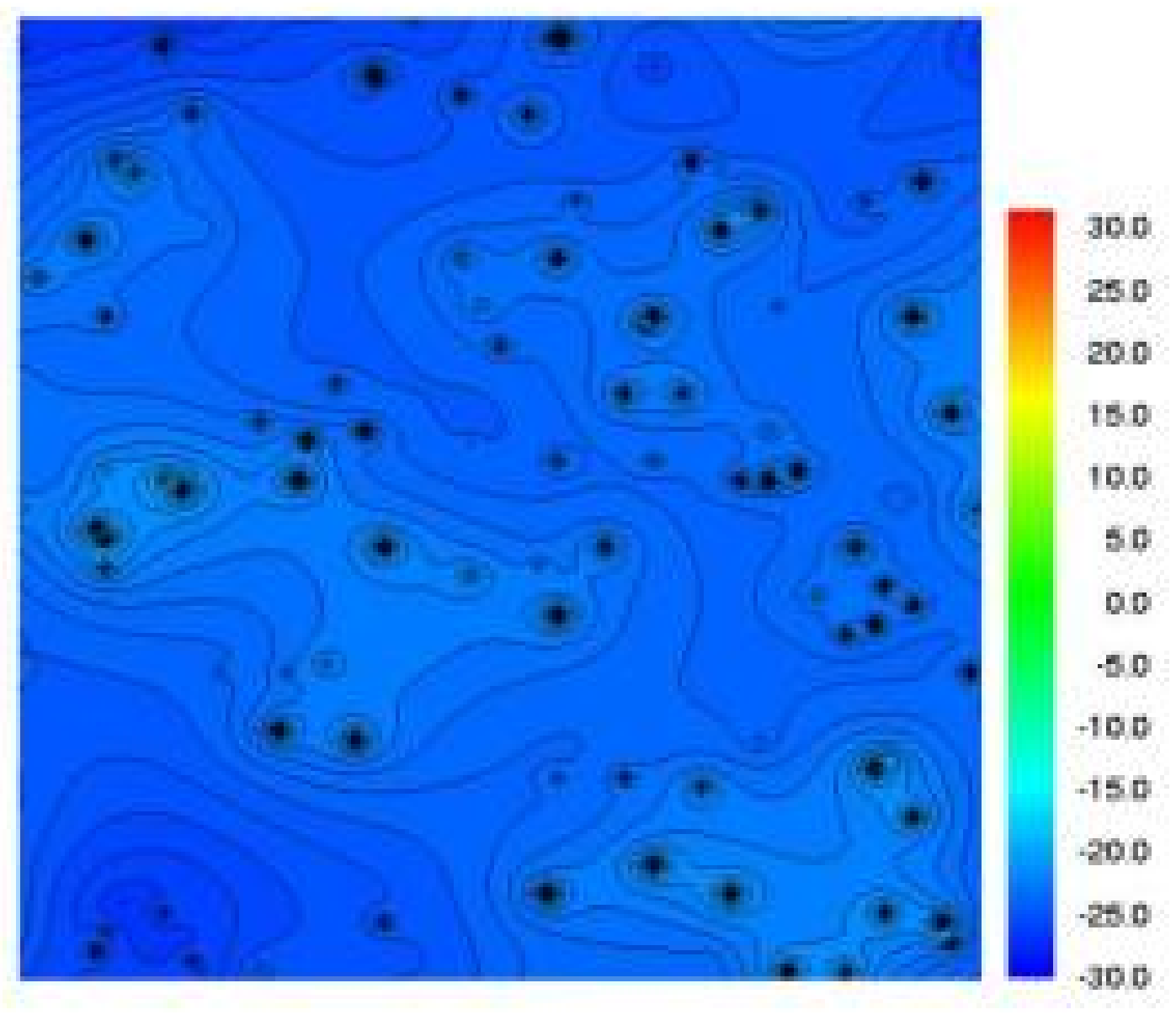,height=0.2\textheight}}
    }}
    \centerline{ \hbox{
    \subfigure[$\tau/\tau_0 = 0.00$]{
    \epsfig{file=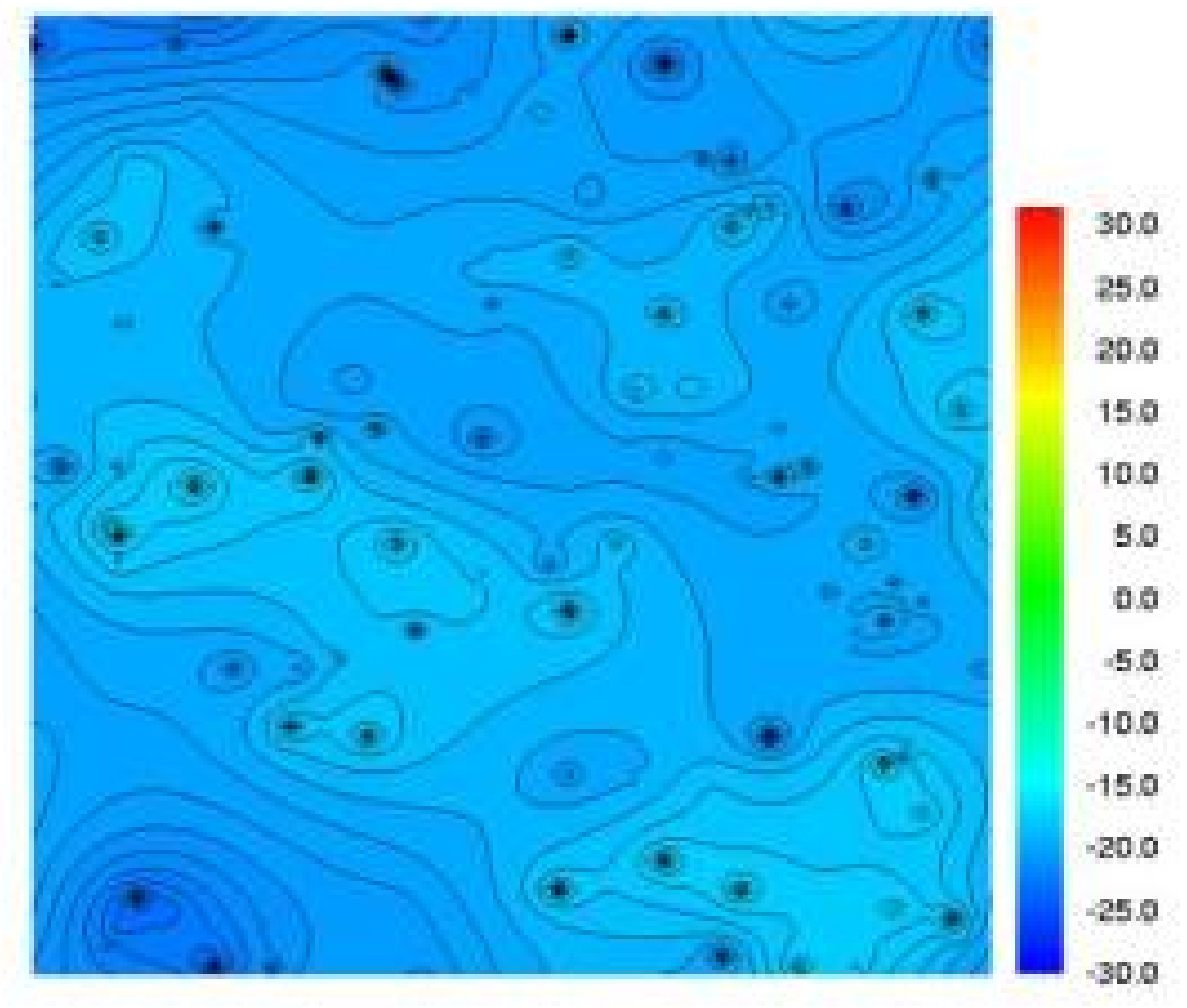,height=0.2\textheight}}
    \hglue -0.2in \subfigure[$\tau/\tau_0 = -0.50$]{
    \epsfig{file=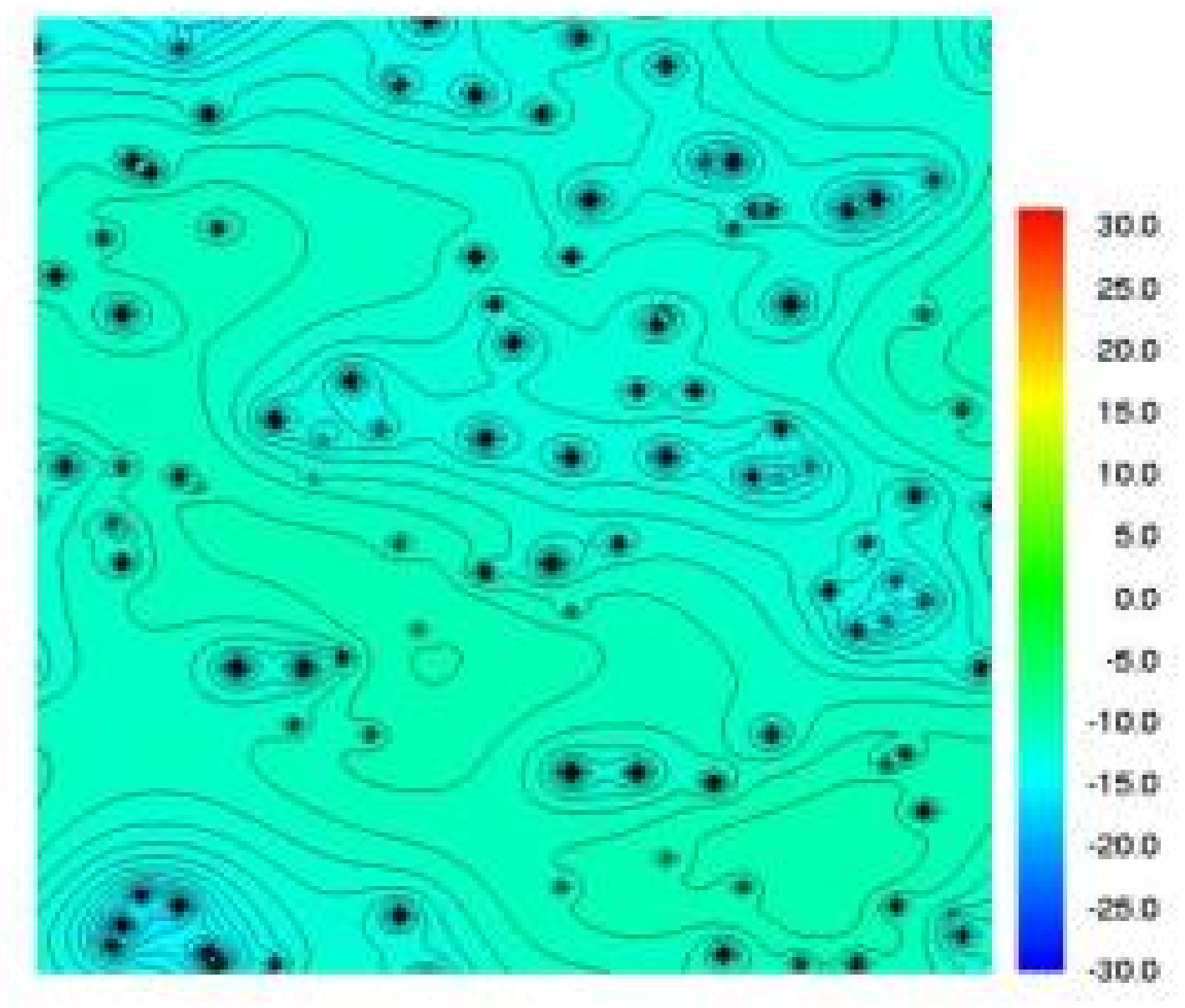,height=0.2\textheight}}
    \hglue -0.2in \subfigure[$\tau/\tau_0 = 0.99$]{
    \epsfig{file=./fig/pattern-nu03-t095.eps,height=0.2\textheight}}
    }}
    \caption{Evolution of the dislocation pattern in response to cyclic
    loading.  Figs. (a)--(i) correspond to $\tau/\tau_0$ = 0.99, 0.50,
    0.00, -0.50, -0.99, -0.50, 0.00, 0.50, 0.99, respectively.}
    \label{fig:loadunload-pattern}
\end{figure}

\begin{figure}
    \centerline{ \hbox{
    \subfigure[$\tau/\tau_0 = 0.00$ ]{
    \epsfig{file=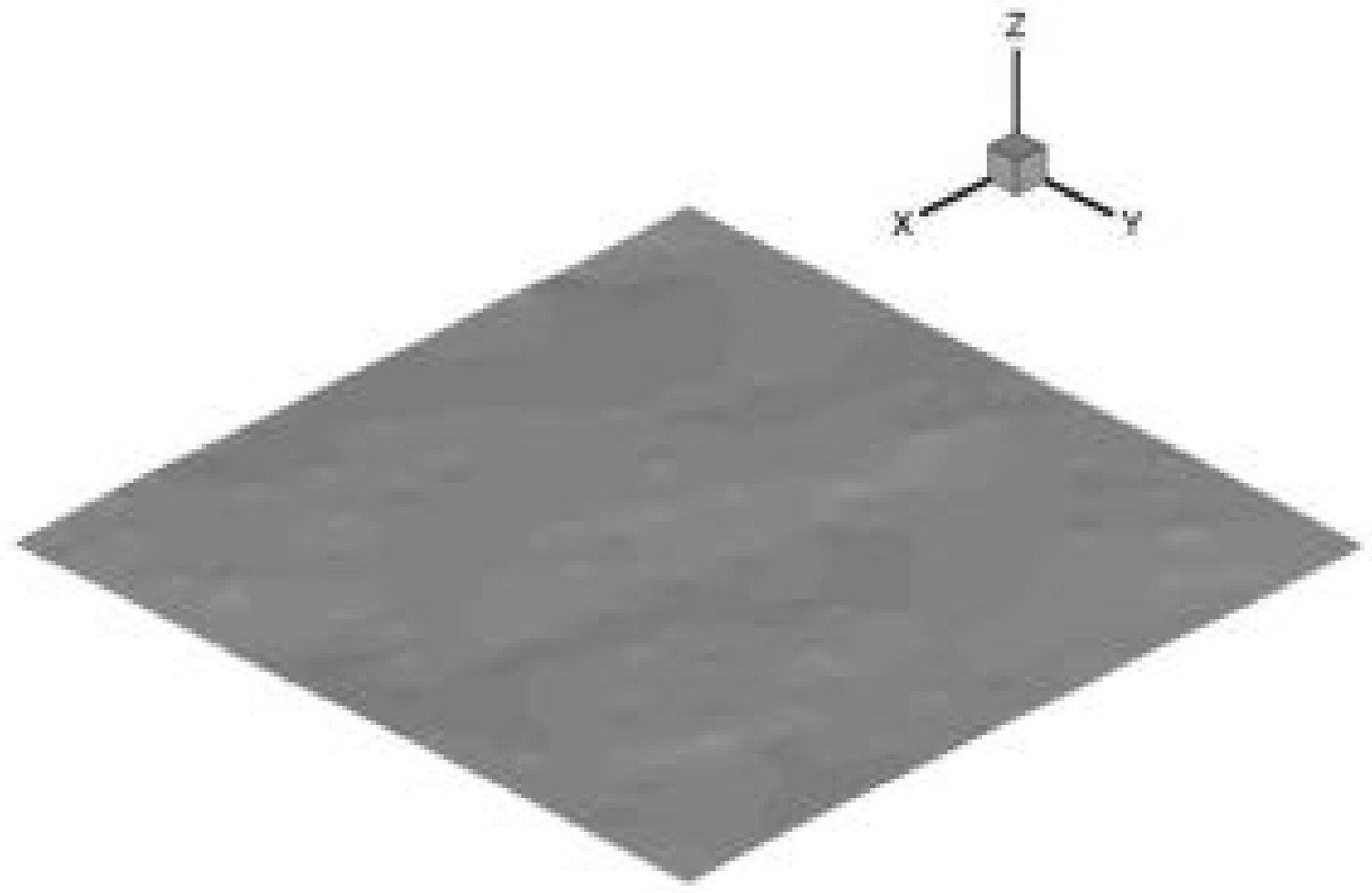,height=0.2\textheight}}
    \hglue -0.2in \subfigure[$\tau/\tau_0 = 0.30$]{
    \epsfig{file=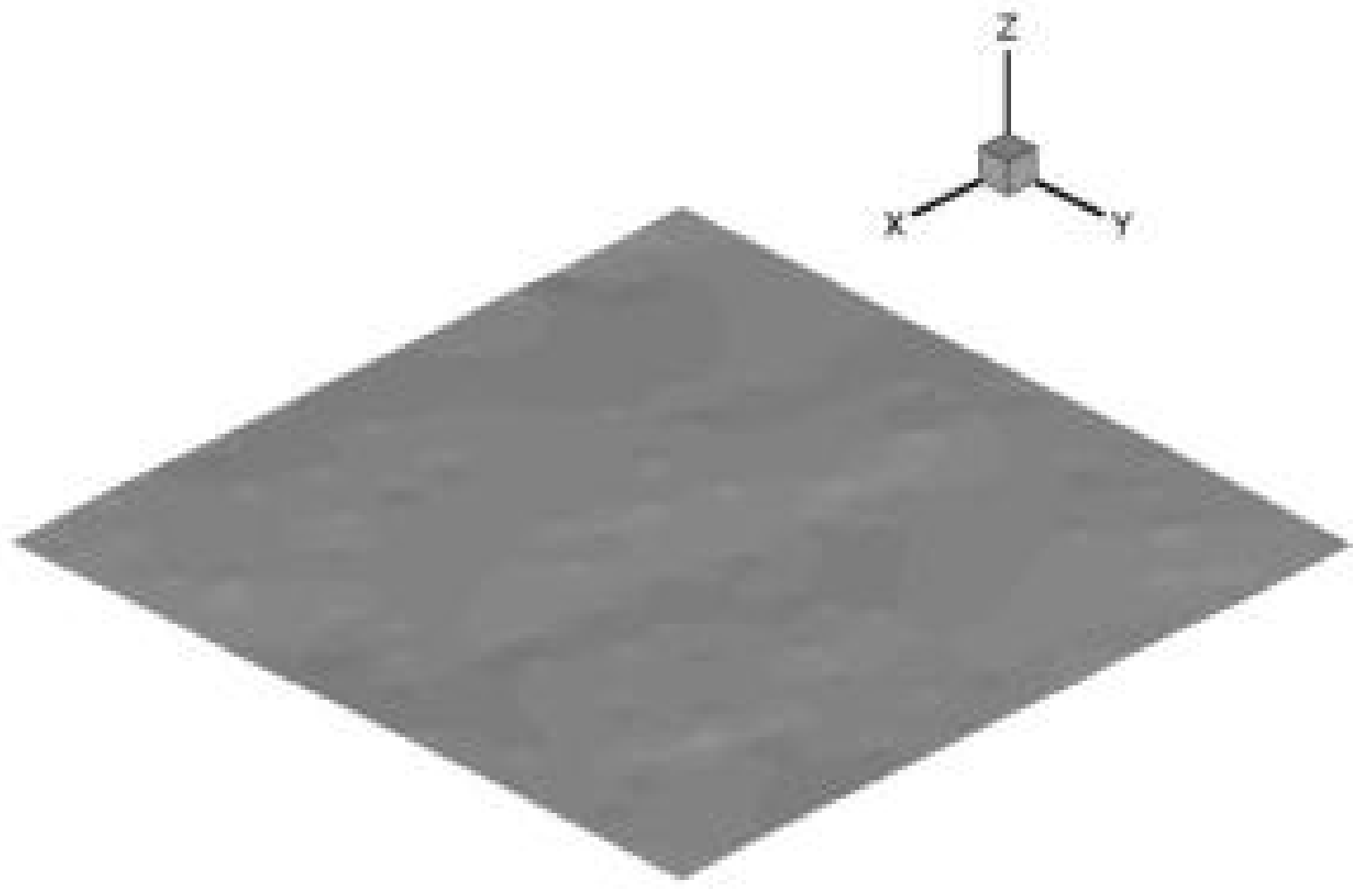,height=0.2\textheight}}
    \hglue -0.2in \subfigure[$\tau/\tau_0 = 0.60$]{
    \epsfig{file=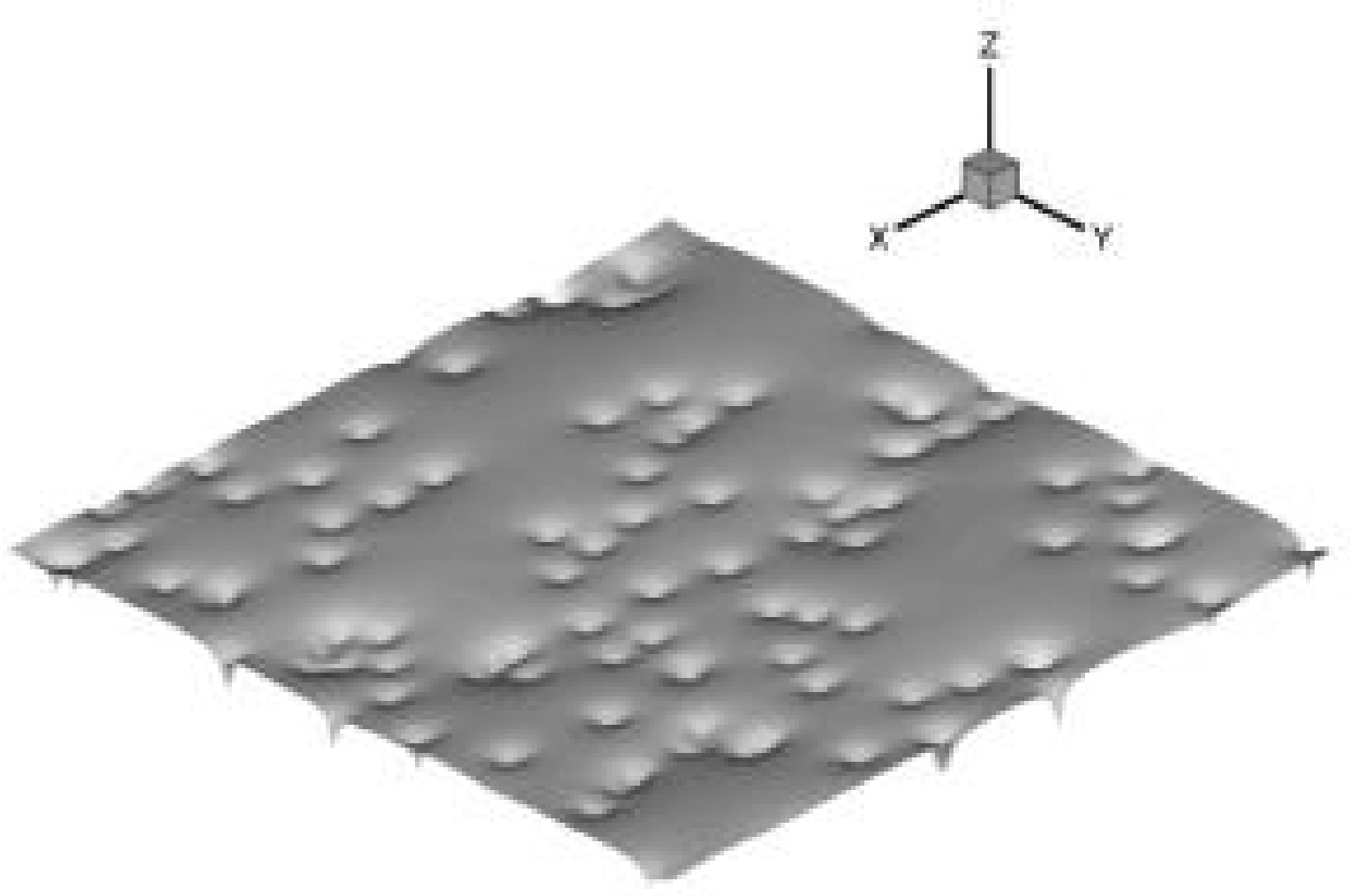,height=0.2\textheight}}
    }}
    \centerline{ \hbox{
    \subfigure[$\tau/\tau_0 = 0.99$]{
    \epsfig{file=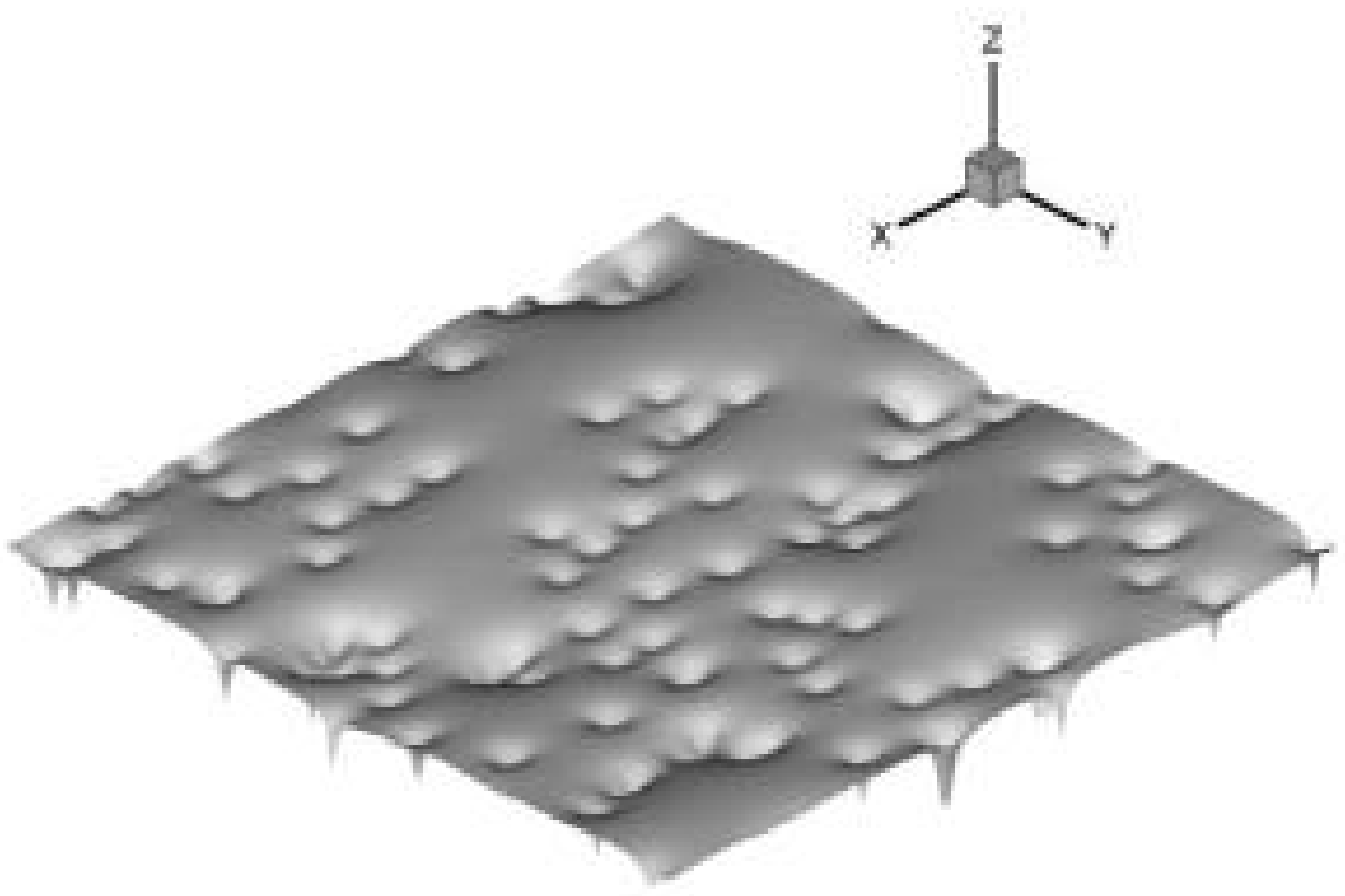,height=0.2\textheight}}
    \hglue -0.2in \subfigure[$\tau/\tau_0 = 0.60$]{
    \epsfig{file=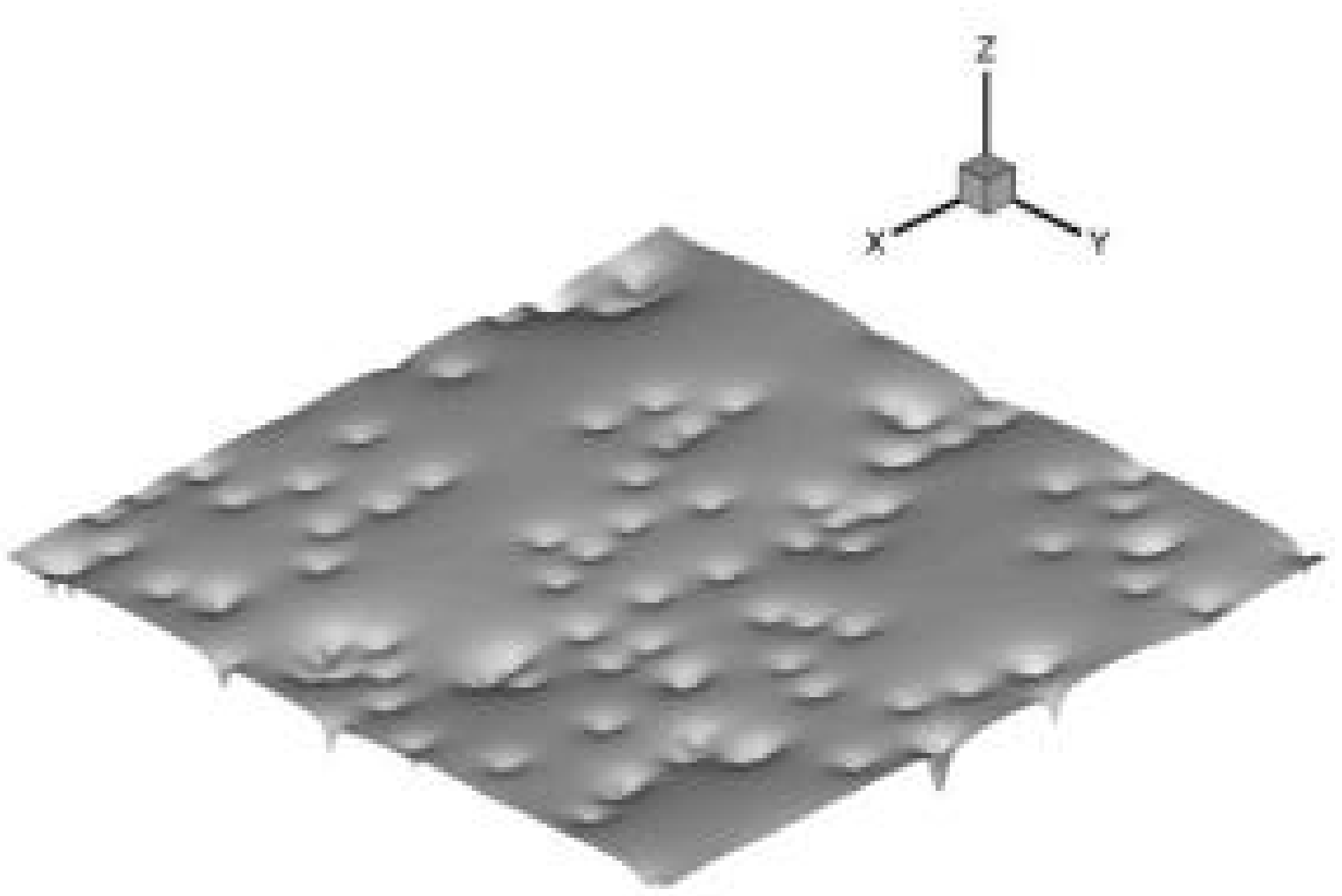,height=0.2\textheight}}
    \hglue -0.2in \subfigure[$\tau/\tau_0 = 0.00$]{
    \epsfig{file=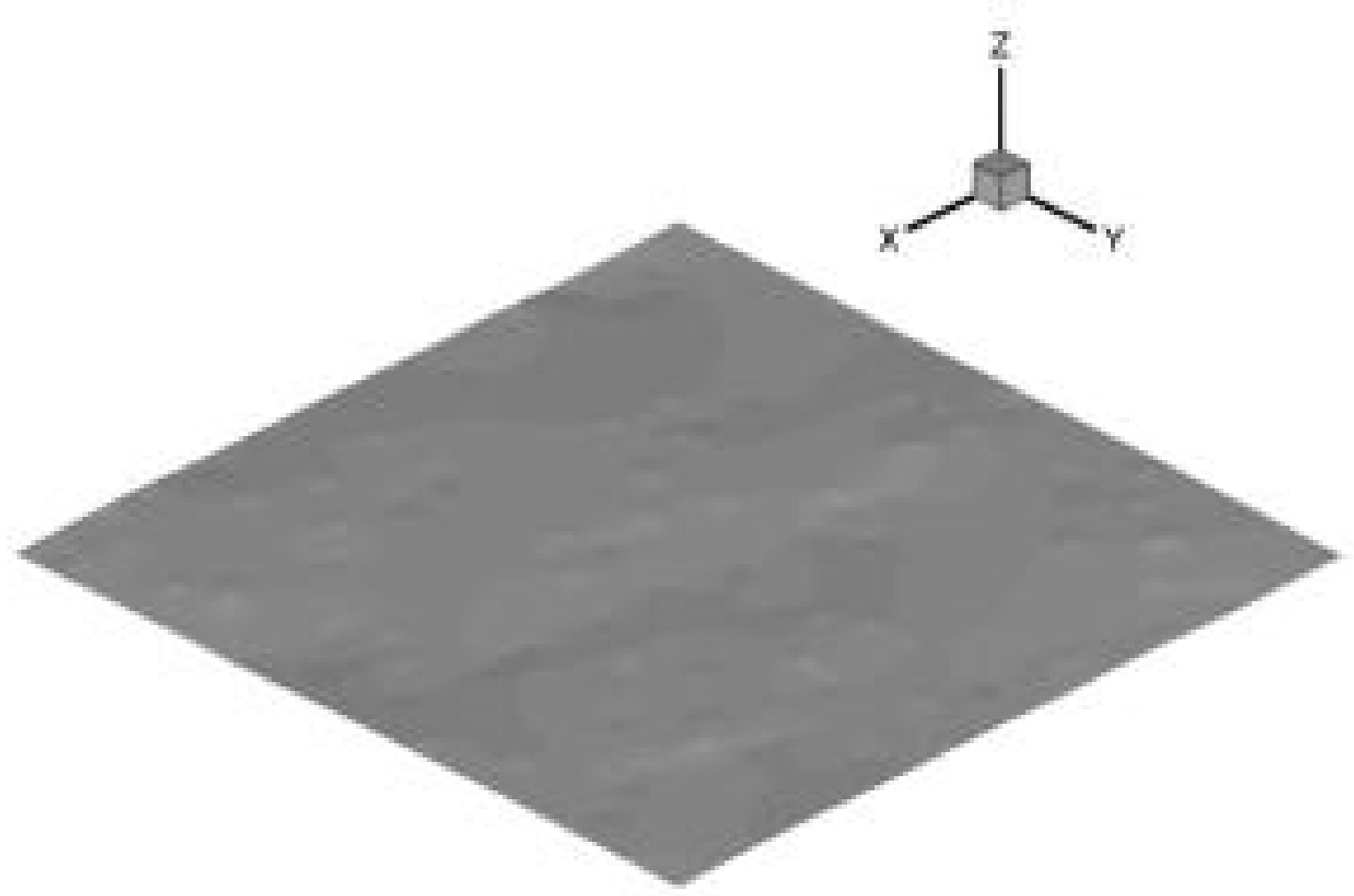,height=0.2\textheight}}
    }}
    \centerline{ \hbox{
    \subfigure[$\tau/\tau_0 = -0.30$]{
    \epsfig{file=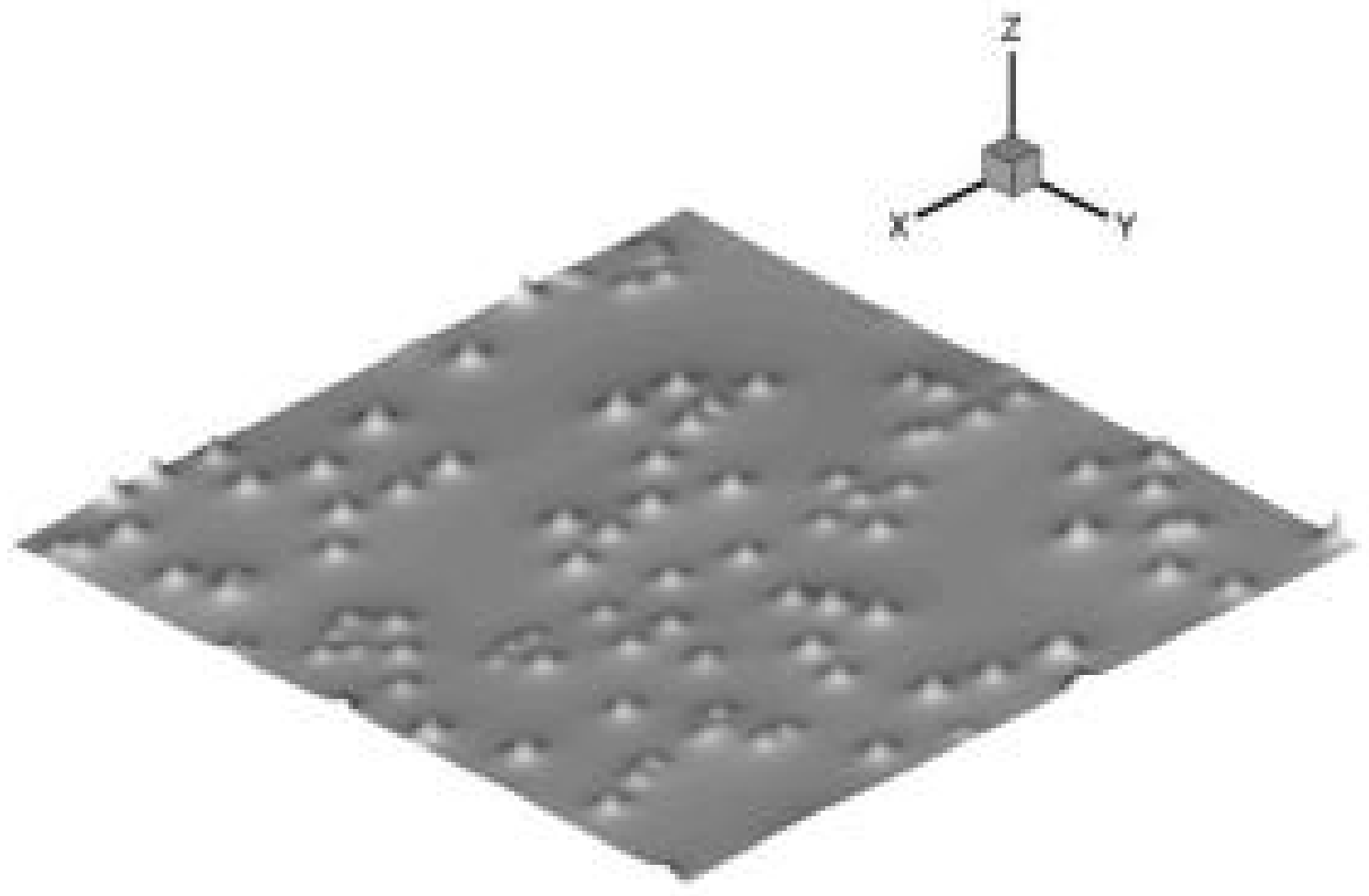,height=0.2\textheight}}
    \hglue -0.2in
    \subfigure[$\tau/\tau_0 = -0.60$]{
    \epsfig{file=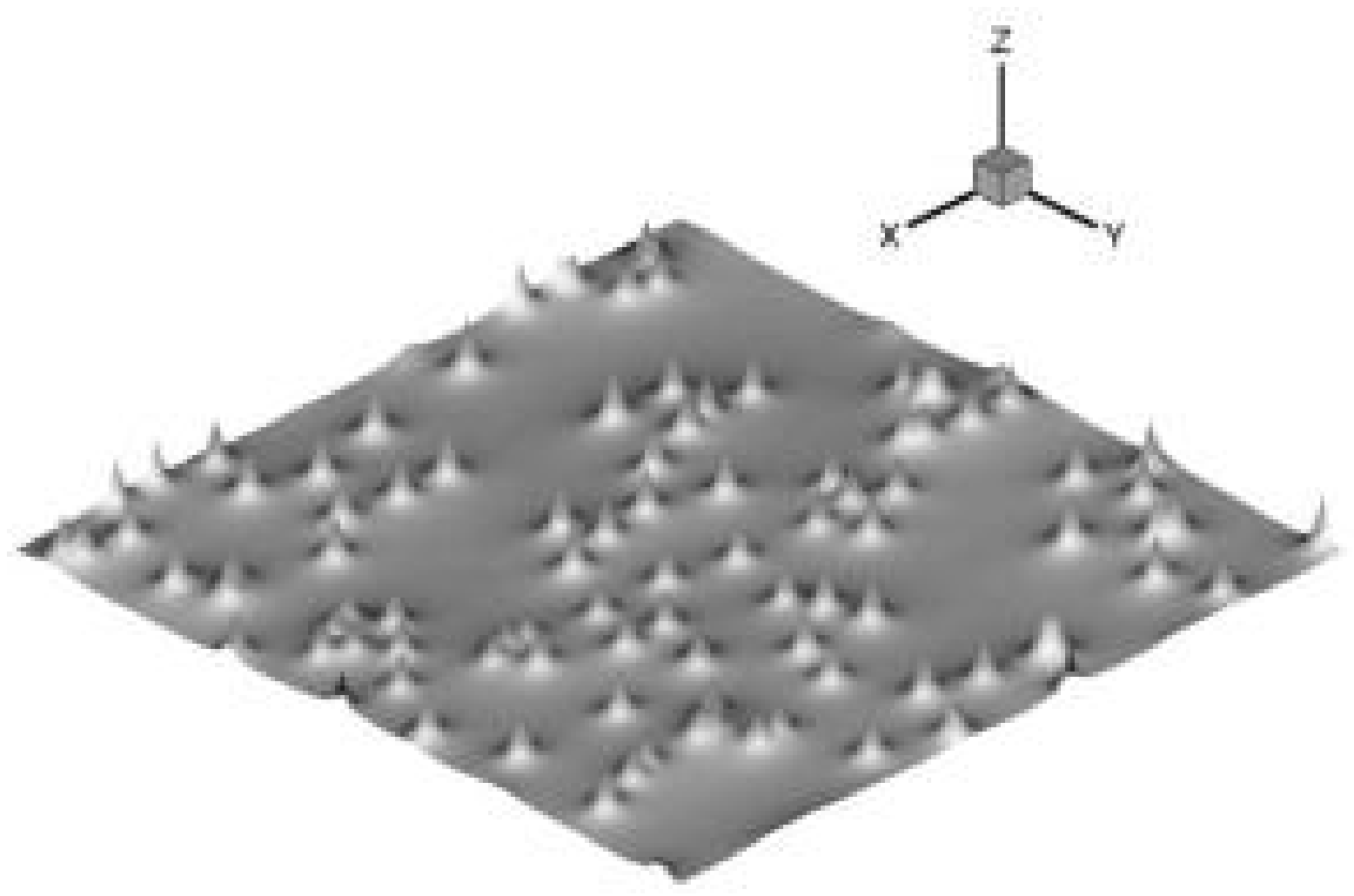,height=0.2\textheight}}
    \hglue -0.2in
    \subfigure[$\tau/\tau_0 = -0.99$]{
    \epsfig{file=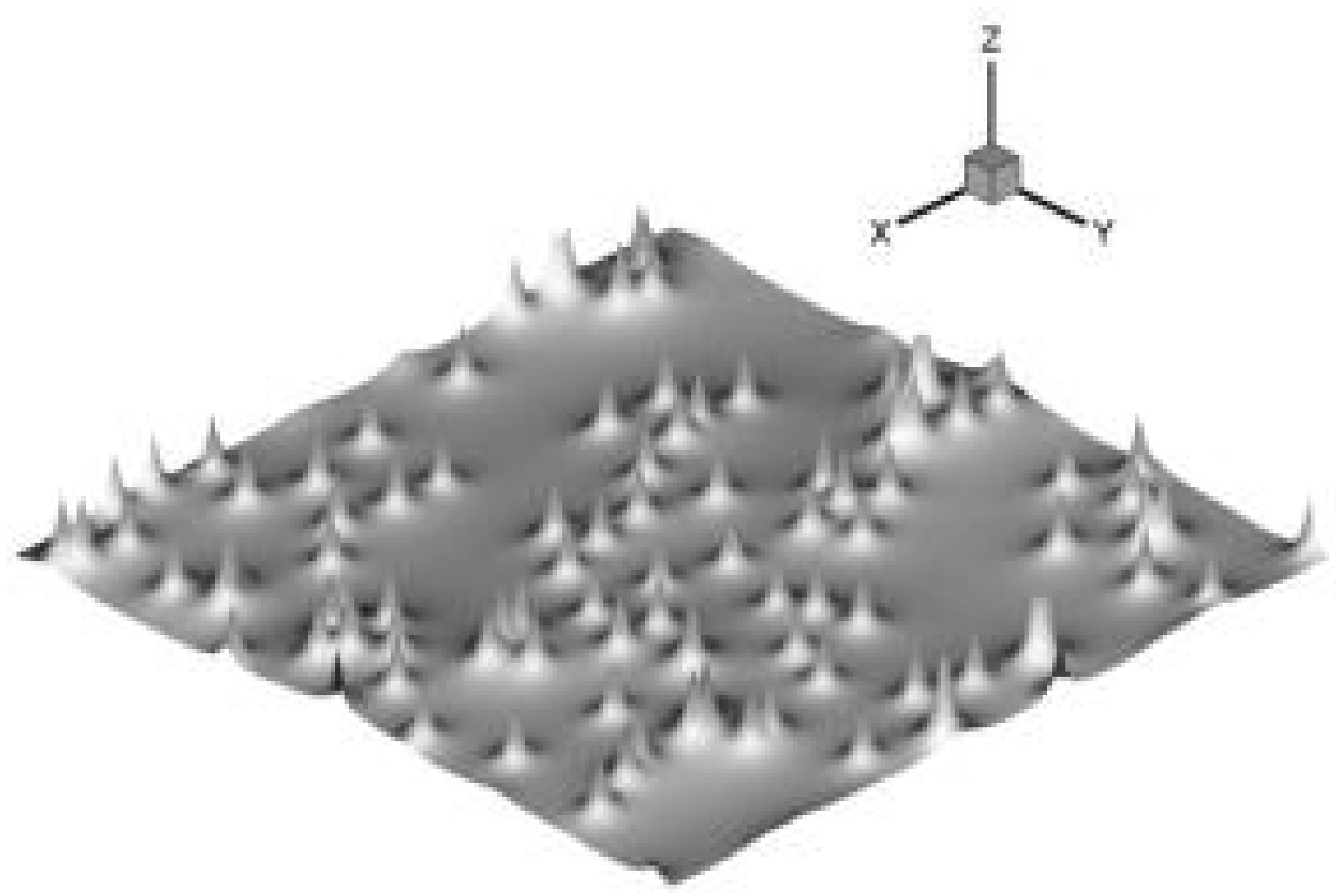,height=0.2\textheight}}
    }}
    \caption{Three-dimensional view of the evolution of the phase
    field during a fully reversed loading cycle, showing the
    switching of the cusps at the obstacles upon unloading and
    reloading. Figs.~(a)--(i) correspond to $\tau/\tau_0$ = 0.00 ,
    0.30, 0.60, 0.99, 0.60, 0.00, -0.30, -0.60, -0.99, respectively.}
    \label{fig:cycle3d}
\end{figure}

\begin{figure}
\begin{center}
    \epsfig{file=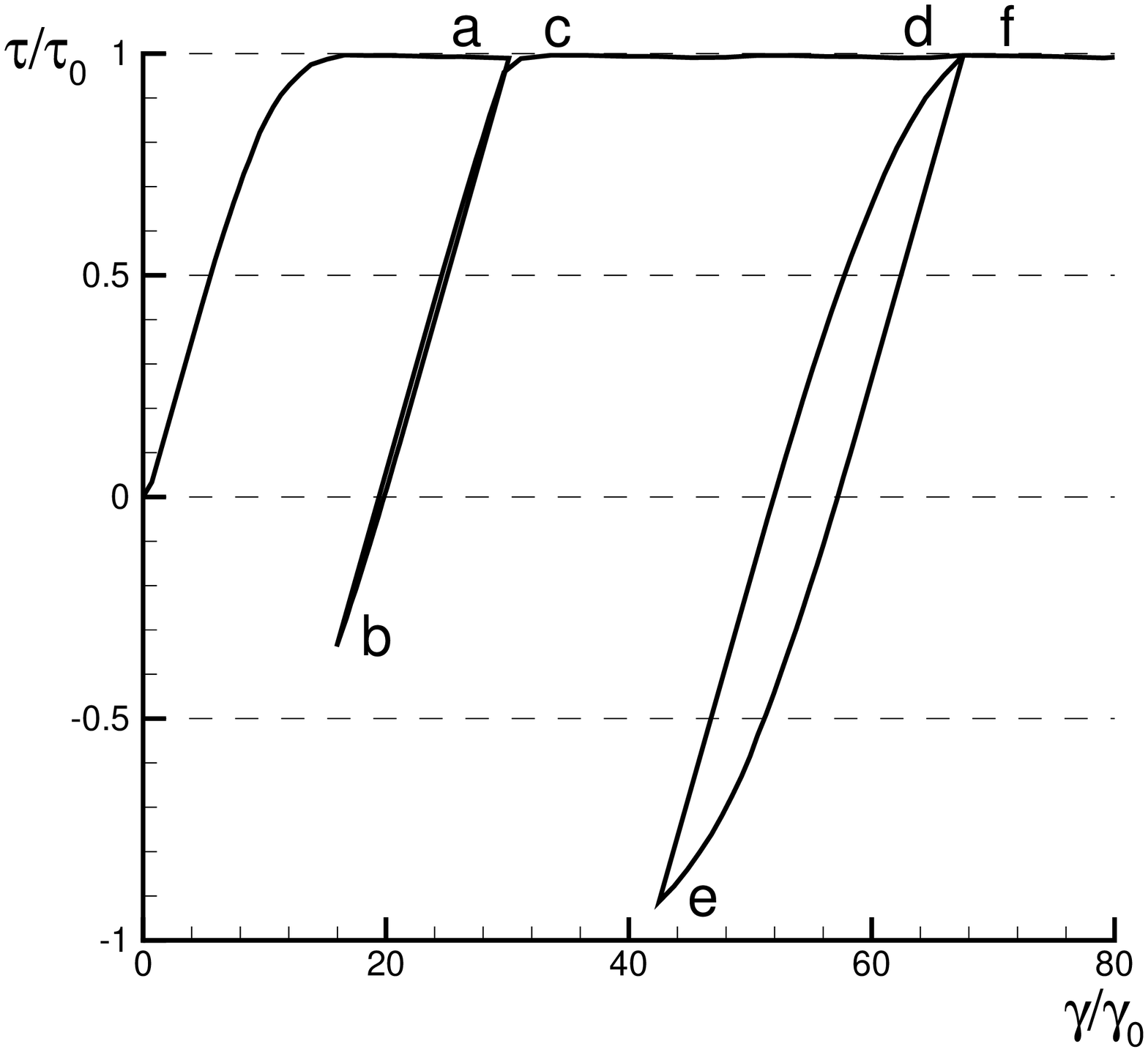,height=0.35\textheight}
    \caption{Cyclic reloading behavior for various extents of
    reverse yielding, exhibiting fading memory effect. Labels
    a--i indicate the loading sequence.}
    \label{fig:Fading}
\end{center}
\end{figure}

\begin{figure}
    \centerline{ \hbox{
    \subfigure[$c b^2 = 10^{-2}$]{
    \epsfig{file=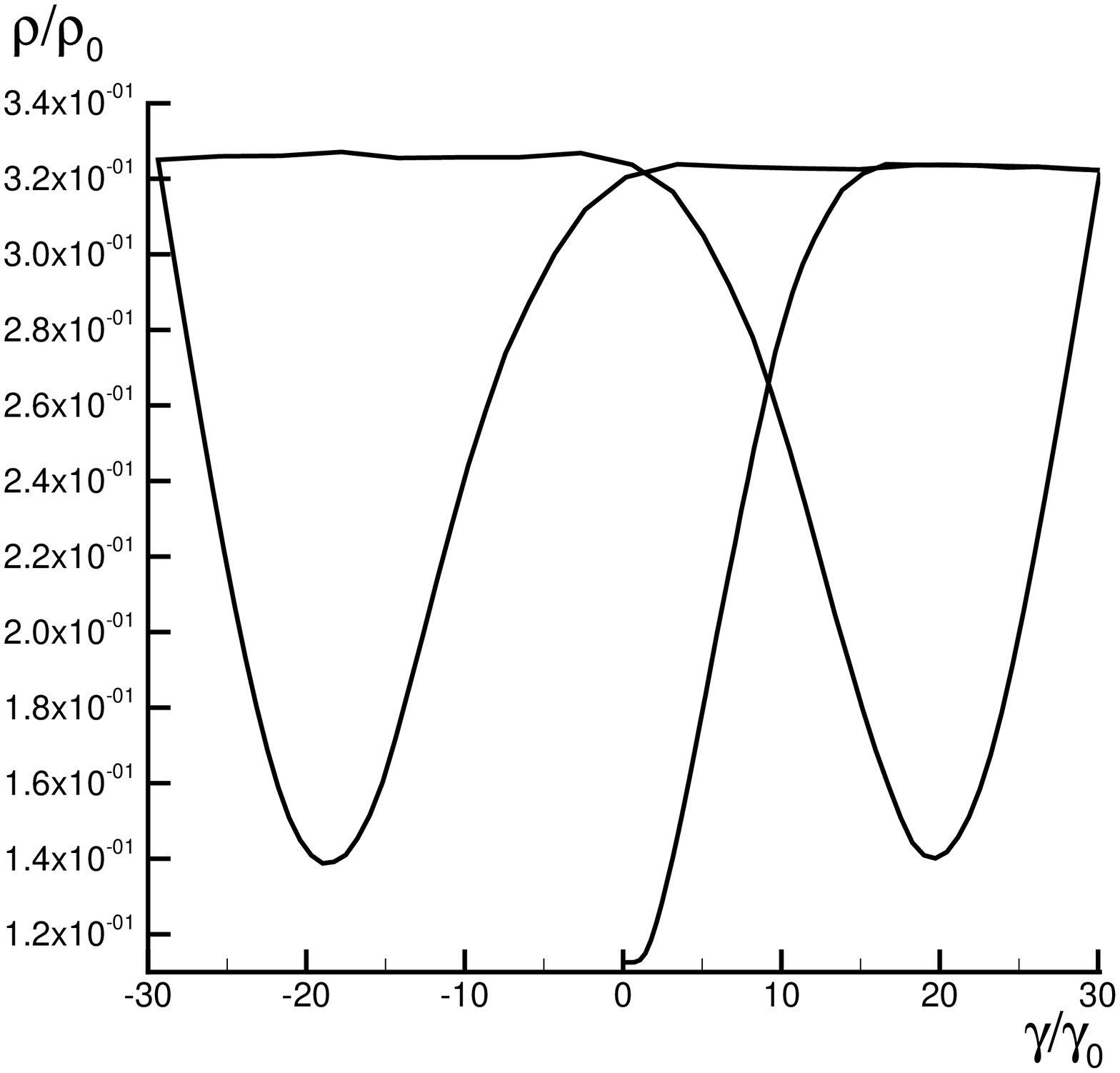,height=0.35\textheight}}
    \hglue -0.2in 
    \subfigure[$c b^2 = 10^{-4}$]{
    \epsfig{file=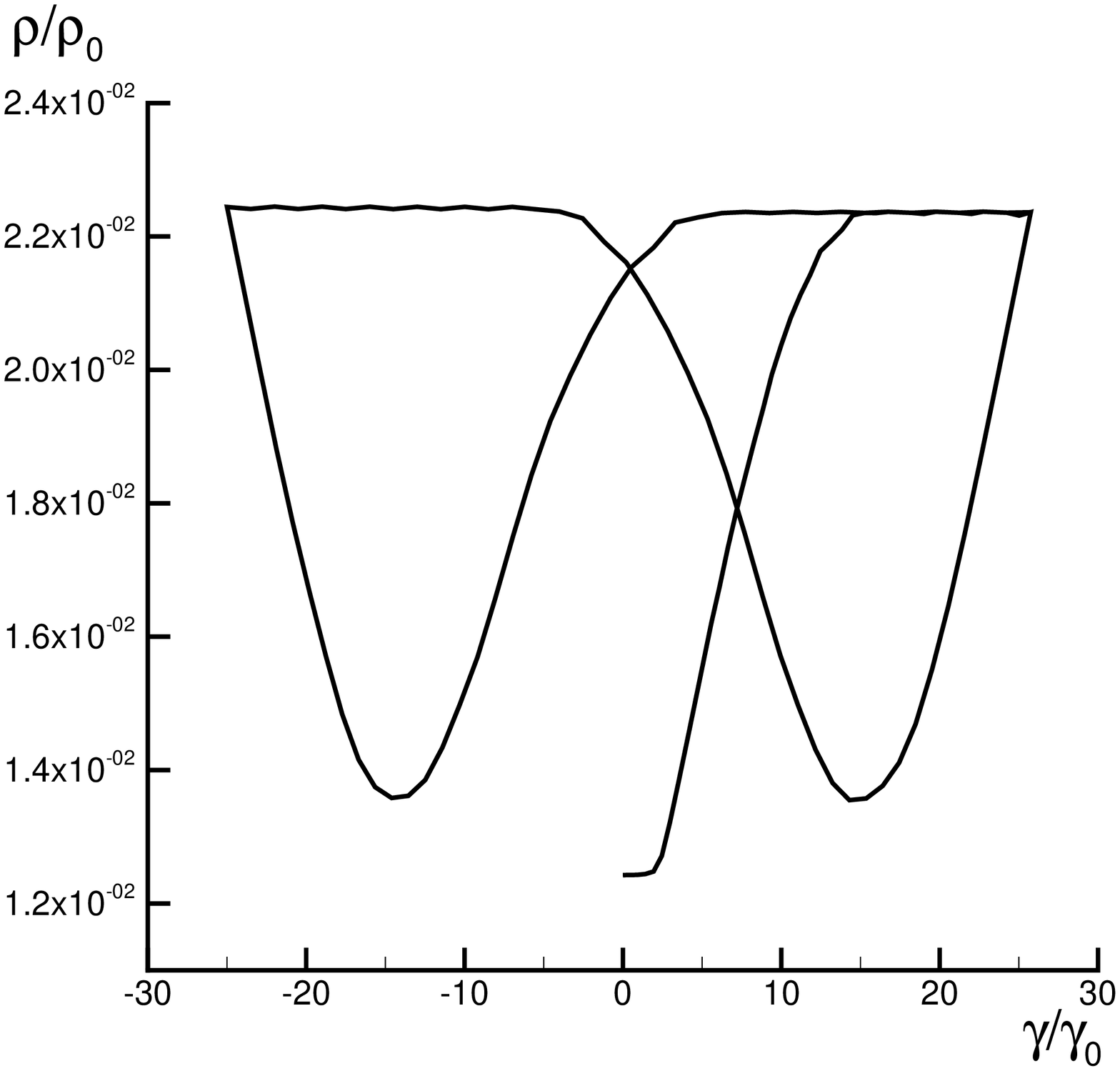,height=0.35\textheight}}
    }}
    \centerline{ \hbox{
    \subfigure[$c b^2 = 10^{-6}$]{
    \epsfig{file=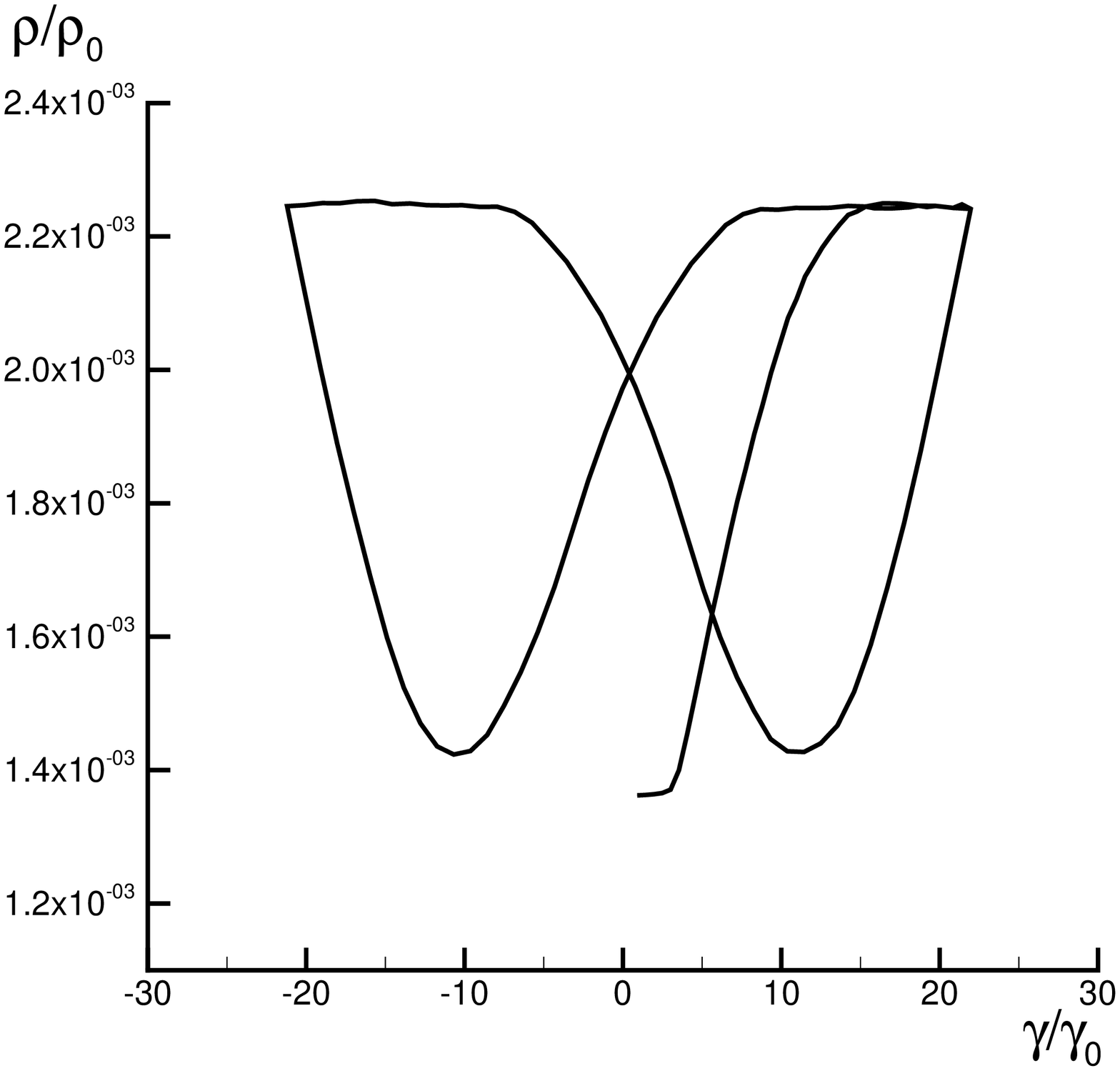,height=0.35\textheight}}
    \hglue -0.2in 
    \subfigure[$c b^2 = 10^{-8}$]{
    \epsfig{file=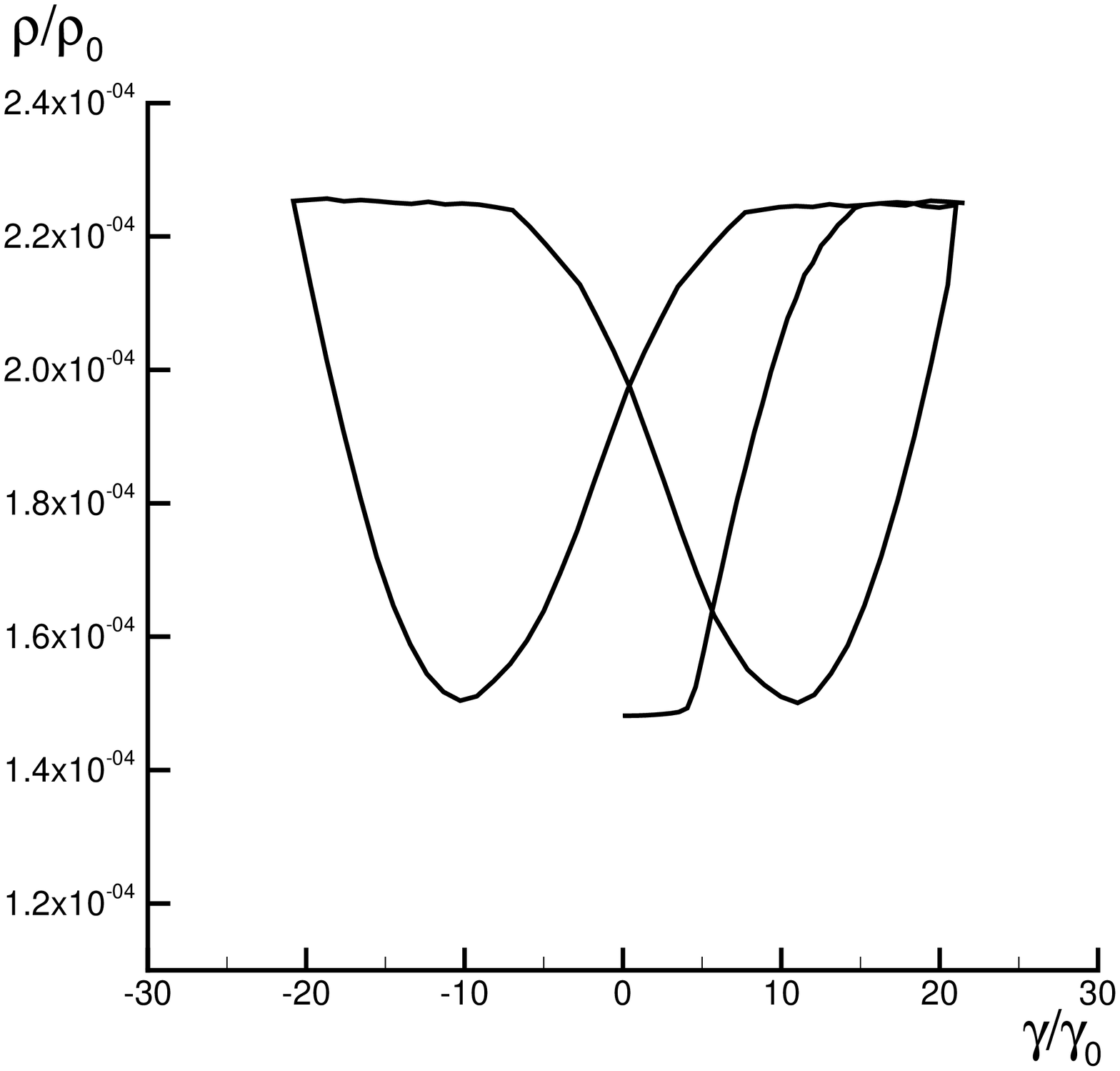,height=0.35\textheight}}
    }}
    \caption{Evolution of the dislocation density with macroscopic
    slip strain during cyclic loading for different values of
    the obstacle density $c$.}
    \label{fig:cycle_rho}
\end{figure}

Next we investigate the behavior predicted by the theory under load
reversal and cyclic loading. As in the preceding example, we assume a
periodicity and take the unit cell $\Omega$ to be a square of
dimension $100 b$.  We randomly select $100$ points within the
periodic cell as the obstacle sites and assign to these obstacles a
uniform strengths $f = 10 \mu b^2$. The Poisson's ratio $\nu$ is set
to $0.3$, and, for simplicity, we take the Peierls stress $\tau^P =
0$. We consider the cases of partial unloading, where the system is
reloaded after a small amount of reverse yielding; and a
fully-reversed loading cycle. In both cases, the unloading and
reloading events bring to the fore the irreversible and dissipative
nature of the system, and necessitate full use of the incremental
theory developed in Section~\ref{sec:Irreversible_processes}.

Figs.~\ref{fig:cycle}a and b show the predicted stress-strain
curve and the evolution of the dislocation density in the cases
of fully-reversed loading. The corresponding evolution of the
dislocation ensemble is shown in
Fig.~\ref{fig:loadunload-pattern} for the case of fully-reversed
loading. The various matching sections of these curves are
labeled for ease of reference. The interval O--a of the curves in
Fig.~\ref{fig:cycle} simply reproduces the monotonic response
described in the preceding section. Upon load reversal at point
a, the system initially unloads elastically,
Fig.~\ref{fig:cycle}a, segment a--c. In this stage, the applied
shear stress continues to push the dislocations against the
obstacles in the loading direction. However, as the applied load
is decreased the force exerted by the dislocations on the
obstacles correspondingly decreases and, consequently, remains
below the obstacle strength. Under these conditions, the
dislocations remain pinned at the obstacles and simply recoil
elastically. This process gives the linear unloading branch a--c
in the stress-strain curve, Fig.~\ref{fig:cycle}a.

It should be carefully noted that, owing to the frictional sliding
of the dislocations over the obstacles that occurs during the
loading state, the interaction forces between the dislocations
and the obstacles do not vanish when the applied shear stress is
reduced to zero. Instead, the system is left in a state of
permanent or residual deformation in which the
dislocation/obstacle force system is self-equilibrated. The
corresponding dislocation pattern for the fully-reversed case is
shown in Fig.~\ref{fig:loadunload-pattern}c. This remanent field
is in analogy to the self-equilibrated residual stress fields
which remain in elastic-plastic solids upon unloading. The
combination of a reversed applied stress and the residual force
field causes the system to yield prematurely in the reserve
direction, Fig.~\ref{fig:cycle}a, point d, and thus the theory
predicts the Bauschinger effect.

Once reverse yielding commences, the obstacles continue to oppose the
motion of the dislocations, which now takes place in the reverse
loading direction. In particular, the frictional forces at the
obstacles switch sign relative to the loading phase. This process of
switching is clearly apparent in Fig.~\ref{fig:cycle3d}, which shows
the evolution of the phase field during the loading cycle. Thus,
during the loading phase the obstacles pull down on the phase field,
Fig.~\ref{fig:cycle3d}b-d, causing it to cusp downward at the
obstacles, whereas during the reverse loading phase the obstacles pull
up on the phase field, Fig.~\ref{fig:cycle3d}g-i, causing it to cusp
upward.

The stress-strain curve shown in Fig.~\ref{fig:Fading} reveals
that the system exhibits `fading memory'. Thus, when the system
is reloaded, the stress-strain curve gradually transitions
towards the virgin loading curve, and, with sufficient reloading,
the system eventually `forgets' the unloading cycle. The extent
of this loss of memory depends on the extent of reverse yielding.
Consider, for instance, the case in which unloading is purely
elastic and no reverse yielding occurs. This  corresponds to
reloading, from point b in Fig.~\ref{fig:Fading}. Because no
reverse sliding over obstacles has occurred at this point, the
dislocation/obstacle configuration remains undisturbed.
Consequently, upon reloading all the obstacles yield
simultaneously and the system exhibits an abrupt yield point
coinciding exactly with the point of unloading, point a in
Fig.~\ref{fig:Fading}. Thus, in this case no loss of memory
occurs and the system exhibits \emph{return-point memory} in the
sense of Sethna {\it et al} \cite{sethna:1993}. This type of
behavior should now be compared with that corresponding to a
limited amount of reverse loading. In this case, the dislocations
slip over the obstacles in the reverse loading direction, and
some of the details of the dislocation/obstacle configuration
established at the unloading point d are lost in the process.
When the system is reloaded, point e in Fig. \ref{fig:Fading},
the obstacles yield anew gradually and not all at once, as in the
case of a purely elastic unloading, and the system exhibits a
gradual transition towards the virgin loading curve,
Fig.~\ref{fig:Fading}, segment e-f.  In the case of
fully-reversed unloading, Fig.~\ref{fig:cycle}, the amount of
reverse yielding, segment d-e, is large enough that the unloading
point is effectively wiped out from the memory of the system,
with the result that the reloading curve e-i is ostensibly
identical to the unloading curve a-e. These trends are in good
agreement with the experimental cyclic stress-strain data for
structural steels reported by Ortiz and Popov \cite{ortiz:1982},
which was obtained from tests specially designed to exhibit the
fading memory effect just described.

The evolution of the dislocation density during a loading cycle is
of considerable interest as well, Fig.~\ref{fig:cycle}b . Thus,
upon unloading the dislocation density decreases as a result of
the elastic relaxation of the dislocation lines.  The dislocation
density bottoms out --- but does not vanish entirely
--- upon the removal of the applied stress,
Fig.~\ref{fig:cycle} b, point c, as some dislocations remain
locked in within the system in the residual state,
Fig.~\ref{fig:loadunload-pattern}c.  The dislocation density
increases again during reverse loading, Fig.~\ref{fig:cycle} b,
segment c-d, and the cycle is repeated during reloading,
Fig.~\ref{fig:cycle}b, segment c-d, segment e-i, giving rise to a
dislocation density {\it vs.} slip strain curve in the form of a
`butterfly'. This type of behavior is indeed observed
experimentally \cite{Morrow:1975}.  It also arises in models of
the stored energy of cold work \cite{BonderLindenfeld:1995}, and
is in analogy to the hysteretic loops exhibited by magnetic
systems \cite{kinderlehrer:1997, sethna:1993, dahmen:1994}.

Finally, we exercise the model over the range of periodic cell sizes
$10^2 b - 10^5 b$, while at the same time keeping the number of
obstacles constant at 100. The corresponding obstacles density
$c=N/|\Omega|$ consequently varies from $10^{-2}/b^2$ to
$10^{-8}/b^2$. Of primary interest here is to ascertain how the
macroscopic behavior of the slip plane depends on the obstacle
density. Our calculations show that the stress-strain curve, when
expressed in terms of the normalized variables $\tau/\tau_0$ and
$\gamma/\gamma_0$ is ostensibly independent of $c$. It should be
carefully noted that, for obstacles of uniform strength, $\tau_0 = f
c$, and thus the saturation strength scales in direct proportion to
the obstacle density. The variation of the dislocation density
evolution with $c$ is shown in Fig.~\ref{fig:cycle_rho}. It is
interesting to note from this figure that the saturation value of the
dislocation density scales as $1/\sqrt{c}$, but that otherwise the
evolution of the dislocation density with macroscopic slip strain
remains ostensibly identical in all cases.

This example also serves to underscore the ability of the theory to
effectively deal with large domains of analysis.  Thus, owing to the
absence of a computational grid the complexity of the calculations
scales with the number of obstacles and is otherwise independent of
the cell size, which enables the consideration of large cells sizes in
the relevant range of observation.

\subsection{Dislocation line-energy anisotropy}

\begin{figure}
    \centerline{ \hbox{
    \subfigure[Stress]{
    \epsfig{file=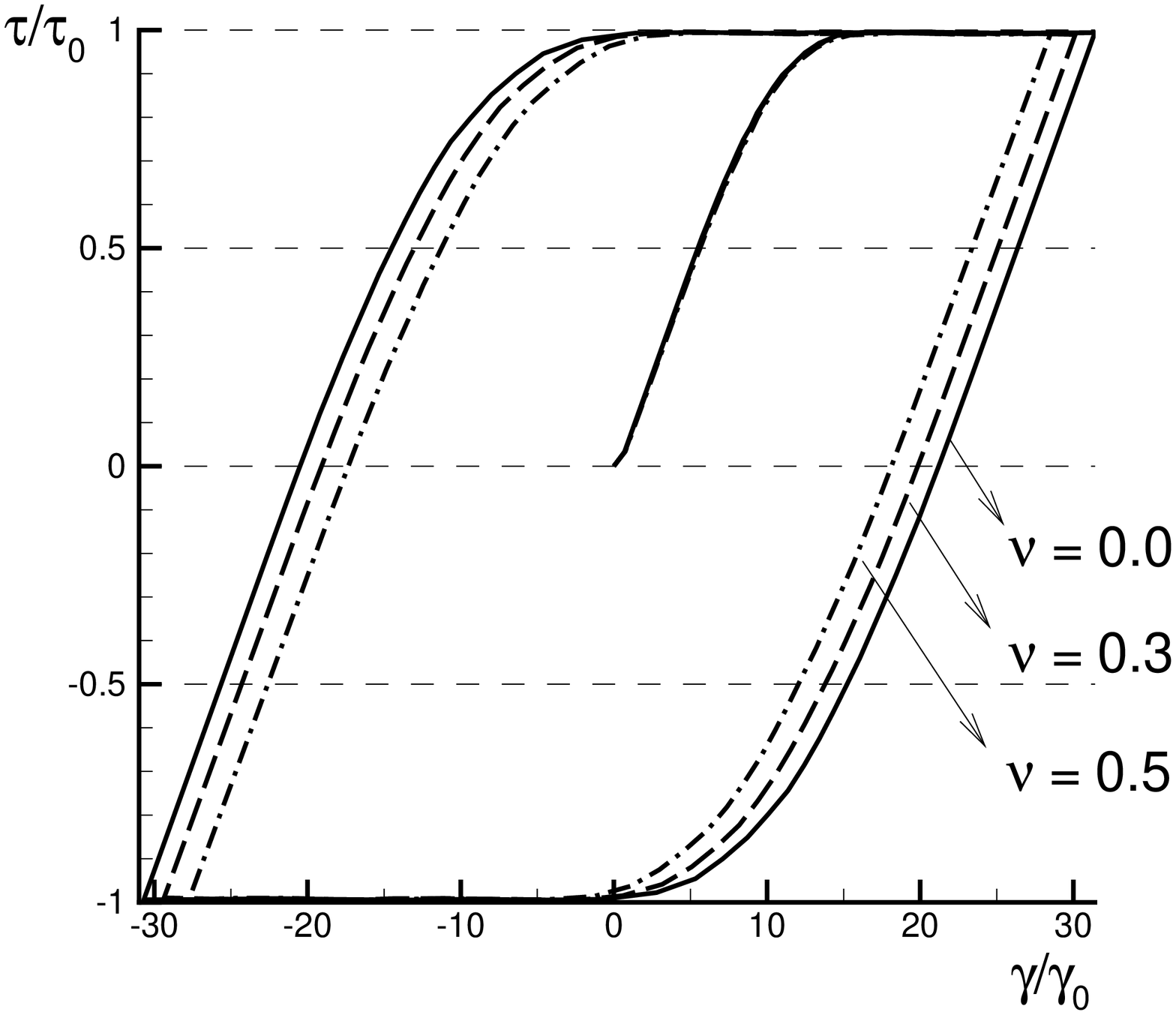,height=0.35\textheight}}
    \hglue -0.2in
    \subfigure[Dislocation Density]{
    \epsfig{file=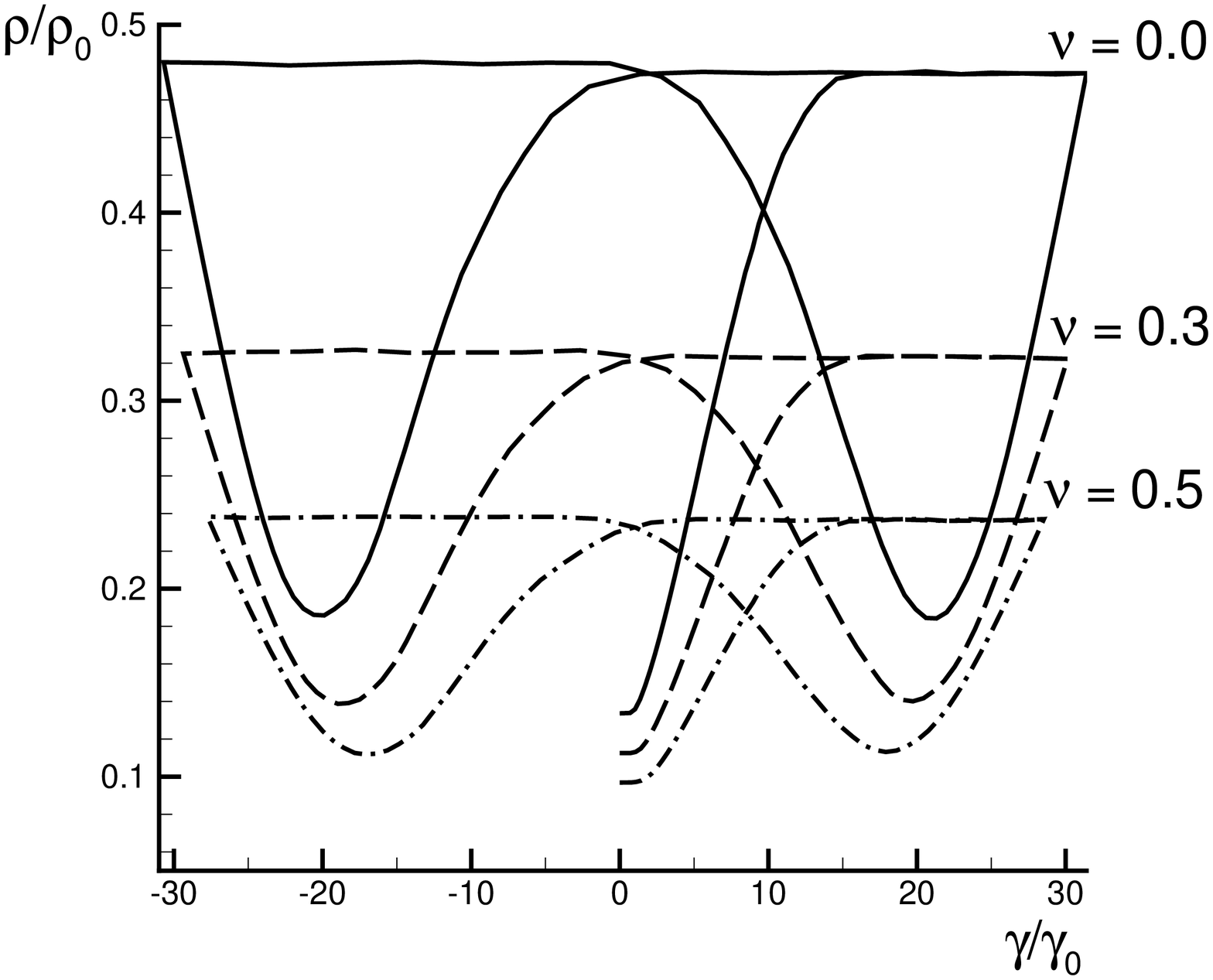,height=0.35\textheight}}
    }}
    \caption{Effect of dislocation line-energy anisotropy on the
    hardening characteristics of the system. (a) Applied shear stress
    {\it vs.} macroscopic slip strain. (b) Evolution of dislocation
    density with macroscopic slip strain.}
    \label{fig:aniso}
\end{figure}

\begin{figure}
    \centerline{ \hbox{
    \subfigure[$\nu = 0.0$] {
    \epsfig{file=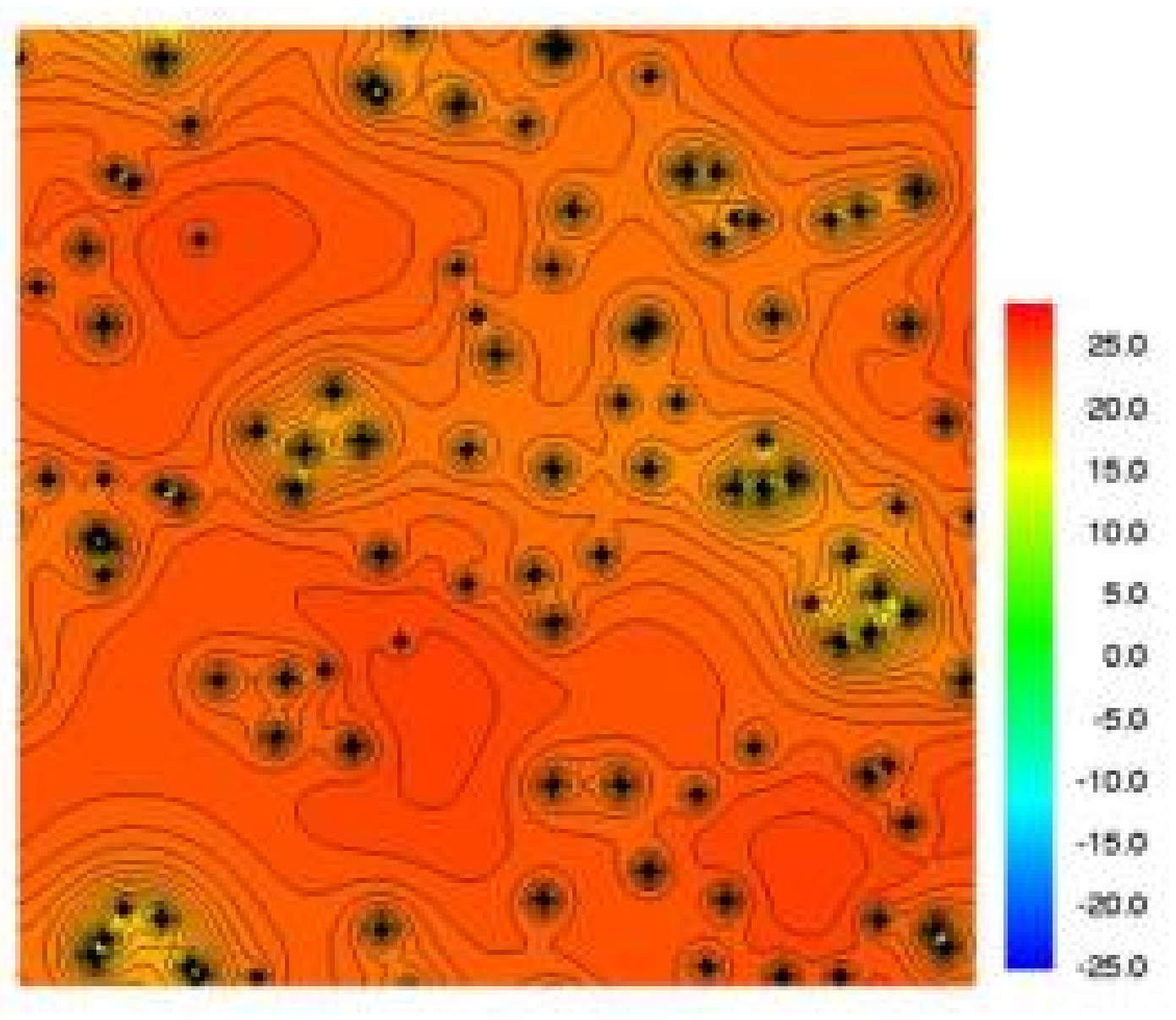,height=0.25\textheight}}
    \hglue -0.73in
    \subfigure[$\nu = 0.3$] {
    \epsfig{file=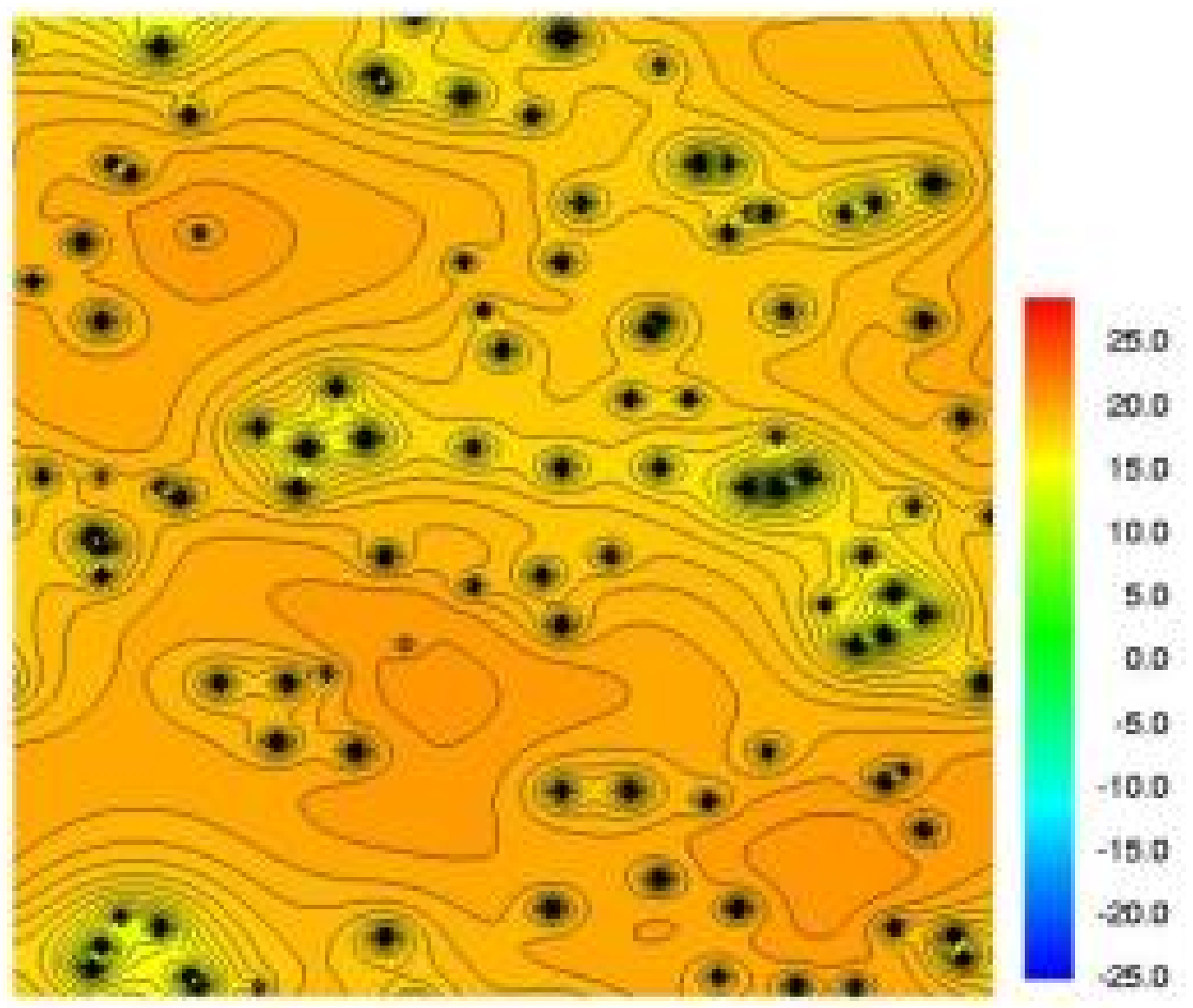,height=0.25\textheight}}
    \hglue -0.7in
    \subfigure[$\nu = 0.5$] {
    \epsfig{file=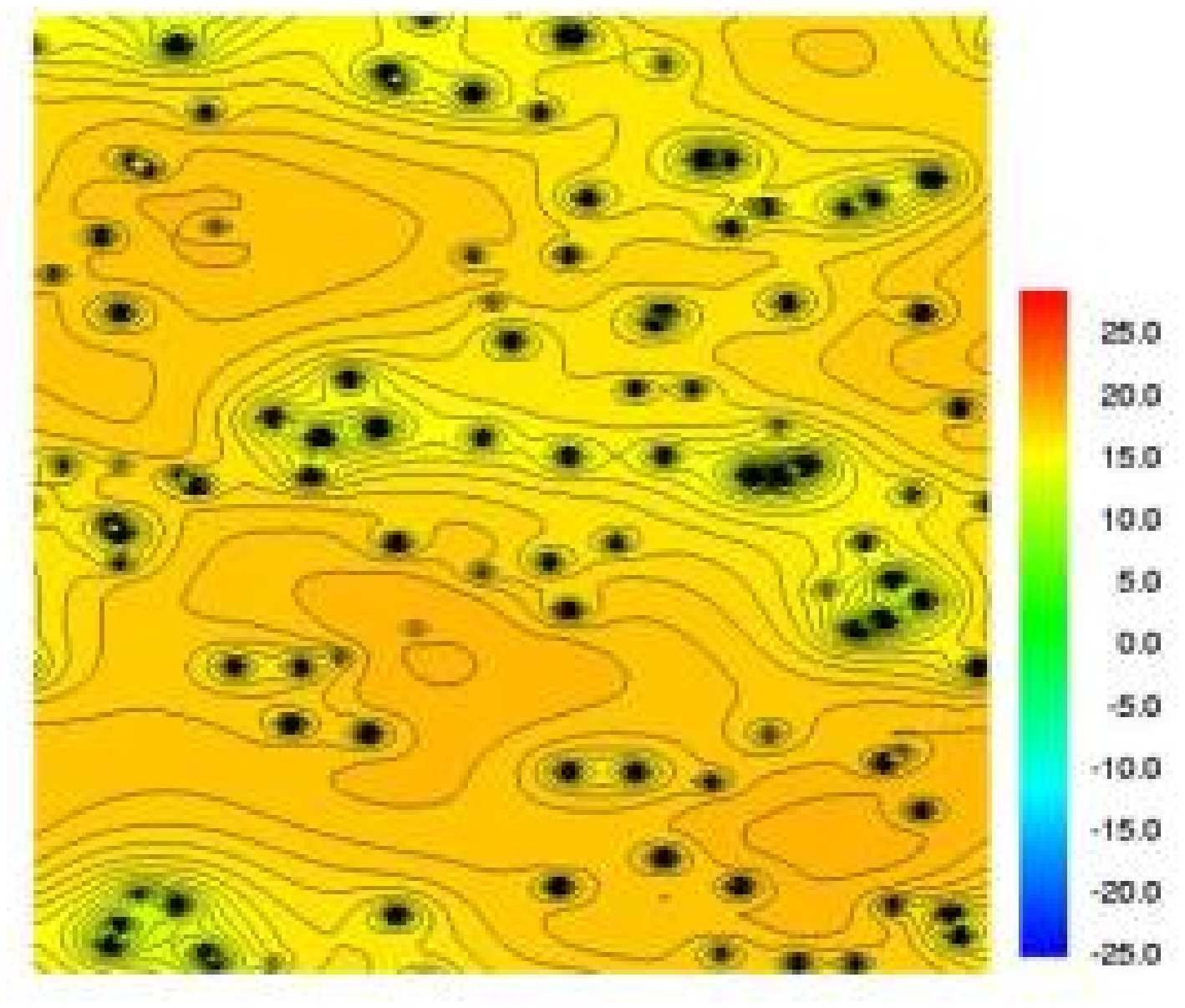,height=0.25\textheight}}
    }}
    \caption{Effect of dislocation line-energy anisotropy on the
    evolution of the dislocation pattern. Figs. (a)--(c) correspond
    to $\tau/\tau_0$ = 0.99, and $\gamma/\gamma_0 $ = 23, 20 and 18,
    respectively.}
    \label{fig:aniso-l}
\end{figure}

We conclude this section with a brief parametric study of the effect
of dislocation line-energy anisotropy on the hardening characteristics
of the system. In particular, we consider systems which are identical
in every respect except for the value of Poisson's ratio. By virtue of
eq.~(\ref{Eq:CoreEnergy}), different choices of Poisson's ratio result
in different ratios between the energy per unit length of edge and
screw segments. We recall that, for $\nu > 0$, the minimum (maximum)
energy per unit length is attained for pure screw (edge) segments. All
remaining parameters are as in the preceding examples. In particular,
the unit cell dimension is $100 b$, the number of obstacles is $100$,
the obstacle strength is $f = 10 \mu b^2$, Poisson's ratio $\nu$ is
$0.3$, and the Peierls stress $\tau^P$ is set to $0$.

Fig.~\ref{fig:aniso} shows the predicted stress-strain curves and
dislocation density evolution for $\nu = $ 0.0, 0.3 and 0.5.
Since the saturation stress $\tau_0$ solely depends on obstacle
strength and concentration, all three stress-strain curves
saturate at the same level. By contrast, as $\nu$ is increased
the dislocation line-energy of both edge and screw segments
increases and the system correspondingly stiffens, with the
result that lower values of the slip strain are attained,
Fig.~\ref{fig:aniso}a. The rate of dislocation multiplication
also decreases with increasing $\nu$, Fig.~\ref{fig:aniso}b.
Perhaps more interestingly, Fig.~\ref{fig:aniso-l} shows the
effect of the anisotropy on the dislocation geometry. Thus, as
Poisson's ratio is increased screw segments become energetically
favorable relative to edge segments, and the dislocation pattern
exhibits a preponderance of elongated screw segments,
Fig.~\ref{fig:aniso-l}c.

\section{Summary and Concluding remarks}
\label{ConcludingRemarks}

We have developed an analytically tractable phase-field theory of
dislocation dynamics, strain hardening and hysteresis in ductile
single crystals at low temperatures. The phase-field
representation furnishes a simple and effective means of tracking
the motion of large numbers of dislocations within discrete slip
planes through random arrays of point obstacles under the action
of an applied shear stress. The theory rests on a variational
framework for dissipative systems and accounts for energetic and
kinetic effects  \cite{OrtizRepetto1999, RadovitzkyOrtiz1999,
OrtizRepettoStainier2000, OrtizStainier1999,
KaneMarsdenOrtizWest:2000}. The energetics accounted for in the
theory include: the core energy of the dislocations, represented
by a piecewise-quadratic Peierls potential
\cite{OrtizPhillips:1999}; the long-range elastic interactions
between primary dislocations and between primary and forest
dislocations; and the energy of interaction with the applied
resolved shear stress. The kinetics of the system stem from the
assumed irreversible interactions between dislocations and
obstacles, and from lattice friction, and result in hardening,
path dependency, and hysteresis in the macroscopic behavior.

The theory predicts a range of behaviors which are in qualitative
agreement with observation. Thus, in the base case of single slip
under monotonic loading, the theory predicts saturation at a
stress which depends on obstacle strength and density. It also
predicts dislocation multiplication at an initial parabolic rate,
during a `micro-slip' regime, followed by saturation at a maximum
dislocation density. The ease with which the phase-field
representation allows for complex geometrical and topological
transitions, as the dislocation ensemble percolates through the
obstacles, is quite remarkable. The cyclic behavior predicted by
the theory is particularly noteworthy. As required, the theory
predicts the essential phenomena of loading/unloading
irreversibility and hysteresis. But the theory also captures more
subtle aspects of the cyclic behavior of crystals, such as the
Bauschinger effect, consisting of the premature yielding of a
slip system under reverse loading; the fading memory effect,
whereby reverse yielding gradually eliminates the influence of
previous loading; and the evolution of the dislocation density,
leading to characteristic `butterfly' curves also observed in
magnetic systems \cite{kinderlehrer:1997, sethna:1993,
dahmen:1994}.

There are other aspects of the theory which help to make contact
with actual materials. For instance, it is possible within the
theory to account for obstacles of different species, and to
accord to them varying densities and strengths. This furnishes an
avenue for building into the theory results from recent work
concerned with dislocation-dislocation interactions
\cite{baskes:1998, RodneyPhillips1999, PhillipsRodneyShenoy1999,
ShenoyKuktaPhillips2000}. However, in order to build additional
physical realism into the theory a number of extensions and
enhancements immediately suggest themselves, to wit:

\begin{enumerate}

\item It should be possible, within the general framework
outlined in Section~\ref{Sec:DislocationEnergies}, to formulate a
fully three-dimensional theory of crystallographic slip
accounting for the full complement of slip systems in a crystal
class and, e.~g., an infinite stack of uniformly-spaced slip
planes within each slip system.

\item A topological restriction of the phase-field representation
of crystallographic slip, as presented here, is that, since the
dislocation lines follow the level contours of a scalar phase
field, they must necessarily trace closed loops or form networks
of zero net Burgers vector. This topological constraint precludes
the effective representation of cross slip, and therefore points
to the need for non-trivial mathematical extensions of the
theory.

\item In the formulation presented in this paper, the inelastic
interaction between dislocations and the crystal lattice, and,
between dislocations and obstacles, has been modeled as a
rate-independent frictional interaction in the interest of
analytical tractability. However, the incremental variational
framework developed in \cite{OrtizRepetto1999,
RadovitzkyOrtiz1999, OrtizRepettoStainier2000, OrtizStainier1999,
KaneMarsdenOrtizWest:2000} is general enough to enable
consideration of more complex kinetic relations, e.~g.,
accounting for the temperature and strain-rate dependence of the
Peierls stress.

\item As presented here, the theory accounts for the
edge/screw line-energy anisotropy predicted by linear elasticity.
However, this type of anisotropy does not suffice to model the
vastly different behavior of screw and edge segments observed in
bcc crystals, which largely owes to differences in the mobility
of both types of segments. Within the present theory, a proper
accounting of this effect can be achieved simply by allowing the
Peierls stress to depend on the gradient $\nabla\xi$ of the phase
field, which describes the local orientation of the dislocation
segments within the slip plane.

\end{enumerate}

Finally, the fact that the theory permits the reconstruction of
the full dislocation ensemble from the value of the phase field
at the obstacle sites can hardly be overemphasized. Indeed, the
geometry of each dislocation line is given by the theory
analytically and in close form. In addition, the same analytical
treatment endows each dislocation with a well-defined core, which
effectively regularizes --- and eliminates the divergences which
afflict --- linear elastic dislocation theory. The practical
effect of this analytical tractability is the elimination of any
need for the use of numerical grids to discretize the slip plane
or, worse, the entire crystal. Since the theory targets the
equilibrium configurations of the dislocation ensemble directly,
no costly resolution of transients, often introduced as numerical
artifacts in order to advance the solution in time and requiring
the use of exceedingly small time steps, is required. These
attributes render the complexity of the calculations required by
the theory, in its present state of development, commensurate
with the number of obstacles. This drastic reduction in
complexity may open the way for embedding the theory within a
finite-deformation formulation of single-crystal elastic-plastic
behavior (e.~g., \cite{OrtizStainier1999,cuitino:1992b}), with a
view to its use in large-scale finite element calculations of
macroscopic samples.

\section*{Acknowledgments}

The support of the DOE through Caltech's ASCI Center for the
Simulation of the Dynamic Response of Materials is gratefully
acknowledged.


\begin{thebibliography}{10}

\bibitem{Ortiz1999}
M.~Ortiz.
\newblock Plastic yielding as a phase transition.
\newblock {\em Journal of Applied Mechanics-Transactions of the ASME},
  66(2):289--298, 1999.

\bibitem{Wang:2001}
{Y.U} Wang, {Y.M} Jin, {A.M.} Cuiti\~no, and {A. G.} Khachaturyan.
\newblock Phase field microelasticity theory and modeling of multiple
  dislocation dynamics.
\newblock {\em Applied Physics Letters}, 78(16):2324--2326, 2001.

\bibitem{Ghoniem:2000}
{N. M.} Ghoniem, {S.H} Tong, and {L.Z.} Sun.
\newblock Parametric dislocation dynamics: A thermodynamics-based approach to
  investigations of mesoscopic plastic deformation.
\newblock {\em Physical Review B}, 61(2):913--927, 2000.

\bibitem{OrtizRepetto1999}
M.~Ortiz and {E.A.} Repetto.
\newblock Nonconvex energy minimization and dislocation structures in ductile
  single crystals.
\newblock {\em Journal of the Mechanics and Physics of Solids}, 47(2):397--462,
  1999.

\bibitem{RadovitzkyOrtiz1999}
R.~Radovitzky and M.~Ortiz.
\newblock Error estimation and adaptive meshing in strongly nonlinear dynamic
  problems.
\newblock {\em Computer Methods in Applied Mechanics and Engineering},
  172(1-4):203--240, 1999.

\bibitem{OrtizRepettoStainier2000}
M.~Ortiz, {E.A.} Repetto, and L.~Stainier.
\newblock A theory of subgrain dislocation structures.
\newblock {\em Journal of the Mechanics and Physics of Solids},
  48(10):2077--2114, 2000.

\bibitem{OrtizStainier1999}
M.~Ortiz and L.~Stainier.
\newblock The variational formulation of viscoplastic constitutive updates.
\newblock {\em Computer Methods in Applied Mechanics and Engineering},
  171(3-4):419--444, 1999.

\bibitem{KaneMarsdenOrtizWest:2000}
C.~Kane, J.~E. Marsden, M~Ortiz, and M.~West.
\newblock Varaitional integrators and the newmark algorithm for conservative
  and dissipative mechanical systems.
\newblock {\em International Journal for Numerical Methods in Engineering},
  49:1295--1325, 2000.

\bibitem{OrtizPhillips:1999}
M.~Ortiz and R.~Phillips.
\newblock Nanomechanics of defects in solids.
\newblock {\em Advances in Applied Mechanics}, 36:1--79, 1999.

\bibitem{baskes:1998}
{M.I.} Baskes, {R.G.} Hoagland, and T.~Tsuji.
\newblock An atomistic study of the strength of an extended-dislocation
  barrier.
\newblock {\em Modelling and Simulation in Materials Science and Engineering},
  6(1):9--18, 1998.

\bibitem{RodneyPhillips1999}
D.~Rodney and R.~Phillips.
\newblock Structure and strength of dislocation junctions: An atomic level
  analysis.
\newblock {\em Physical Review Letters}, 82(8):1704--1707, 1999.

\bibitem{ShenoyKuktaPhillips2000}
{V.B.} Shenoy, {R.V.} Kukta, and R.~Phillips.
\newblock Mesoscopic analysis of structure and strength of dislocation
  junctions in fcc metals.
\newblock {\em Physical Review Letters}, 84(7):1491--1494, 2000.

\bibitem{xu:1993}
G.~Xu and M.~Ortiz.
\newblock A variational boundary integral method for the analysis of
  three-dimensional cracks of arbitrary geometry modeled as continuous
  distributions of dislocation loops.
\newblock {\em International Journal for Numerical Methods in Engineering},
  36:3675--3701, 1993.

\bibitem{Xu:2000}
G.~Xu.
\newblock A variational boundary integral method for the analysis of
  three-dimensional cracks of arbitrary geometry in anisotropic elastic solids.
\newblock {\em Journal of Applied Mechanics, ASME}, 67:403--408, 2000.

\bibitem{Kroner:1958}
E.~Kr\"oner.
\newblock {Berechnung der elastischen Konstanten des Vielkristalls aus den
  Konstanten des Einkristalls}.
\newblock {\em Zeitung der Physik}, 151:504--518, 1958.

\bibitem{mura:1987}
T.~Mura.
\newblock {\em Micromechanics of defects in solids}.
\newblock Kluwer Academic Publishers, Boston, 1987.

\bibitem{Nye:1953}
J.~F. Nye.
\newblock {Some geometrical relations in dislocated crystals}.
\newblock {\em Acta Metallurgica}, 1:153--162, 1953.

\bibitem{Rice1992}
{J.R.} Rice.
\newblock Dislocation nucleation from a crack tip - an analysis based on the
  peierls concept.
\newblock {\em Journal of the Mechanics and Physics of Solids}, 40(2):239--271,
  1992.

\bibitem{SunBeltzRice1993}
{Y.M.} Sun, {G.E.} Beltz, and {J.R.} Rice.
\newblock Estimates from atomic models of tension shear coupling in dislocation
  nucleation from a crack tip.
\newblock {\em Materials Science and Engineering}, A170(1-2):67--85, 1993.

\bibitem{YamaguchiVitekPope1981}
{M.} Yamaguchi, {V.} Vitek, and {D.} Pope.
\newblock Planar faults in the li$_2$ lattice, stability and structure.
\newblock {\em Philosophical Magazine}, 43:1027, 1981.

\bibitem{SunRiceTruskinovsky1991}
{Y.} Sun, {J.~R.} Rice, and {L.} Truskinovsky.
\newblock {\em High-Temperature ordered intermetallic alloys}, volume 213,
  chapter Dislocation nucleation {\it versus} cleavage in Ni$_3$Al and Ni,
  pages 243--248.
\newblock Materials Research Society, 1991.

\bibitem{JuanKaxiras1996}
{Y.M.} Juan and E.~Kaxiras.
\newblock Generalized stacking fault energy surfaces and dislocation properties
  of silicon: A first- principles theoretical study.
\newblock {\em Philosophical Magazine}, A74(6):1367--1384, 1996.

\bibitem{JuanSunKaxiras1996}
{Y.M.} Juan, {Y.M.} Sun, and E.~Kaxiras.
\newblock Ledge effects on dislocation emission from a crack tip: A
  first-principles study for silicon.
\newblock {\em Philosophical Magazine Letters}, 73(5):233--240, 1996.

\bibitem{hirth:1968}
J.~P. Hirth and J.~Lothe.
\newblock {\em Theory of Dislocations}.
\newblock McGraw-Hill, New York, 1968.

\bibitem{Mathai:1973}
A.~M. Mathai and R.~K. Saxena.
\newblock {\em Generalized Hypergeometric Functions with Applications in
  Statistics and Physical Sciences}.
\newblock Kluwer Academic Publishers, Boston, 1987.

\bibitem{Clarke:1983}
F.~H. Clarke.
\newblock {\em Optimization and Nonsmooth Analysis}.
\newblock John Wiley \& Sons, New York, 1983.

\bibitem{Rockafellar1970}
{R.Y.} Rockafellar.
\newblock {\em Convex Analysis}.
\newblock Princeton University Press, Princeton, {N.J.}, 1970.

\bibitem{Hardikar:2001}
K.~Hardikar, V.~Shenoy, and R.~Phillips.
\newblock {Reconciliation of atomic-level and continuum notions concerning the
  interaction of dislocations and obstacles}.
\newblock {\em Journal of the Mechanics and Physics of Solids}, 49:1951--1967,
  2001.

\bibitem{sethna:1993}
J.~P. Sethna, K.~Dahmen, S.~Kartha, Krumhansl~J. A., B.~W. Rpberts, and
  Shore~J. D.
\newblock Hysteresis and hierarchies: dynamics of disorder-driven first-order
  phase transformations.
\newblock {\em Physical Review Letters}, 70:3347--3350, 1993.

\bibitem{ashby:1972}
M.~F. Ashby.
\newblock The deformation of plastically non-homogeneous alloys.
\newblock In A.~Kelly and R.~B. Nicholson, editors, {\em Strengthening Methods
  in Crystals}, pages 137--192. Wiley, 1972.

\bibitem{livingston:1962}
J.~D. Livingston.
\newblock The density and distribution of dislocations in deformed copper
  crystals.
\newblock {\em Acta Metallurgica}, 10:229--239, 1962.

\bibitem{ortiz:1982}
M.~Ortiz and E.~P. Popov.
\newblock A statistical theory of polycrystalline plasticity.
\newblock {\em Proceedings of the Royal Society of London}, A379:439--458,
  1982.

\bibitem{Morrow:1975}
{J.~D.} Morrow.
\newblock Unpublished test results, 1975.

\bibitem{BonderLindenfeld:1995}
{S. R.} Bodner and {A.} Lindenfeld.
\newblock Constitutive modeling of the stored energy of cold work under cyclic
  loading.
\newblock {\em European Journal of Mechanics, A/Solids}, 14(3):333--348, 1995.

\bibitem{kinderlehrer:1997}
D.~Kinderlehrer and L.~Ma.
\newblock The hysteric event in the computation of magnetization.
\newblock {\em Journal of Nonlinear Science}, 23:101--128, 1997.

\bibitem{dahmen:1994}
K.~Dahmen, S.~Kartha, A.~Krumhansi, B.~W. Roberts, Sethna~J. P., and Shore~J.
  D.
\newblock Disorder-driven first-order phase transformations: A model of
  hysteresis.
\newblock {\em Journal of Applied Physics}, 75:5946--5948, 1994.

\bibitem{PhillipsRodneyShenoy1999}
R.~Phillips, D.~Rodney, V.~Shenoy, E.~Tadmor, and M.~Ortiz.
\newblock Hierarchical models of plasticity: dislocation nucleation and
  interaction.
\newblock {\em Modelling and Simulation in Materials Science and Engineering},
  7(5):769--780, 1999.

\bibitem{cuitino:1992b}
A.~M. Cuiti\~no and M.~Ortiz.
\newblock Computational modeling of single crystals.
\newblock {\em Modelling and Simulation in Materials Science and Engineering},
  1:255--263, 1992.

\end{thebibliography}
\end{document}